\documentclass[modern]{aastex631} 

\newcommand{\ldl}{$\lambda/\Delta\lambda$}
\newcommand{\teff}{$T_{\rm eff}$}
\newcommand{\logg}{$\log{g}$}
\newcommand{\logk}{$\log\kappa_{zz}$}

\usepackage{chemformula}
\accepted{May 23, 2025}

\submitjournal{ApJ}

\shorttitle{Brown Dwarf Model Fitting Methods}
\shortauthors{Lueber \& Burgasser}

\graphicspath{{./}{figures/}}

\begin{document}

\title{Comparing Grid Model Fitting Methodologies for Low-Temperature Atmospheres: Markov Chain Monte Carlo versus Random Forest Retrieval}

\correspondingauthor{Anna Lueber}
\email{anna.lueber@physik.uni-muenchen.de}

\author[0000-0001-6960-0256]{Anna Lueber}
\affiliation{Ludwig Maximilian University, Faculty of Physics, University Observatory, Scheinerstr. 1, Munich D-81679, Germany}
\affiliation{University of Bern, Center for Space and Habitability, Gesellschaftsstrasse 6, CH-3012, Bern, Switzerland}

\author[0000-0002-6523-9536]{Adam J.\ Burgasser}
\affiliation{Department of Astronomy \& Astrophysics, UC San Diego, La Jolla, CA, USA}

\begin{abstract}
The atmospheres of low-temperature stars, brown dwarfs, and exoplanets are challenging to model due to strong molecular features and complex gas and condensate chemistry. Self-consistent atmosphere models are commonly used for spectral fitting, but computational limits restrict the production of finely-sampled multi-dimensional parameter grids, necessitating interpolation methods to infer precise parameters and uncertainties. Here, we compare two grid-model fitting approaches: a Markov Chain Monte Carlo (MCMC) algorithm interpolating across spectral fluxes, and a Random Forest Retrieval (RFR) algorithm trained on a grid model set. We test these with three low-temperature model grids—Sonora Diamondback, Sonora Elf Owl, and Spectral ANalog of Dwarfs (SAND)—and a sample of eleven L and T dwarf companions to FGKM stars with known distances, compositions, and ages. Diamondback models are optimal for early- and mid-type L dwarfs, Elf Owl for mid- and late T dwarfs, and SAND for young L dwarfs and L/T transition objects. The MCMC approach yields higher fit quality and more precise parameters, though best-fit parameters are generally consistent between approaches. RFR analysis is orders of magnitude faster after training. Both approaches yield mixed results when comparing fit parameters to expected values based on primary (metallicity and surface gravity) or evolutionary models (temperature and radius). We propose modeling low-temperature spectra efficiently by first fitting multiple model sets using RFR, followed by a more accurate MCMC assessment, to accelerate improved grid development.
\end{abstract}


\section{Introduction} \label{sec:intro}

The atmospheres of the coldest stars, brown dwarfs, and exoplanets
({\teff} $\lesssim$ 2500~K) give rise to complex spectra, shaped by strong molecular bands and continuum features shaped by complex gas and condensate chemistry modulated by a wide range of elemental abundances and non-equilibrium processes.
These sources pose unique atmospheric characterization challenges that blend planetary and stellar approaches \citep{1997ARA&A..35..137A,2001RvMP...73..719B,2015ARA&A..53..279M}, and the increased availability of high fidelity spectra of these sources, particularly low-temperature brown dwarfs, have driven advances in 
treatments of condensation and cloud formation \citep{2001ApJ...556..872A,2001ApJ...556..357A,2001A&A...376..194H,2008ApJ...675L.105H},
vertical mixing and non-equilibrium chemistry \citep{1985ApJ...299.1067F,1999ApJ...519L..85G,2006ApJ...647..552S},
non-solar abundances \citep{2012ApJ...758...36M,2024ApJ...961..139G},
and modifications to convective heat transport \citep{2015ApJ...804L..17T,2016ApJ...817L..19T},
among others.
Nevertheless, there remain distinct challenges to accurately reproducing the spectra of cool brown dwarfs,
including the rapid evolution of clouds at the L dwarf/T dwarf transition \citep{2006ApJ...640.1063B,2018ApJ...854..172C},
molecular abundances that exceed or fall short of model predictions (e.g., CO$_2$ and PH$_3$; \citealt{2014ApJ...793...47S,2024ApJ...973...60B}),
the complex spectra of cold Y dwarfs \citep{2021ApJ...918...11L}, 
and the untested chemistry of metal-poor brown dwarfs \citep{2009ApJ...697..148B,2020ApJ...898...77S,2022A&A...663A..84L}.

Modeling low temperature spectra and inferring atmospheric parameters is 
commonly achieved through comparison to pre-computed, self-consistent grids of atmosphere models,
either through the identification of a single ``best-fit'' model (e.g., \citealt{2008ApJ...678.1372C,2009ApJ...702..154S,2012A&A...540A..85P})
or robust inference of parameter distributions by interpolation across model parameters 
by principal component analysis (PCA; \citealt{2021ApJ...916...53Z}) 
or Markov Chain Monte Carlo methods (MCMC; \citealt{2016AJ....151...46A,2024ApJ...971L..25B}).
Depending on the source examined and models deployed, these approaches can lead to varying degrees of accuracy in the
reproduction of observed spectra, with corresponding systematic uncertainties and biases in
inferred parameters \citep{2021ApJ...921...95Z,2024arXiv241101378B}.
An alternative approach is retrieval modeling, in which radiative transfer models are computed continuously with varying pressure-temperature profiles and molecular abundances \citep{2008JQSRT.109.1136I,2009ApJ...707...24M,2014ApJ...793...33L,2017MNRAS.470.1177B, 2020ApJ...890..174K, 2024ApJ...970...71X}. 
This resource-intensive approach results in accurate fits to observed spectra, but the lack of self-consistent energy balance or chemistry can make it difficult to interpret the atmospheric processes underlying the fit. 
A third approach that is growing in popularity is the application of various machine-learning (ML) models trained on atmosphere grids, which we refer to as ML-retrieval \citep{2016ApJ...820..107W, 2018NatAs...2..719M, 2019AJ....158...33C, 2020AJ....159..192F, 2022A&A...662A.108A, 2023A&A...672A.147V, 2024A&A...681A...3G}. 
This approach is highly efficient, making it well-suited to large and high-resolution spectral surveys that may be prohibitive with direct fit or retrieval methods.
However, ML retrievals can also be subject to potential biases inherent in the underlying grid model or the ML algorithm deployed. 
The challenges associated with deriving inferences from sparsely sampled, high-dimensional parameter spaces have been extensively studied in computer experiment design and analysis. Methodologies like Latin Hypercube Sampling and emulator-based interpolation, including Gaussian Process regression, have been developed to navigate these spaces while minimizing model evaluations \citep{santner2003design, fang2005design, kleijnen2015design}. Applications thereof can be found in e.g., \cite{2022ApJ...934...31F}.

While numerous studies have compared direct-fit and retrieval approaches to the fitting of brown dwarf and exoplanet spectra, few comparisons have been made between direct-fit and ML-retrieval methods with a focus on precision, accuracy, and efficiency in the spectral fits and inferred parameters. In this study, we compare an MCMC-based direct-fitting approach to a ML-retrieval approach based on the supervised machine learning method known as Random Forest \citep{2001MachL..45....5B}. In Section~\ref{sec:methods} we describe our empirical spectral sample of eleven benchmark cool stars and brown dwarfs with independently-constrained ages and compositions, the three atmospheric model grids used in our fits, and the two model-fit approaches explored in this study.
In Section~\ref{sec:results} we review the results of our fits, evaluating both the quality of reproduction of the spectra and the robustness of the inferred model parameters (uncertainties and physical reliability). 
We also examine in detail three case studies that span the temperature range of our sample.
In Section~\ref{sec:discussion} we discuss our outcomes through the lens of model fitting precision, accuracy, and efficiency.
Our main results are summarized in Section~\ref{sec:summary}.

\section{Methods} \label{sec:methods}

\subsection{Empirical Spectral Sample} \label{sec:sample}

\begin{figure}
\centering
\includegraphics[width=0.49\textwidth]{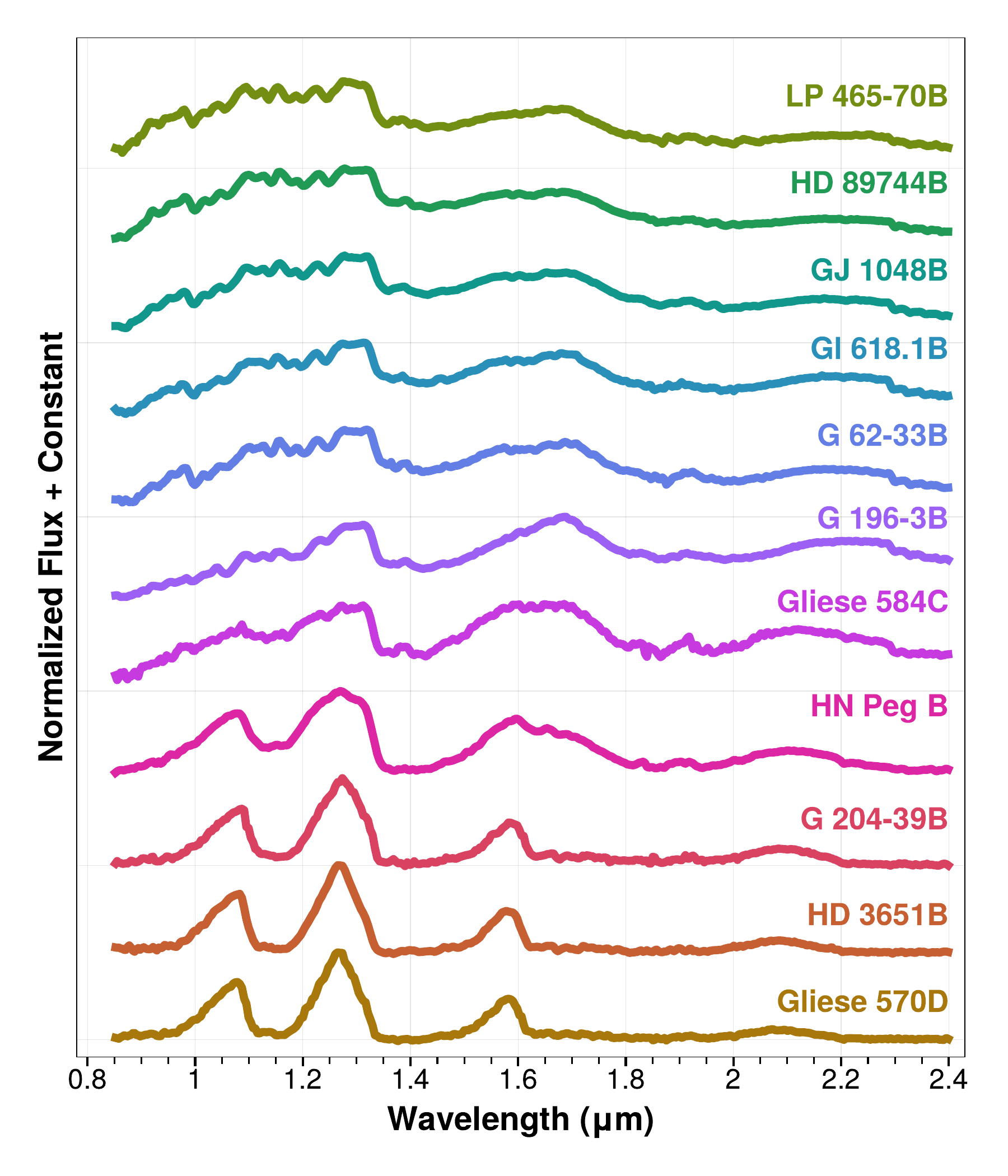}
\includegraphics[width=0.49\textwidth]{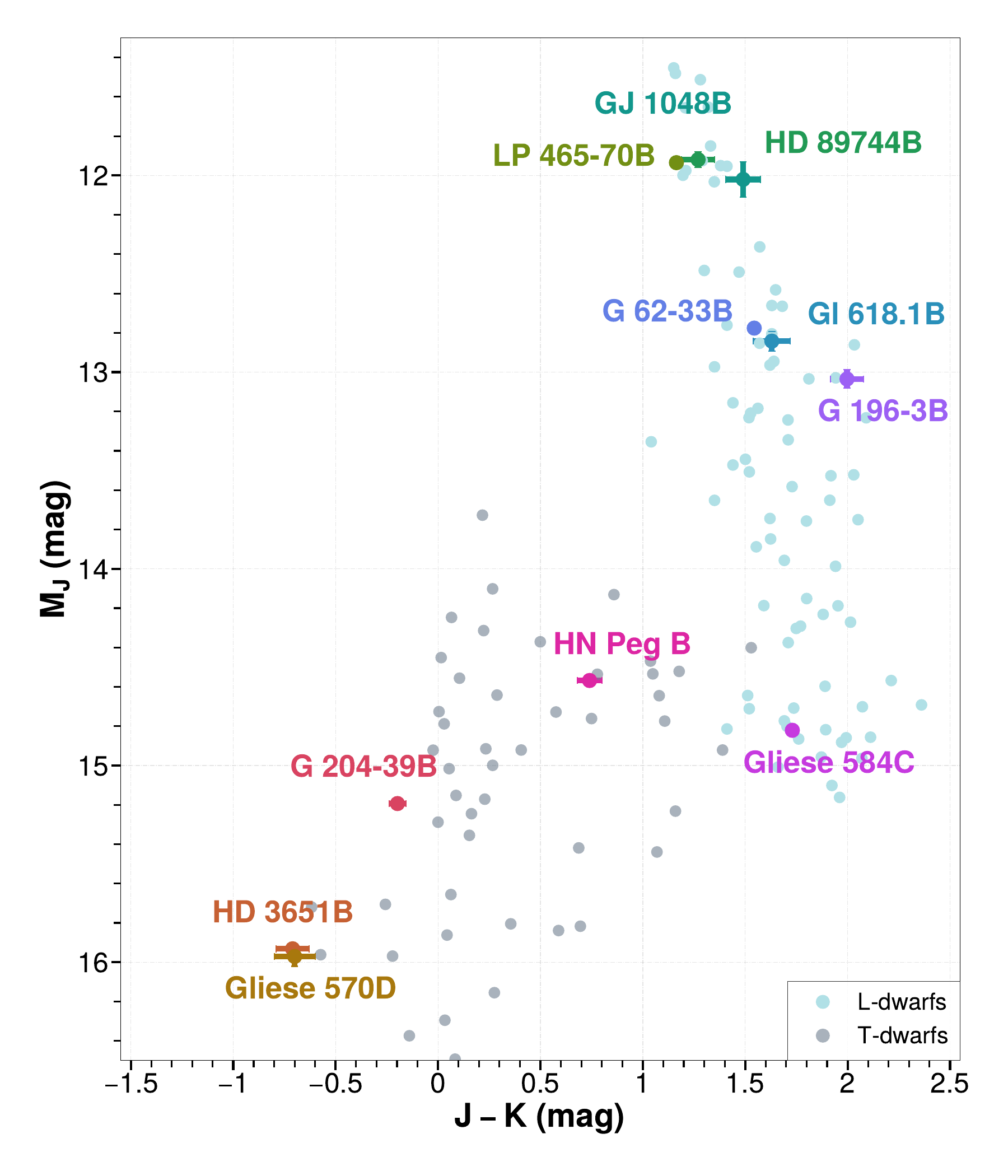}
\caption{Left: Near-infrared spectra of our curated sample of eleven L and T dwarf companions to main sequence stars (see Table~\ref{tab:sample}). All spectra are normalized at their maximum flux value and offset by a constant. The associated data uncertainties are included in the plot but are too small to be visible. Right: Color–magnitude diagram for our curated sample of L and T dwarfs, compared to a broader sample of L dwarfs (light blue circles) and T dwarfs (gray circles) with measured parallaxes compiled by \cite{2012ApJS..201...19D}. \label{fig:sequence}}
\end{figure}

Our analysis focused on a sample of eleven L and T dwarfs spanning spectral types L0 to T8 that are all widely-separated singular companions to main sequence stars. The sample is summarized in Table~\ref{tab:sample}, which includes prior estimates of effective temperatures ({\teff}), surface gravities ({\logg}) and radii from \citet{2023ApJ...959...63S} based on bolometric luminosities, system ages, and evolutionary models. We also list the physical properties of the primaries, notably metallicity ([Fe/H]) and age, as inferred by various methods in the cited references.  
All of the L and T dwarfs in the sample have previously published low-resolution ({\ldl} $\approx$ 150), near-infrared (0.8--2.5~$\mu$m) spectra obtained with SpeX instrument on the 3m NASA Infrared Telescope Facility \citep{2003PASP..115..362R},
with high values of signals-to-noise ranging over S/N = 40--270 in the 1.25--1.30~$\mu$m region, with an average S/N $\approx$ 100.
We retrieved these spectra from the SpeX Prism Library Analysis Toolkit (SPLAT; \citealt{2017ASInC..14....7B}), and flux calibrated them using each source's absolute $M_J$ magnitude based on photometry from the Two Micron All Sky Survey (2MASS; \citealt{2006AJ....131.1163S}) and the parallax of the primary from Gaia Data Release 3 \citep{2023AA...674A...1G}. One source, GJ~1048B, is obscured by its primary in the 2MASS $J$-band, so we used its 2MASS $K_s$-band photometry instead and computed a synthetic $J$ magnitude from the spectrum. 
Figure~\ref{fig:sequence} displays the sequence of these spectra in normalized $F_\lambda$ flux densities.

\begin{deluxetable}{llcccccllccl}
\rotate
\tablecaption{Benchmark Spectral Sample. \label{tab:sample}} 
\tabletypesize{\scriptsize}
\tablehead{ 
\multicolumn{7}{c}{Secondary} & \multicolumn{4}{c}{Primary} \\
\cline{1-7} \cline{8-12}
\colhead{Name} & 
\colhead{SpT} & 
\colhead{M$_J$} &
\colhead{M$_K$} &
\colhead{\teff\tablenotemark{a}} &
\colhead{\logg\tablenotemark{a}} &
\colhead{R\tablenotemark{a}} &
\colhead{Name} & 
\colhead{SpT} &
\colhead{[Fe/H]} &
\colhead{Age} &
\colhead{Ref} \\ 
\colhead{} & 
\colhead{} & 
\colhead{(mag)} &
\colhead{(mag)} &
\colhead{(K)} &
\colhead{(cm/s$^2$)} &
\colhead{(R$_{Jup}$)} &
\colhead{} &
\colhead{} &
\colhead{(dex)} &
\colhead{(Gyr)} &
\colhead{} \\ 
}
\startdata
LP~465-70B & L0 & 11.935$\pm$0.003 & 10.771$\pm$0.003 & 2148$\pm$11 & 5.34$\pm$0.01 & 0.99$\pm$0.01 & LP~465-70 & M4 & $-$0.120$\pm$0.006 & 4.5--10 & [1-6] \\
HD~89744B & L0 & 11.92$\pm$0.04 & 10.65$\pm$0.04 & 2302$\pm$19 & 5.35$\pm$0.01 & 0.98$\pm$0.01 & HD 89744 & F7 & $-$0.15$\pm$0.01 & 1.5-3 & [4,7-9] \\
GJ~1048B & L1 & 12.02$\pm$0.09\tablenotemark{b} & 10.53$\pm$0.08 & 2332$\pm$24 & 5.33$\pm$0.02 & 1.00$\pm$0.01 & GJ~1048 & K3.5 & 0.14$\pm$0.01 & 0.6--2 & [4,10-13] \\
Gl~618.1B & L2.5 & 12.84$\pm$0.05 & 11.21$\pm$0.04 & 1907$\pm$27 & 5.34$\pm$0.05 & 0.97$\pm$0.02 & MCC 760 & K7/M0 & 0.37$\pm$0.09 & 0.5-12 & [4,7,9,11,13] \\
G~62-33B & L2.5 & 12.776$\pm$0.005 & 11.232$\pm$0.005 & 1939$\pm$16 & 5.34$\pm$0.01 & 1.08$\pm$0.16 & HD~116012 & K0.5 & 0.061$\pm$0.005 & 3.3-5.1 & [2,4,6,14-15] \\
G~196-3B & L3$\beta$ & 13.03$\pm$0.05 & 11.04$\pm$0.03 & 1725$\pm$52 & 4.77$\pm$0.15 & 1.18$\pm$0.07 & G~196-3 & M3 & \nodata & 0.05--0.10 & [4,16-19] \\
Gliese~584C & L8 & 14.82$\pm$0.02 & 13.09$\pm$0.02 & 1237$\pm$24 & 5.11$\pm$0.09 & 0.95$\pm$0.03 & Gliese 584AB & G2+G2 & $-$0.01$\pm$0.12 & 1--2.5 & [4,9,20-23] \\
HN~Peg~B & T2.5 & 14.57$\pm$0.03 & 13.83 $\pm$0.03 & 1056$\pm$38 & 4.53$\pm$0.22 & 1.14$\pm$0.08 & HN~Peg & G0 & $-$0.43$\pm$0.01 & 0.24$\pm$0.03 & [4,24-26] \\
G~204-39B & T6.5 & 15.193$\pm$0.011 & 15.39$\pm$0.03 & 898$\pm$26 & 4.87$\pm$0.16 & 0.99$\pm$0.05 & G~204-39 & M2.5 & $-$0.04$\pm$0.06 & 0.5-3 & [4,16,27,28] \\
HD~3651B & T7 & 16.08$\pm$0.03 & 16.64$\pm$0.05 & 785$\pm$12 & 5.22$\pm$0.06 & 0.85$\pm$0.02 & HD~3651 & K0.5 & $-$0.082$\pm$0.005 & 6.9$\pm$2.8  & [4,24,29,30] \\
Gliese~570D & T8 & 16.47$\pm$0.05 & 16.67$\pm$0.05 & 771$\pm$18 & 4.99$\pm$0.12 & 0.93$\pm$0.04 & Gliese~570A & K4 & $-$0.04$\pm$0.03 & 3.3$\pm$1.9 & [4,27,31-33]  \\ 
\enddata
\tablenotetext{a}{Estimated {\teff}, {\logg} and radii for L and T dwarf secondaries are adopted from \citet{2023ApJ...959...63S}, and are based on measured bolometric luminosities, system ages, and the evolutionary models of the \cite{2008ApJ...689.1327S} and \cite{2015AA...577A..42B}.}
\tablenotetext{b}{Estimated from spectrophotometric $J-K_s$ color synthesized from the infrared spectrum.}
\tablerefs{
[1] \citet{2002AJ....123.3409H};
[2] UKIDSS \citep{2007MNRAS.379.1599L};
[3] \citet{2013MNRAS.431.2745G};
[4] Gaia DR3 \citep{2023AA...674A...1G};
[5] \citet{2022AJ....163..152S};
[6] \citet{2010AJ....139..176F};
[7] \citet{2001ApJ...551L.163G}
[8] \citet{2012ApJS..201...19D};
[9] \citet{2024ApJ...960..105Z};
[10] \citet{2001AJ....121.2185G};
[11] \citet{2012ApJ...752...56F};
[12] \citet{2006AJ....132..161G};
[13] \citet{1986AJ.....92..139S};
[14] \citet{2014ApJ...794..143B};
[15] \citet{1999MSS...C05....0H};
[16] \citet{2009AJ....137.3345C};
[17] \citet{2021AJ....161...42B};
[18] \citet{2009ApJ...699..649S};
[19] \citet{2023AA...674A..66Z};
[20] \citet{2000AJ....120..447K};
[21] \citet{2023AJ....166..103S};
[22] \citet{1955ApJ...121..337S};
[23] \citet{2007AA...474..653V};
[24] \citet{2007ApJ...654..570L};
[25] \citet{2001AJ....121.2148G};
[26] \citet{2007ApJ...669.1167B};
[27] \citet{2006ApJ...637.1067B};
[28] \citet{2004AJ....127.3553K};
[29] \citet{2021AA...656A.162M};
[30] \citet{1989ApJS...71..245K};
[31] \citet{2016AA...585A...5B};
[32] \citet{2010ApJ...710.1627L};
[33] \citet{2021ApJ...921...95Z}.}
\end{deluxetable}

\subsection{Spectral Models} \label{sec:models}

We evaluated three state-of-the-art grids of atmosphere models that encompass differing treatments of condensation and cloud formation, non-equilibrium chemistry, and abundance variations, as summarized in Table~\ref{tab:models}.
The Sonora Diamondback model grid (DBack; \citealt{2024ApJ...975...59M}) 
is based on the radiative-convective equilibrium framework described in \citet{1989Icar...80...23M} and \citet{2008ApJ...689.1327S}, and assumes radiative–convective and chemical equilibrium throughout the atmosphere.
Sonora Diamondback incorporates condensation and cloud formation in both chemistry and opacity using the approach of \cite{2001ApJ...556..872A}, parameterizing the efficiency of condensation through a sedimentation efficiency parameter ($f_{sed}$). These models include a small range of solar-scaled metallicity variations ($-$0.5 $\leq$ [M/H] $\leq$ +0.5).
The Sonora Elf Owl model grid (EOwl; \citealt{ELFOWL}) is based on the PICASO model framework \citep{2023ApJ...942...71M} and is a cloud-free model that includes the effects of non-equilibrium chemistry (``quenching'') through vertical mixing as parameterized by the vertical diffusion coefficient ($\kappa_{zz}$).
This grid also allows for solar-scaled and non-solar abundances, that latter by varying the C/O ratio relative to an assumed solar (C/O)$_\odot$ = 0.458 \citep{2009LanB...4B..712L}.
The Spectral ANalog of Dwarfs (SAND) model grid \citep{SAND,2022ApJ...930...24G, 2024ApJ...961..139G} is based on the 
PHOENIX code \citep{1997ApJ...483..390H, 2011ASPC..448...91A, 2012RSPTA.370.2765A, 2020RNAAS...4..214G}, and is a chemical equilibrium model that incorporates condensation and cloud formation through a  prescribed gravitational settling algorithm (``hybrid''; \citealt{2012RSPTA.370.2765A}). SAND includes a broad range of metallicities and non-solar abundances, the latter quantified by alpha enrichment ([$\alpha$/Fe]).
For ease of computation, we down-selected these model grids to streamline our fits.
We reduced the number of models
by constraining them to ranges
600~K $\le$ {\teff} $\le$ 2400~K (extended to 3000~K for SAND), 
4.5 $\le$ {\logg} $\le$ 6.0 (units of cm/s$^2$), 
$-$0.5 $\le$ [M/H] $\le$ +0.5, and
and 0.5 $\le$ C/O $\le$ 1.5. These constraints are on top of the parameter limits in the original model grids. 
These constraints result in 864, 5175, and 952 models for Diamondback, Elf Owl, and SAND respectively.
All model spectra were Gaussian-smoothed to the resolution of the SpeX data and mapped onto a common wavelength scale by linear interpolation, and we 
conducted our fits in surface $F_\lambda$ flux densities.

\begin{deluxetable}{llccccccccc}
\tabletypesize{\scriptsize}
\tablecaption{Parameter Ranges for Spectral Model Grids
\label{tab:models}}
\tablehead{
\colhead{Model} & 
\colhead{\teff} & 
\colhead{\logg} &
\colhead{[$M/H$]} &
\colhead{(C/O)} &
\colhead{[$\alpha$/Fe]} &
\colhead{$\log\kappa_{zz}$} &
\colhead{$f_{sed}$} &
\colhead{\# models} &
\colhead{Ref.} \\ 
\colhead{} & 
\colhead{(K)} & 
\colhead{(cm/s$^2$)} &
\colhead{(dex)} &
\colhead{($\odot$)\tablenotemark{a}} &
\colhead{(dex)} & 
\colhead{(cm$^2$/s)} &
\colhead{} &
\colhead{} &
\colhead{}
}
\startdata
Diamondback & 900 -- 2400 & 4.5 -- 5.5 & $-$0.5 -- +0.5 & \nodata & \nodata & \nodata & 1 -- 8 & 864 & [1] \\
Elf Owl & 600 -- 2400 & 4.5 -- 5.5 & $-$0.5 -- +0.5 & 0.5 -- 1.5 & \nodata & 2 -- 9 & \nodata & 5175 & [2] \\
SAND & 700 -- 3000 & 4.5 -- 6.0 & $-$0.35 -- +0.3 & \nodata & $-$0.05 -- +0.4 & \nodata & \nodata & 952 & [3] \\
\enddata
\tablenotetext{a}{C/O ratio relative to solar abundance, with (C/O)$_\odot$ = 0.458 \citep{2009LanB...4B..712L}.}
\tablerefs{
[1] \citet{2024ApJ...975...59M}, 
[2] \citet{ELFOWL}, 
[3] \citet{SAND}}
\end{deluxetable}

\subsection{Modeling Approaches}

\subsubsection{Markov Chain Monte Carlo}

Our baseline modeling approach was a direct comparison of data and grid models to identify the best-fitting parameters and a robust estimate of their uncertainties. We deployed a Metropolis-Hastings MCMC algorithm \citep{1953JChPh..21.1087M,HASTINGS01041970}, starting from an initial estimate of the best model parameters by first comparing the spectrum to each of the individual models in the grid using a $\chi^2$ fitting statistic,
\begin{equation}
    \chi^2 = \sum_{i=1}^N\frac{(O[\lambda_i]-{\beta}M[\lambda_i])^2}{\sigma[\lambda_i]^2}.
    \label{eqn:chi1}
\end{equation}
Here, $N$ is the number of spectral data points, 
$O[\lambda_i]$ is the observed spectrum scaled to absolute flux density units,
$M[\lambda_i]$ is a model spectrum in surface flux density units,
$\sigma[\lambda_i]$ is the observed uncertainty spectrum, 
and $\beta$ is the optimal scale factor computed as
\begin{equation}\label{eqn:chiscale}
    \beta = \frac{\sum_{i=1}^N\frac{O[\lambda_i]M[\lambda_i]}{\sigma[\lambda_i]^2}}{\sum_{i=1}^N\frac{M[\lambda_i]^2}{\sigma[\lambda_i]^2}}
\end{equation}
(cf.\ \citealt{2008ApJ...678.1372C}).
Note that the spectral flux density scalings allow us to estimate the source radius from the scale factor as
\begin{equation}
    R_\beta = 10\sqrt\beta~{\rm pc} = 2.256\times10^{-7}\sqrt\beta~{\rm R_\odot}.
\end{equation}
The parameters of the model with the lowest $\chi^2$ residual were used as the starting point for a single chain of 5,000 steps.  
In each iteration, we drew new parameters using a Gaussian offset with standard deviations of $\sigma_{T_{eff}}$ = 25~K, $\sigma_{\log{g}}$ = 0.2 dex, $\sigma_{[M/H]}$ = 0.2 dex, $\sigma_{[\alpha/H]}$ = 0.2 dex, $\sigma_{C/O}$ = 0.05 dex, $\sigma_{\log\kappa_{zz}}$ = 0.2 dex, and $\sigma_{f_{sed}}$ = 0.25. 
Models between the grid points were linearly interpolated in logarithmic flux units on a logarithmic parameter grid (i.e., {\teff} $\Rightarrow$ $\log$ {\teff}).
Sequential fits ($i\Rightarrow{i+1}$) were compared using the criterion
\begin{equation}\label{eqn:mcmc}
    \frac{\chi^2(i+1)-\chi^2(i)}{\textrm{MIN}[\chi^2]} < \mathcal{U}(0,0.5)
\end{equation}
where MIN is the minimum of all $\chi^2$ values in the chain and $\mathcal{U}(0,0.5)$ is a number drawn from a uniform distribution between 0 and 0.5. 
If the new fit satisfied this criterion, these parameters were added to the chain; otherwise the previous parameters were added. We also enforced a limit on $\chi^2$ values exceeding 2$\times$MIN[$\chi^2$], at which point the chain reverted to the minimum $\chi^2$ parameter set. We verified that all fits converged using the convergence diagnostic defined by \citet{geweke1992}, requiring that the means of the variable parameters in the first 10\% and last 50\% of each chain differ by less than three times their combined variance. We also visually confirmed convergence, and examined each fit to ensure it provided an accurate reproduction of the data.

\subsubsection{Random Forest Retrieval}

Our second modeling approach uses the supervised machine-learning random forest algorithm \citep{Ho1998random, Breiman2001random} integrated in the
HELA framework \citep{2018NatAs...2..719M}, an approach we refer to here as random forest retrieval (RFR). 
The RF algorithm constructs an ensemble of decision trees, each trained on random subsets of pre-labeled training data, in this case a pre-existing grid of atmospheric model spectra labeled by the model parameters.
This ensemble approach allows for the modeling of complex, non-linear relationships between the input spectral data and the target atmospheric parameters with improved accuracy and robustness. This methodology represents a type of Approximate Bayesian Computation \citep{Sisson2018handbook}. 
Two of the key advantages of the RFR method are its computational efficiency and its ability to determine feature importance, a metric that ranks the significance of each spectral data point (referred to as ``features'') in constraining a model parameter. 
The HELA RFR framework has been applied in multiple retrieval studies, including the analysis of high-resolution spectra of the ultra-hot Jupiter KELT-9b \citep{2020AJ....159..192F}, medium-resolution spectra of benchmark brown dwarfs \citep{2020AJ....159....6O, 2023ApJ...954...22L}, the evaluation of information content in James Webb Space Telescope spectra \citep{2020AJ....160...15G, 2024A&A...690A.357L, 2024A&A...687A.110L}, and investigations into model grid sampling strategies \citep{2022ApJ...934...31F}.

The RF algorithm reduces the risk of over-fitting by aggregating predictions across multiple decision trees, enhancing the model's generalization capabilities and making it particularly effective for dealing with high-dimensional datasets. To address potential distortions in the predictions caused by sparse sampling, a bias correction was applied \citep{Cochran1977, Galassi2011}. Within our ML retrieval framework, frequency weights (\( w_i \geq 0, \forall i \)) are calculated as the number of occurrences of training set samples predicted by each decision tree.
The weighted mean ($\mu_w$) of a specific parameter is given by:
\begin{equation}
    \mu_w = \frac{\sum_{i=1}^{N} w_i x_i}{\sum_{i=1}^{N} w_i},
\end{equation}
where $\sum_{i=1}^{N} w_i = N^*$ represents the effective sample size. The weighted variance across the dataset is calculated as the sum of the squared deviations of each data point from the weighted mean, normalized by the effective sample size:
\begin{equation}
    \sigma_w^2 = \frac{\sum_{i=1}^{N} w_i (x_i - \mu_w)^2}{\sum_{i=1}^{N} w_i}.
\end{equation}
This method is particularly suitable for datasets with unequal variances or sampling biases, as it ensures that more reliable data points substantially influence the variability estimate, thereby reducing the impact of less representative samples.

In addition to varying the model parameters, we included the flux scaling factor $f$ which scales the model surface flux densities scaled at that expected brightness at Earth to the observed flux density of the source:
\begin{equation}\label{eq:radius-distance relation}
    O[\lambda]=f\left(\frac{{R}}{d}\right)^2M[\lambda]
\end{equation}
where $d$ and $R$ represent the distance and radius of the source.
As all observed spectra are scaled to their absolute flux densities, we initially assume $(R/d)^2 = (R_{\mathrm{Jup}}/10~\mathrm{pc})^2$ with a uniform distribution for $f$ ranging from 0.5 to 2.0. This scale factor is reported in terms of the inferred radii in solar units, $R = 0.0103\sqrt{f}~R_\odot$. For training, we randomly assigned an individual $f$ value to each spectrum and scaled the fluxes of the corresponding atmospheric model accordingly.

We trained three separate RFR models, one for each of our model grids.
To train the RFR model, the atmosphere model grid was partitioned into training and testing sets with an approximate ratio of 80:20 following prior research \citep{2018NatAs...2..719M, 2020AJ....159..192F, 2023ApJ...954...22L}. The efficacy of this partitioning can be assessed through real versus predicted (RvP) evaluations, which evaluates the coefficient of determination as a quantitative measure of efficacy,
\begin{equation}
    R^{2} = 1 - \frac{\sum_{i=1}^{n} (y_i -f_i)^2}{\sum_{i=1}^{n} (y_i - \bar{y})^2} = 1 - \frac{S_{res}}{S_{tot}},
\end{equation}
with $y_i$ and $f_i$ denoting the real and predicted values for a given parameter, respectively, while $\bar{y_i}$ represents the mean of the given real values. Therefore, $S_{res}$ is the sum of squares of the residual errors, and $S_{tot}$ is the total sum of the errors. $R^{2}$ can vary between negative and positive values, depending on the level of correlation and disparity between inferred and actual values, with perfect alignment yielding $R^{2}$ = 1 and no correlation yielding $R^{2}$ = 0. The presence of noise in data usually results in $R^2$ values below unity, even when the noise-free real versus predicted (RvP) relationship yields an $R^2$ of 1. Note that $R^{2}$ $<$ 0 indicates parameter relationships that are worse than random chance.

The hyperparameters of our RF models were adopted from \cite{2018NatAs...2..719M}. The models were constructed to contain 3,000 regression trees with no maximum tree depth constraint. Each tree grows by partitioning the model space until the decrease in variance with further splits is smaller than a fractional value of 0.01. No tree pruning was performed. Additionally, the maximum number of features considered for splitting at each node was limited to the square root of the number of spectral data points used.

\section{Results} \label{sec:results}

\subsection{Monte Carlo Markov Chain}

\begin{figure}[h]
\centering
\includegraphics[width=0.32\textwidth]{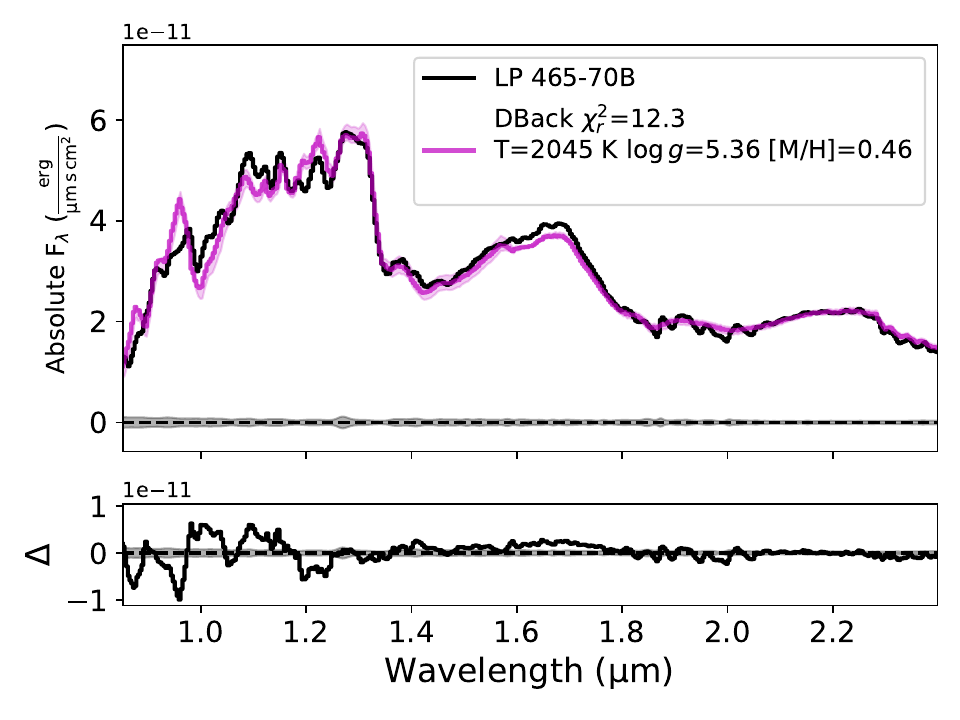}
\includegraphics[width=0.32\textwidth]{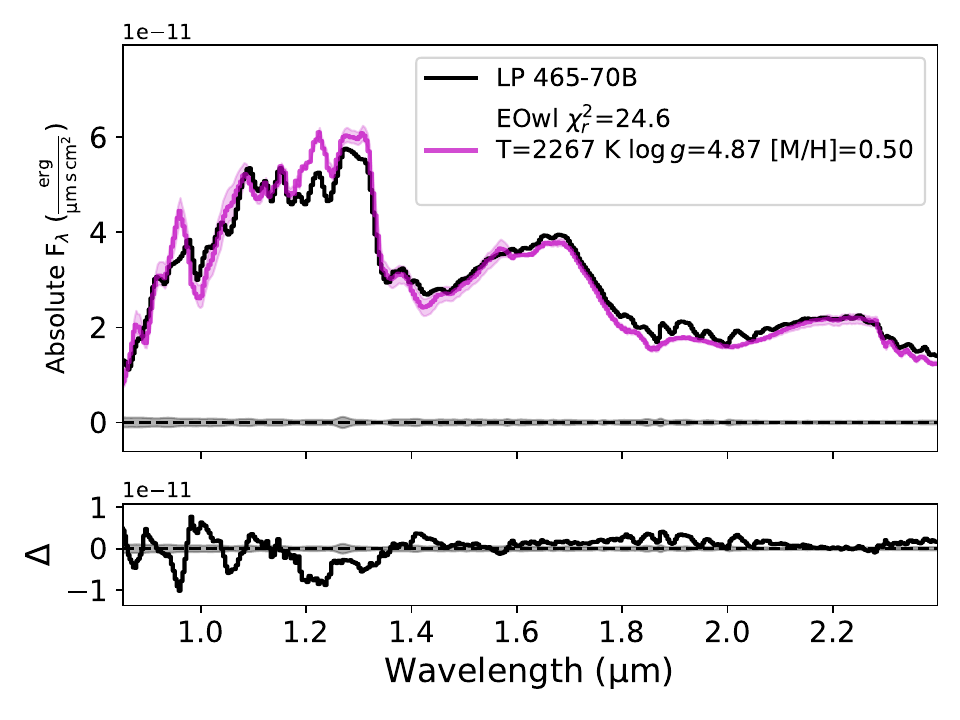}
\includegraphics[width=0.32\textwidth]{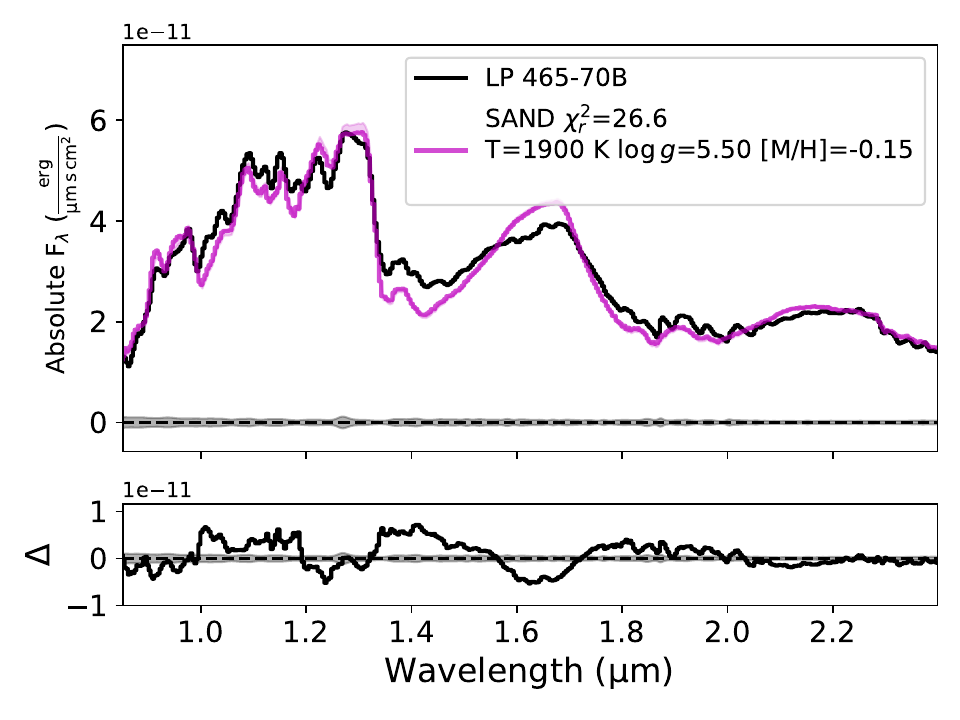} \\
\includegraphics[width=0.32\textwidth]{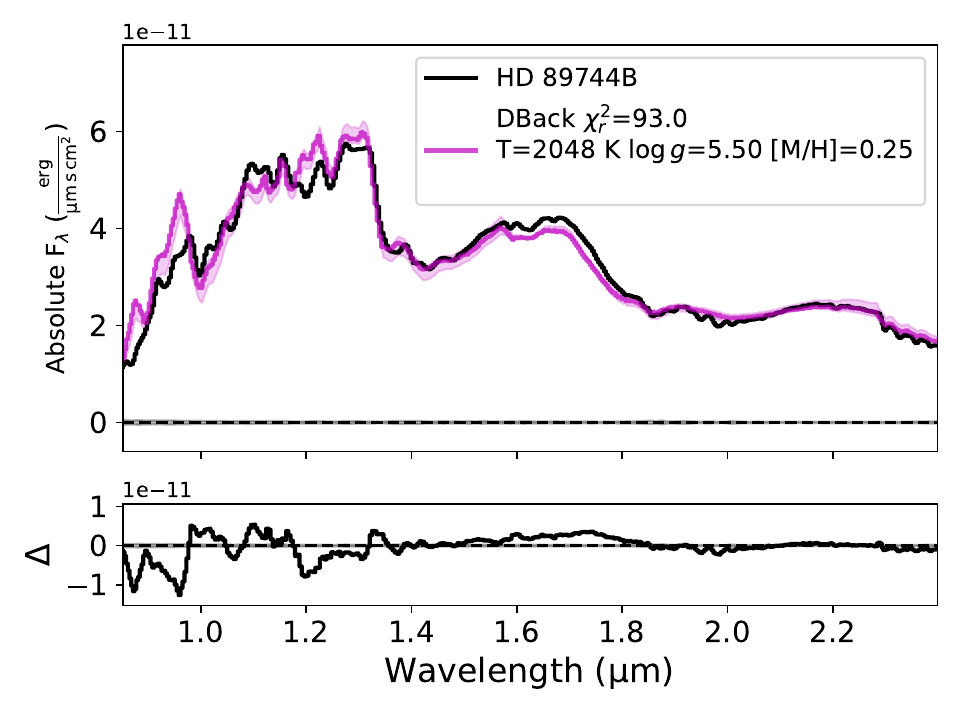} 
\includegraphics[width=0.32\textwidth]{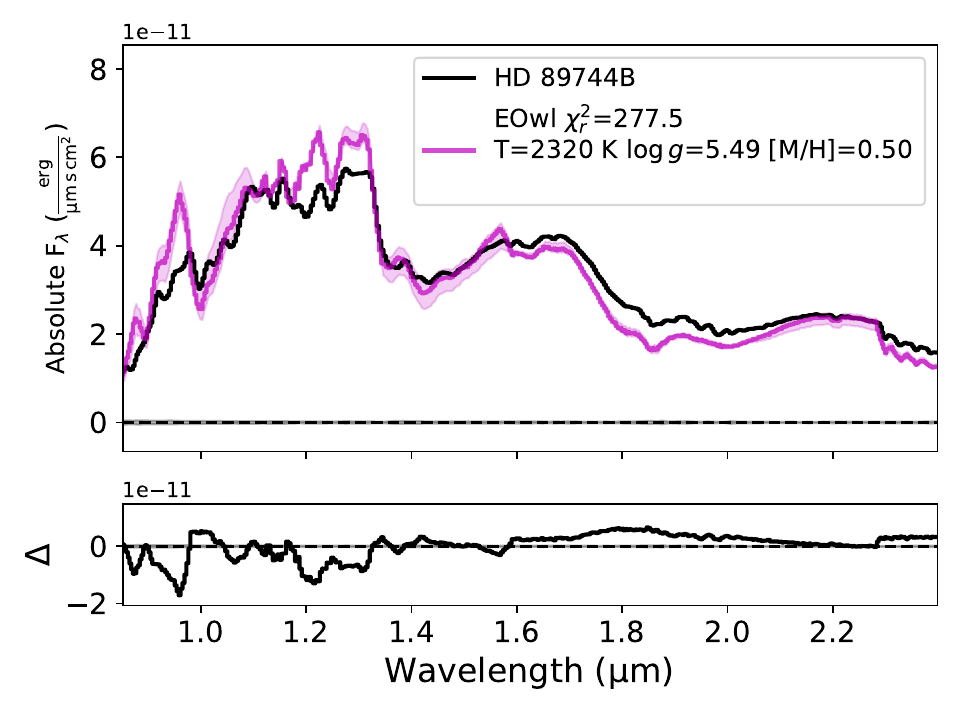} 
\includegraphics[width=0.32\textwidth]{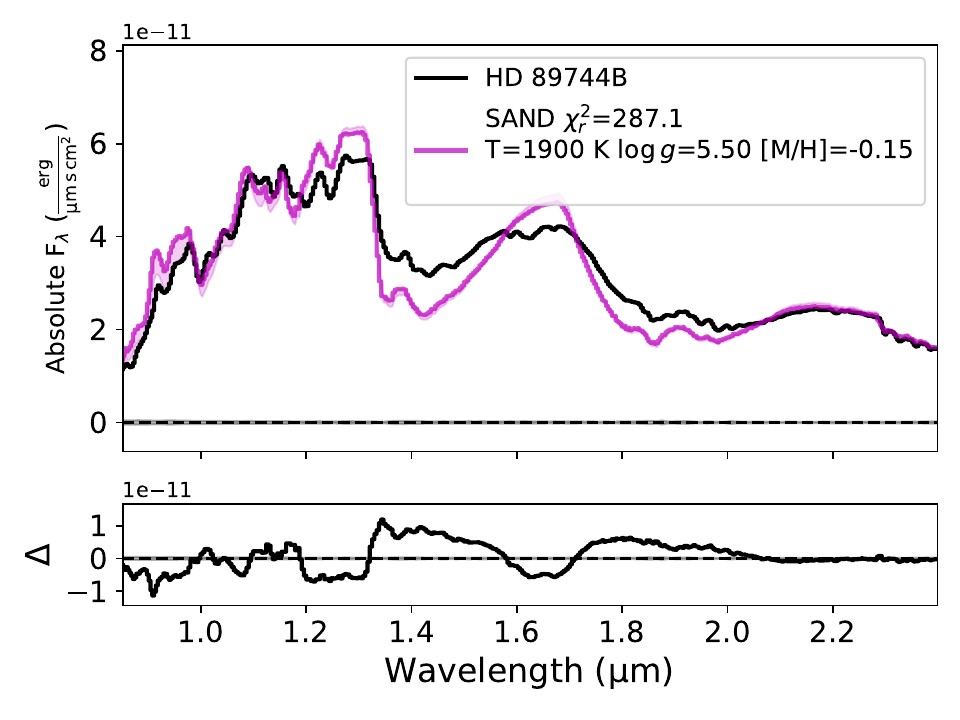} \\
\includegraphics[width=0.32\textwidth]{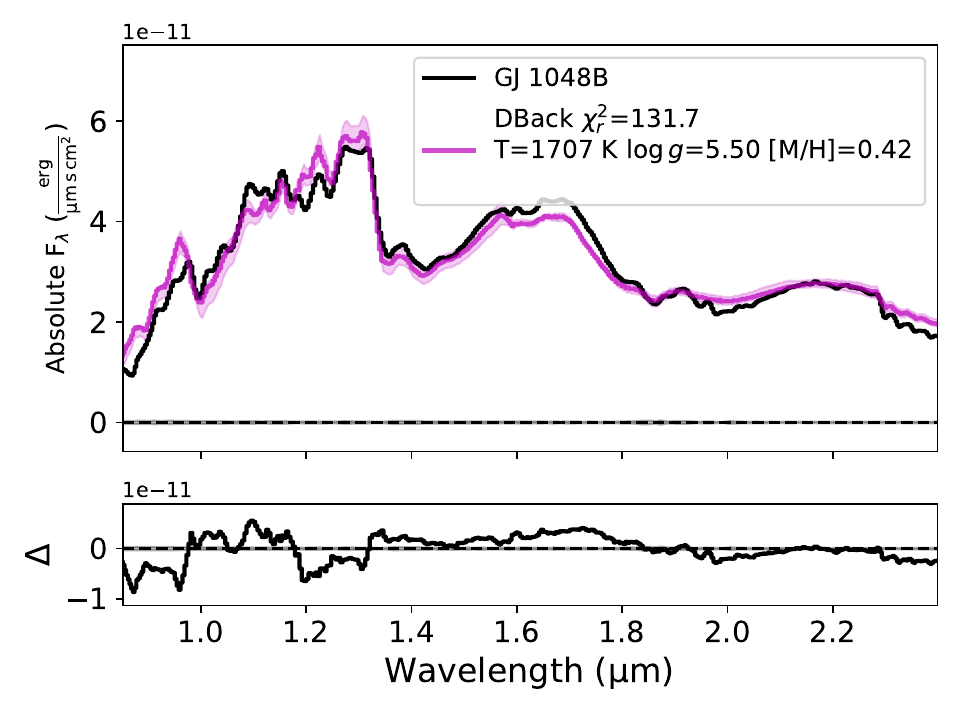} 
\includegraphics[width=0.32\textwidth]{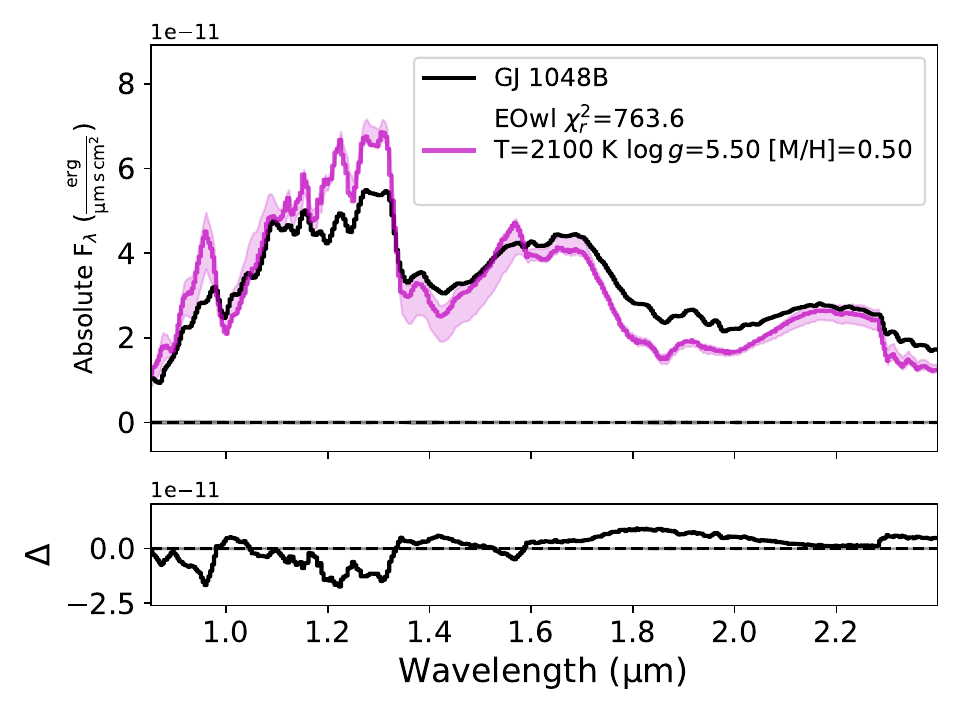} 
\includegraphics[width=0.32\textwidth]{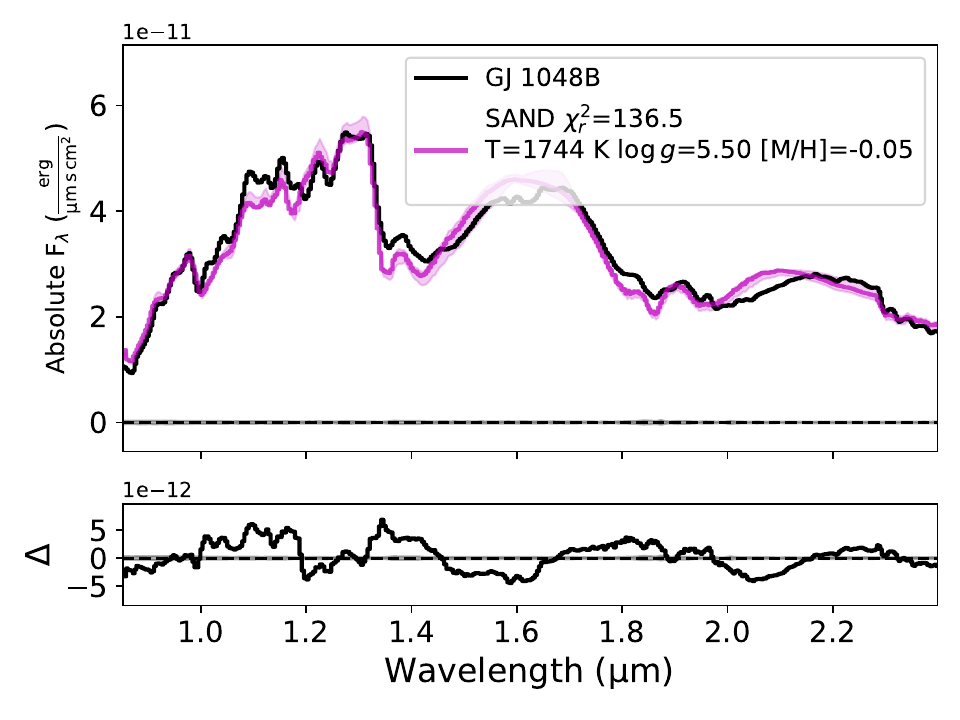} \\
\includegraphics[width=0.32\textwidth]{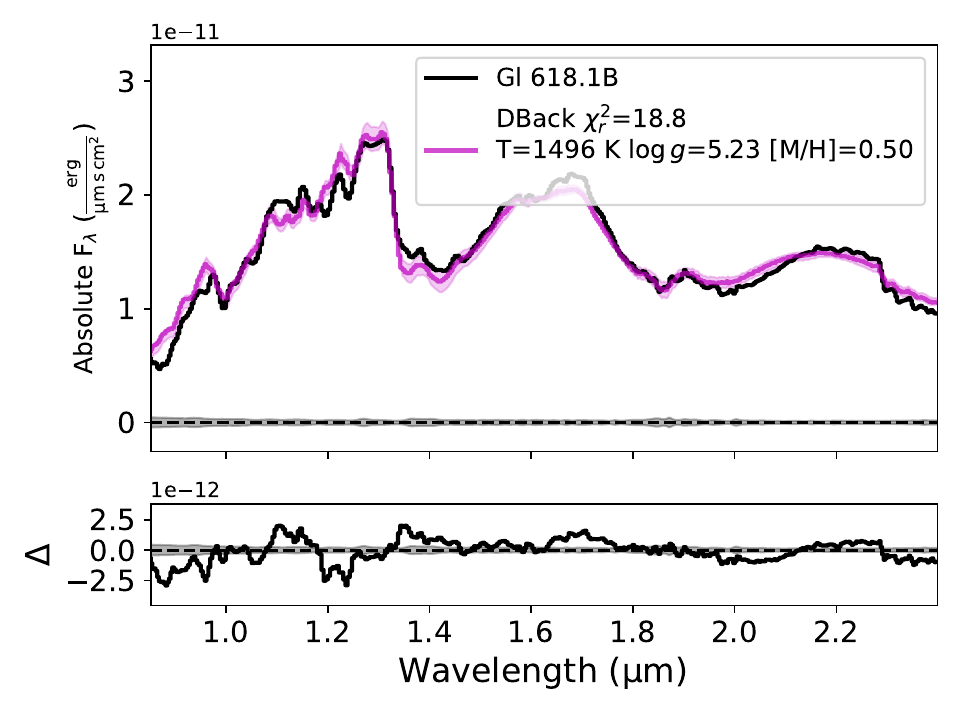} 
\includegraphics[width=0.32\textwidth]{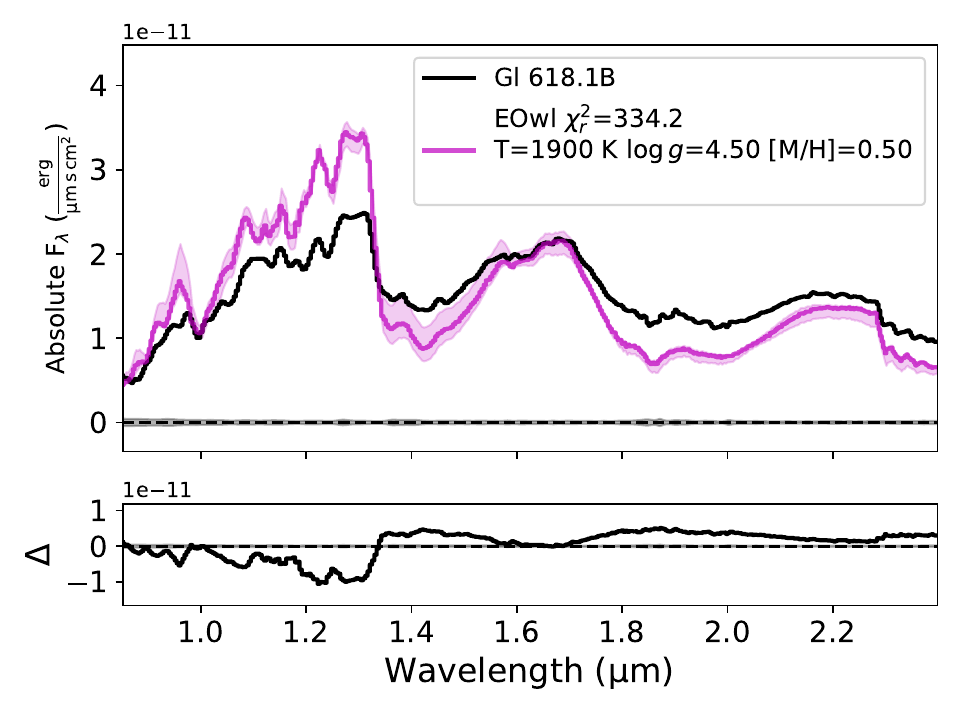} 
\includegraphics[width=0.32\textwidth]{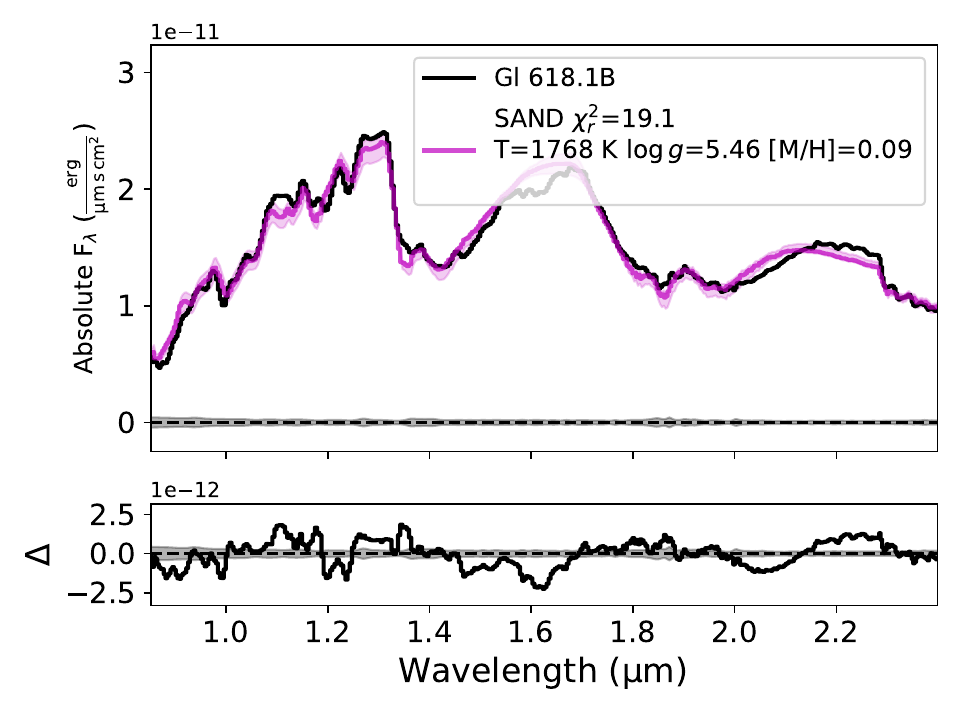}  \\
\includegraphics[width=0.32\textwidth]{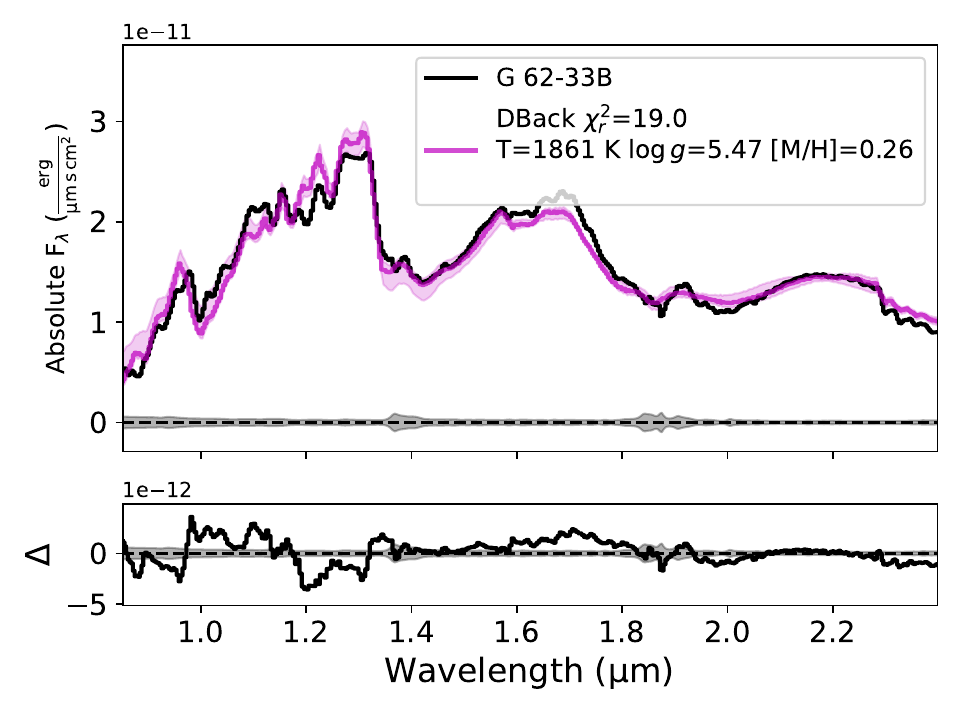}
\includegraphics[width=0.32\textwidth]{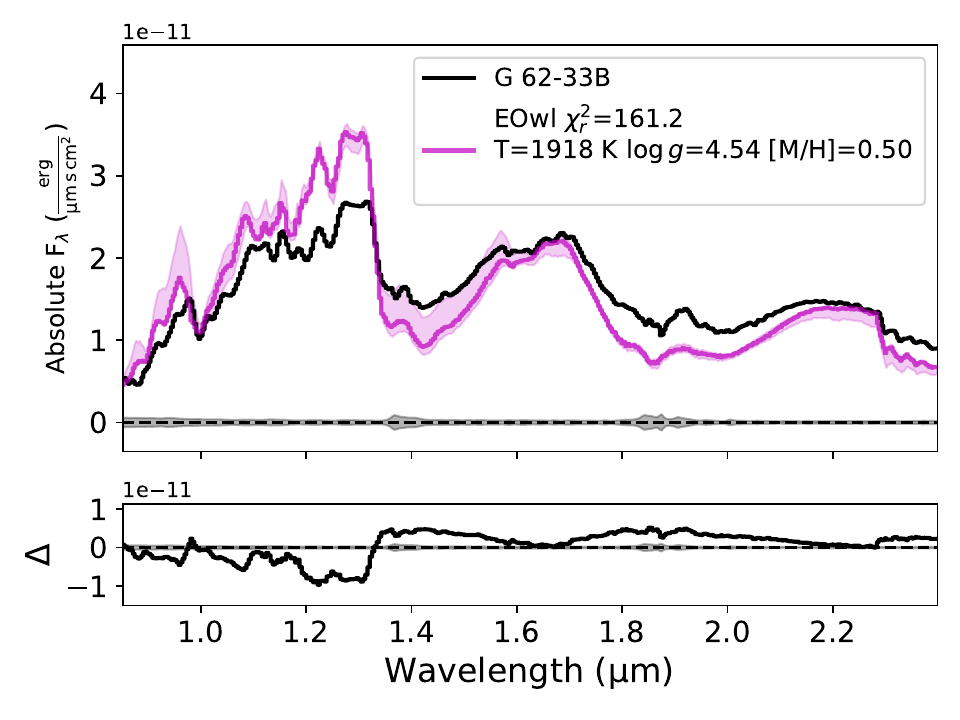}
\includegraphics[width=0.32\textwidth]{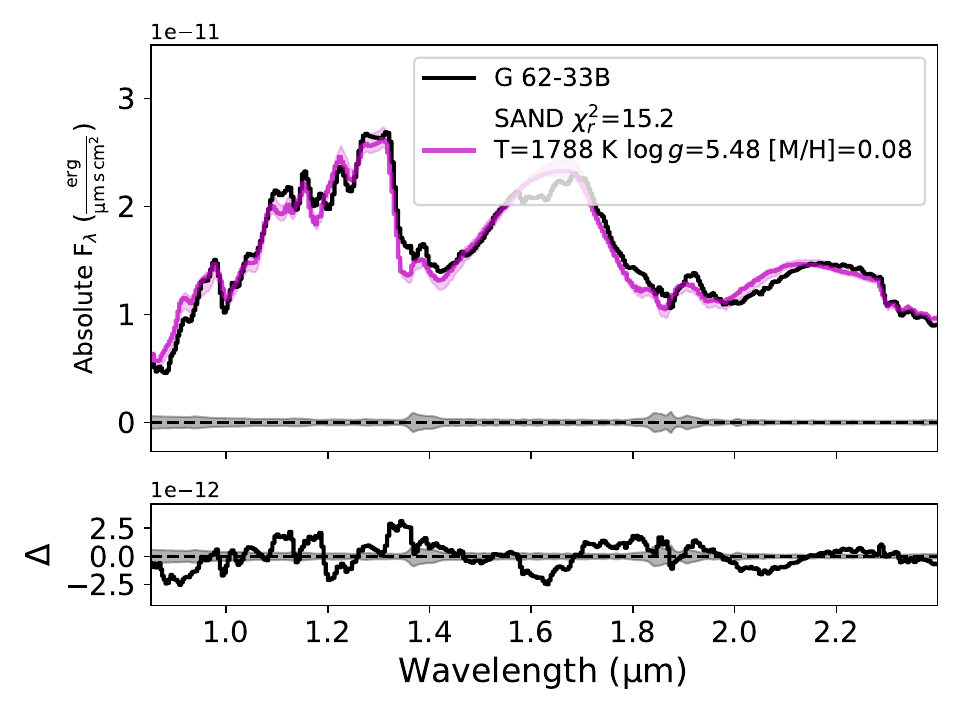} \\
\includegraphics[width=0.32\textwidth]{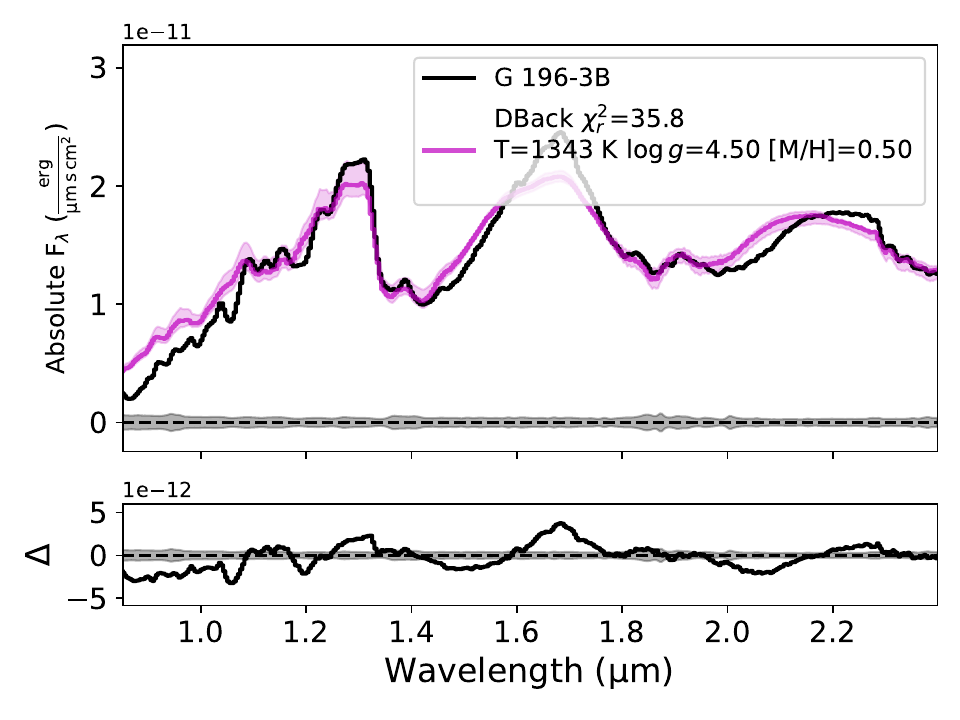} 
\includegraphics[width=0.32\textwidth]{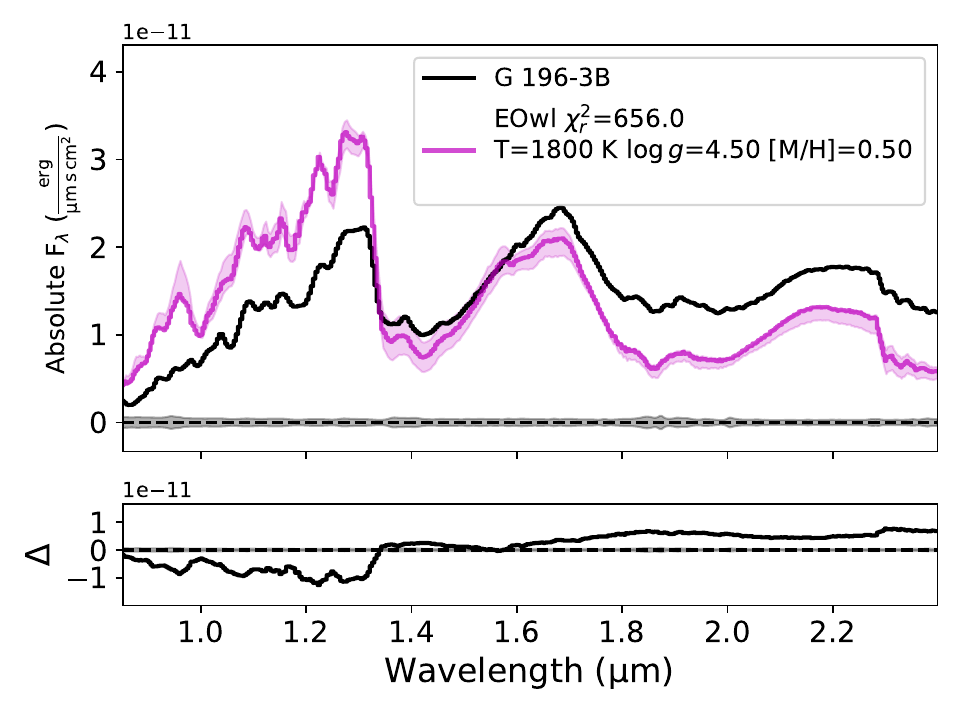} 
\includegraphics[width=0.32\textwidth]{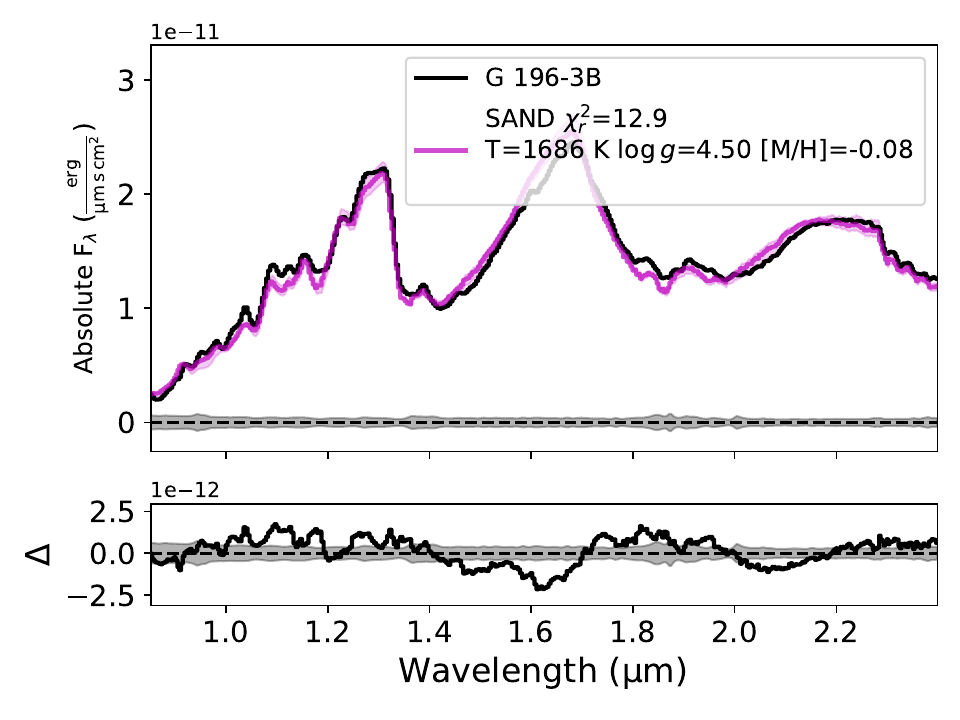}  \\
\caption{Absolute flux-calibrated spectra of our benchmark brown (black lines) compared to best-fit models (thick magenta lines) and posterior draws (magenta shading) for Diamondback (left), Elf Owl (center), and SAND models (right) based on our MCMC analysis. The bottom panel in each plot compares the difference ($\Delta$ = model - data) to the $\pm1\sigma$ uncertainty spectrum (gray bands).
Best-fit parameter values for {\teff}, {\logg}, and [M/H] are indicated in the legend; see full parameters values in Table~\ref{tab:MCMCresults}.
\label{fig:mcmc-comp1}}
\end{figure}

\begin{figure}[h!]
\centering
\includegraphics[width=0.32\textwidth]{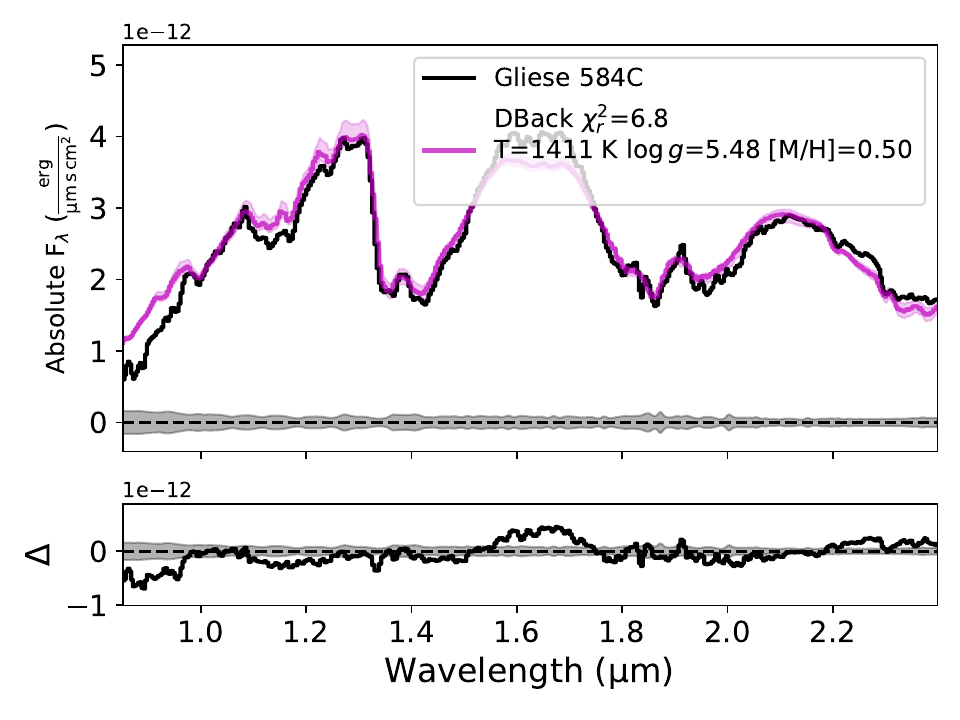}
\includegraphics[width=0.32\textwidth]{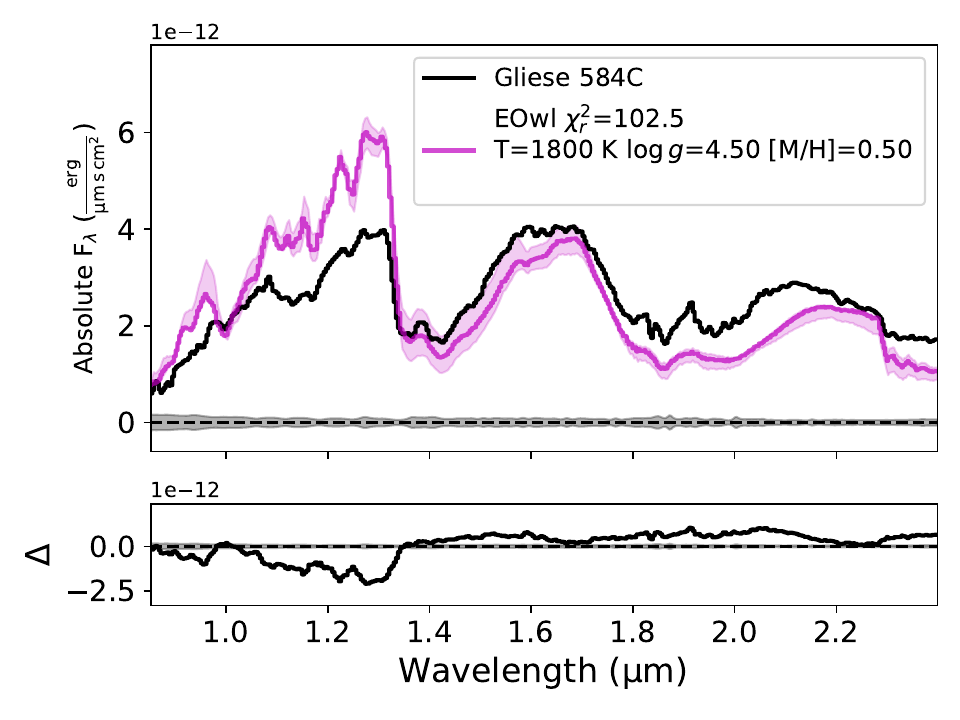}
\includegraphics[width=0.32\textwidth]{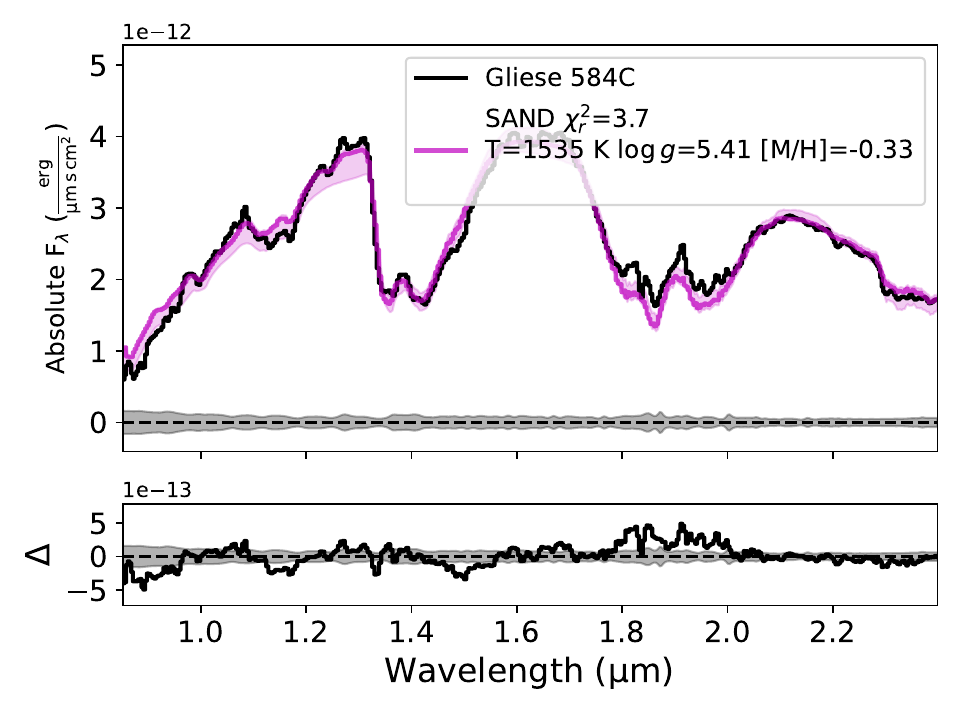} \\
\includegraphics[width=0.32\textwidth]{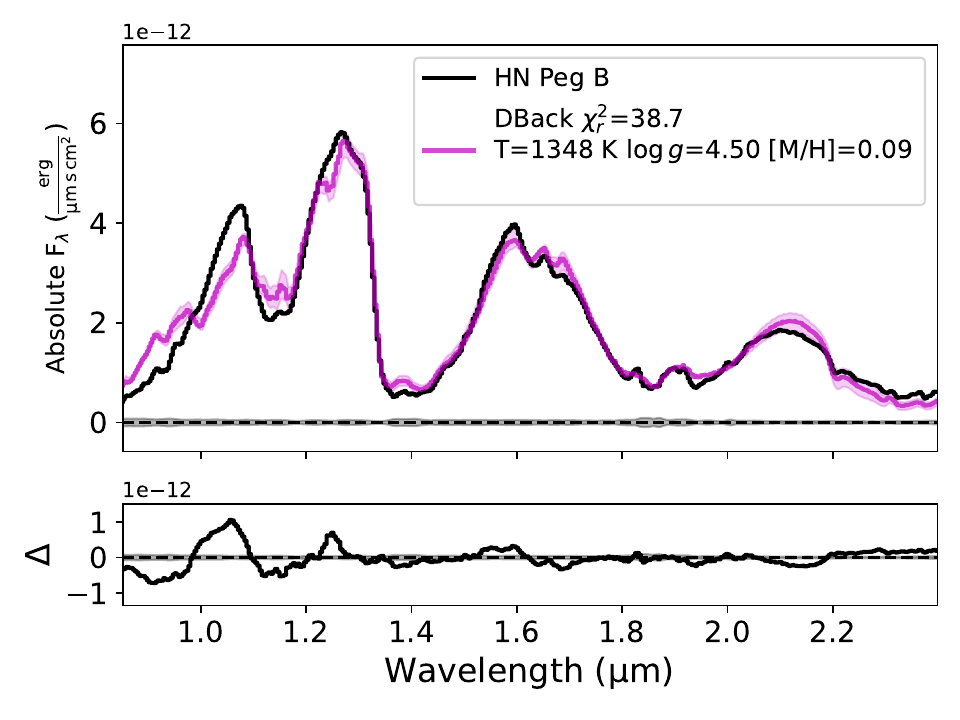} 
\includegraphics[width=0.32\textwidth]{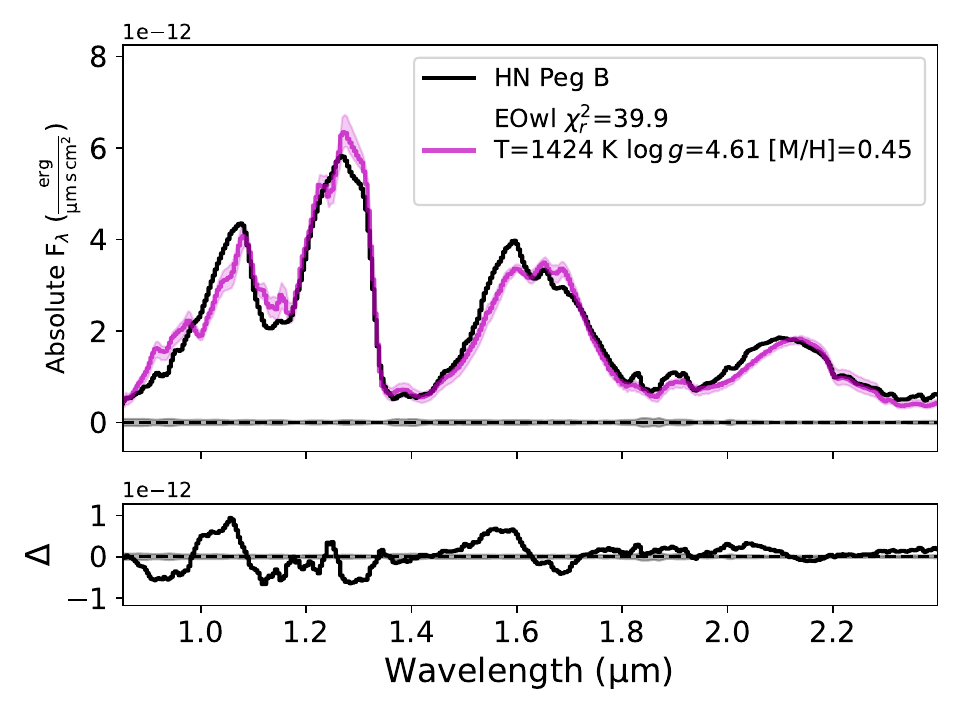} 
\includegraphics[width=0.32\textwidth]{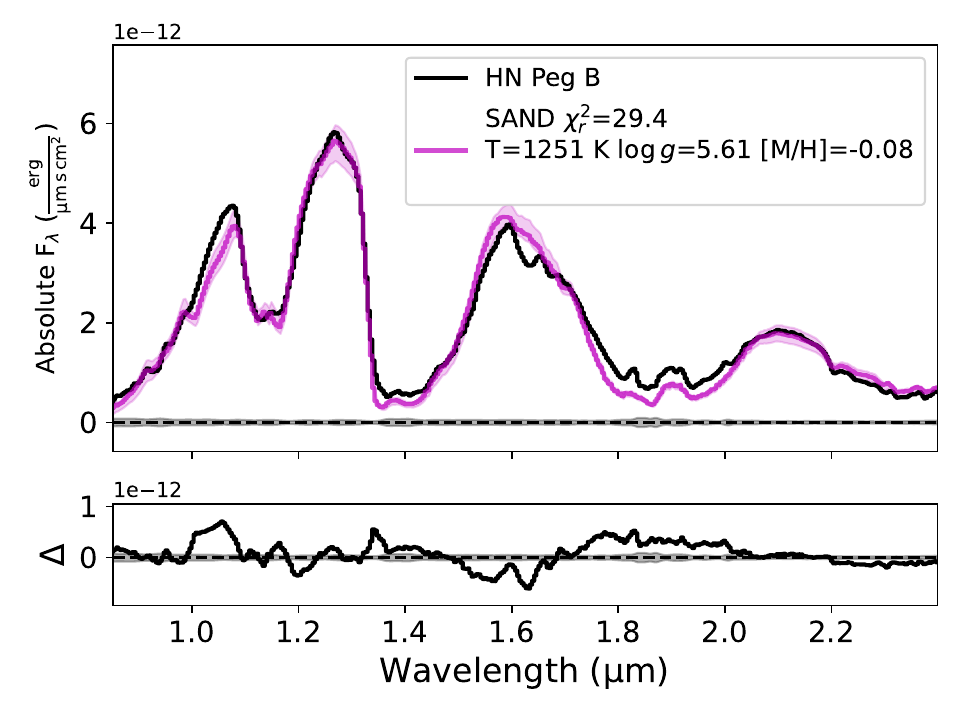} \\
\includegraphics[width=0.32\textwidth]{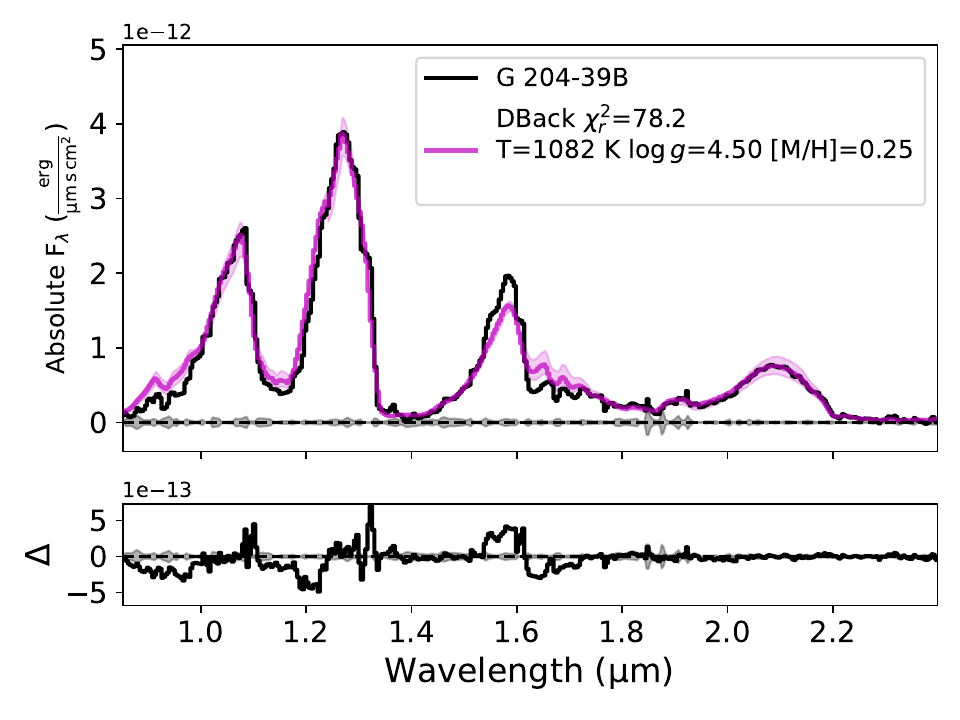} 
\includegraphics[width=0.32\textwidth]{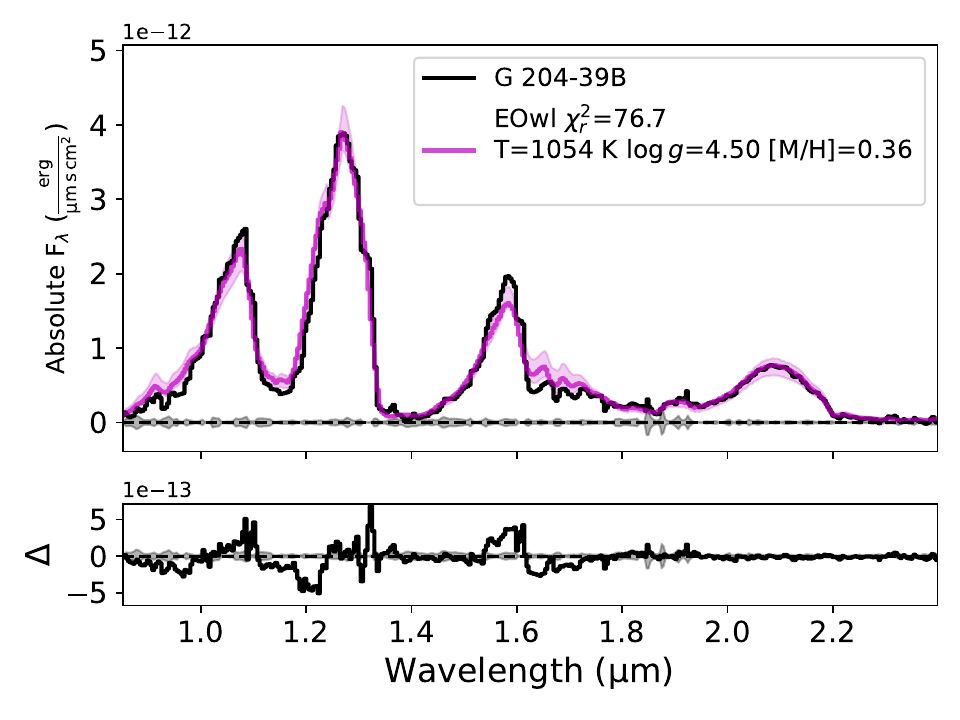} 
\includegraphics[width=0.32\textwidth]{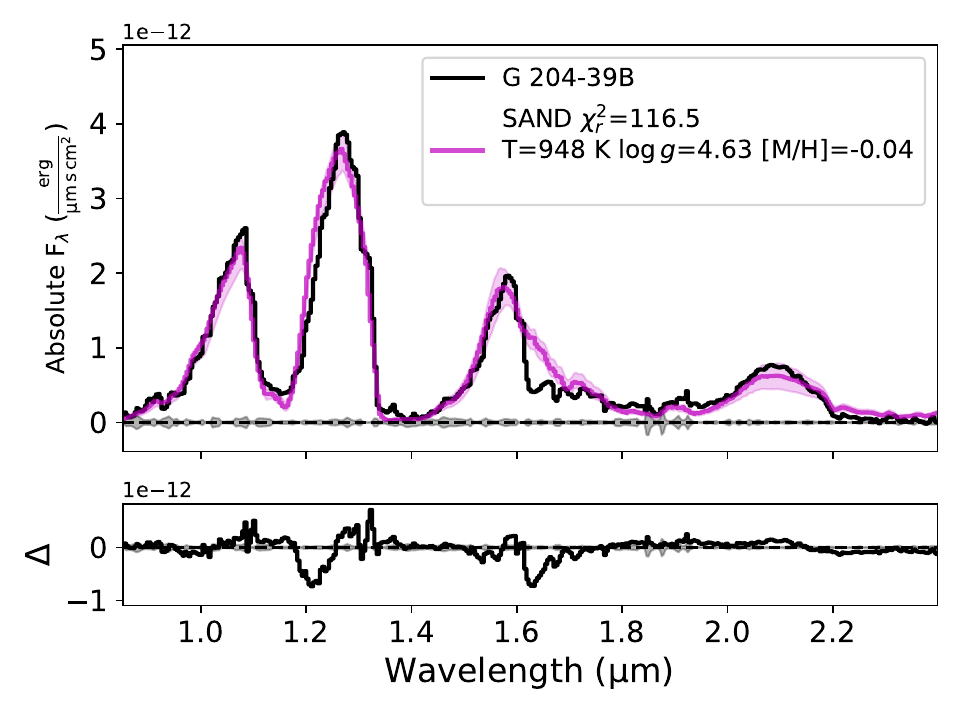} \\
\includegraphics[width=0.32\textwidth]{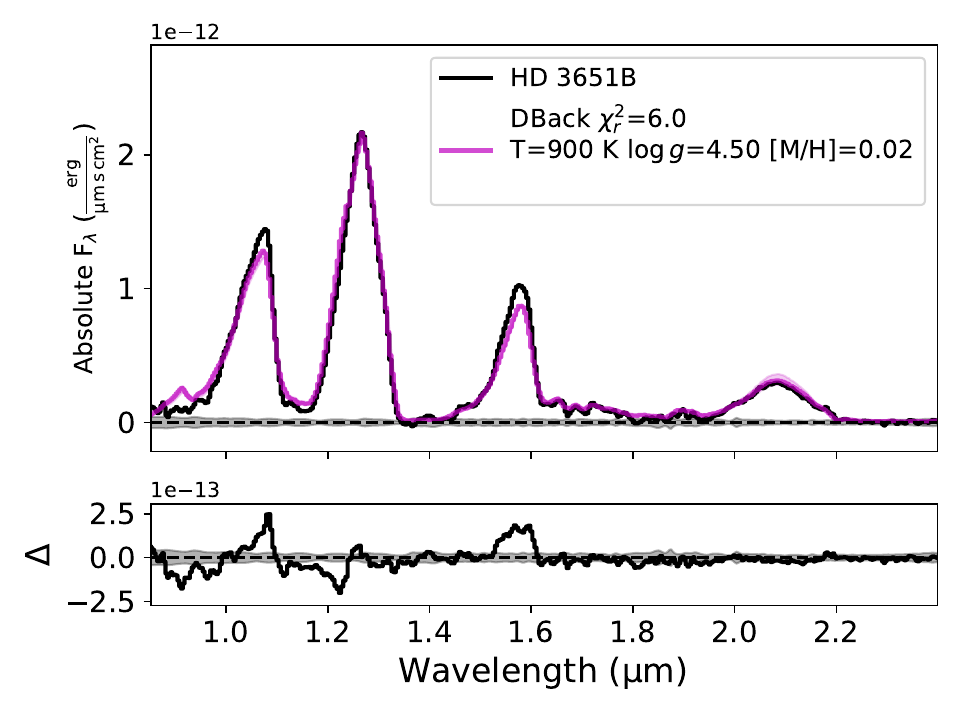} 
\includegraphics[width=0.32\textwidth]{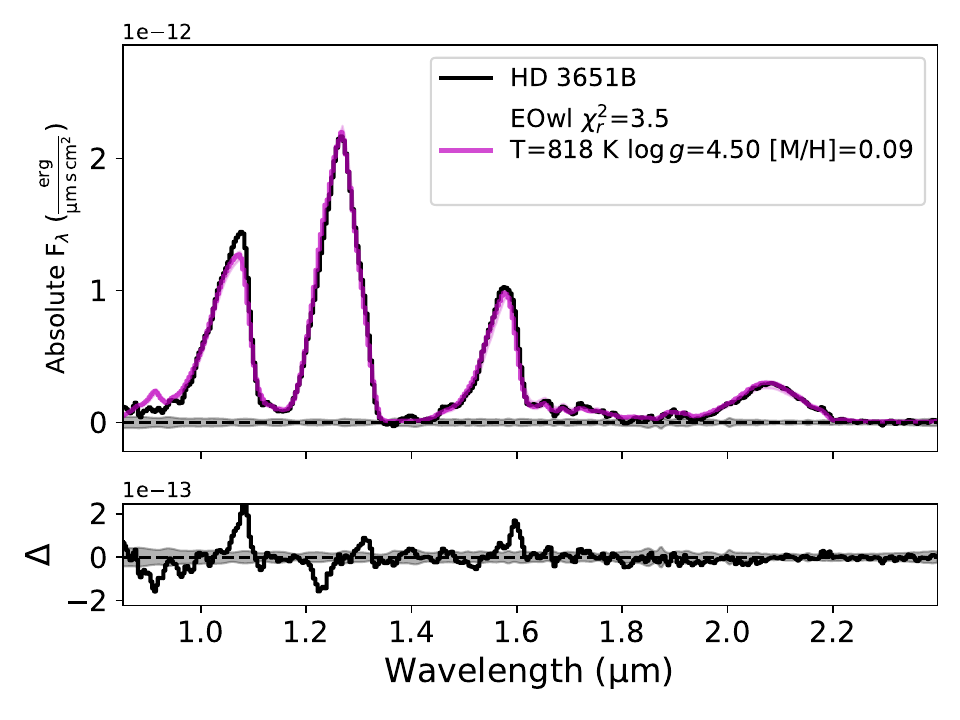} 
\includegraphics[width=0.32\textwidth]{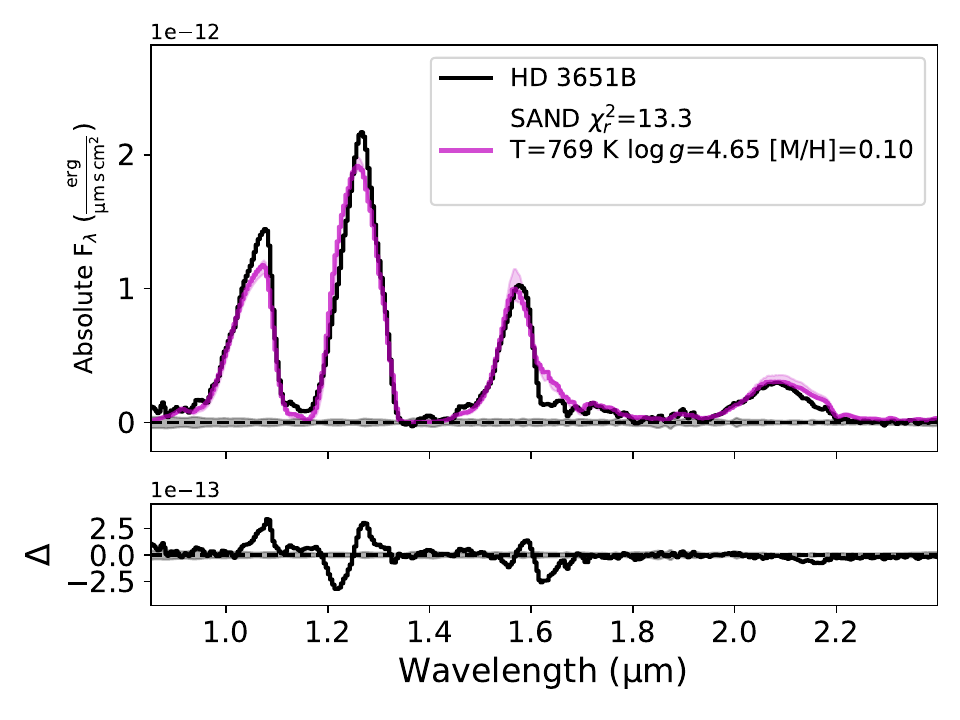} \\ 
\includegraphics[width=0.32\textwidth]{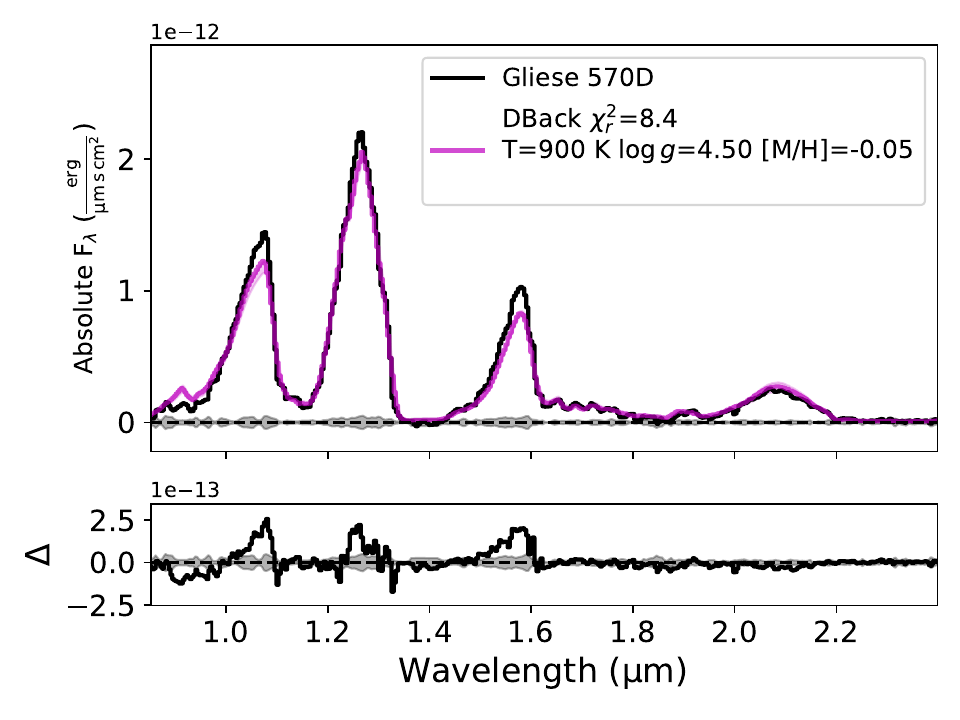}
\includegraphics[width=0.32\textwidth]{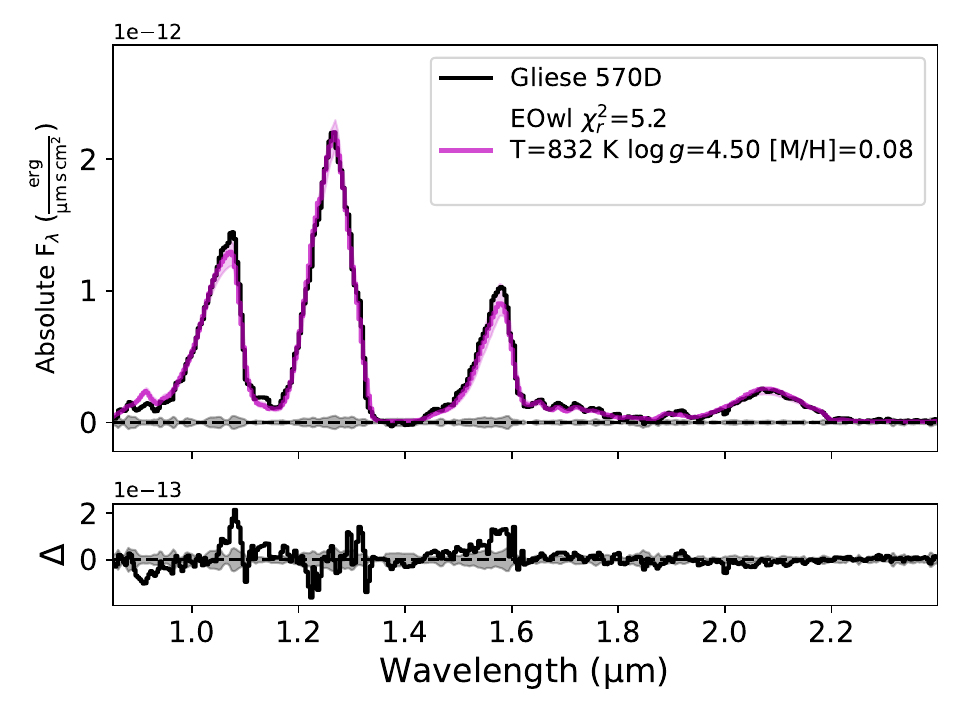}
\includegraphics[width=0.32\textwidth]{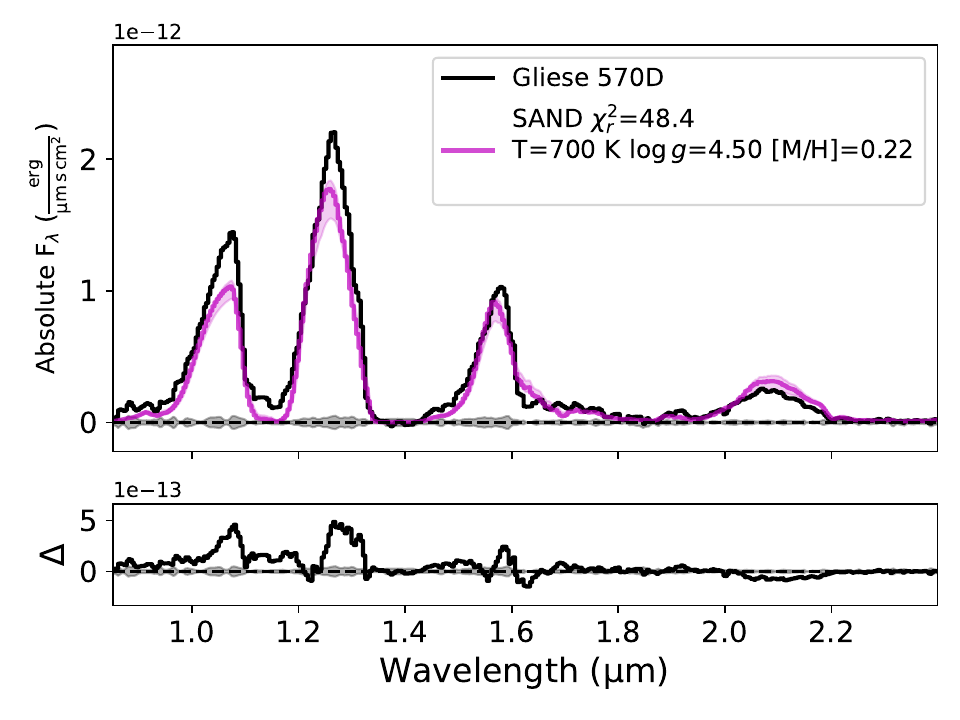} \\
\caption{Figure~\ref{fig:mcmc-comp1} continued. \label{fig:mcmc-comp2}}
\end{figure}

Figures~\ref{fig:mcmc-comp1} and~\ref{fig:mcmc-comp2} display the best-fit models for each of the spectra based on the Diamondback, Elf Owl, and SAND models, while Table~\ref{tab:MCMCresults} summarizes the corresponding posterior parameters.
We find that the Diamondback models generally perform best for the early L dwarfs, reproducing the overall spectral energy distributions but not necessarily detailed molecular features shortward of 1.25~$\mu$m.
For the mid- and late-L dwarfs, the SAND models produce equivalent or better fits than Diamondback, better reproducing the redder spectral slopes of these sources, particularly for the young L dwarf G~196-3B. 
The early T dwarf HN~Peg~B is also best fit by the SAND model despite the poor reproduction of the 1.6~$\mu$m CH$_4$ band in all of the T dwarf fits for this model. 
Elf Owl models yield poor fits to L dwarf and early T dwarf spectra due to the absence of condensate opacity, but become the preferred model 
for the mid- and late-T dwarfs, particularly as we reach the {\teff} parameter limits for the Diamondback and SAND models.

The inferred parameters for these fits largely follow expected trends, with declining effective temperatures with later spectral types, and lower surface gravities for lower-mass objects (e.g., the young source G~196-3B). For the Diamondback models, the $f_{sed}$ parameter is small for L dwarfs (condensates retained in the photosphere) and large for T dwarfs (condensates depleted in the photosphere), also as expected. We also find the radii of the best-fit models to be generally in the expected range of 0.075~R$_\odot$ $\lesssim$ R $\lesssim$ 0.105~R$_\odot$ based on evolutionary models \citep{2001RvMP...73..719B,2003A&A...402..701B}.
There are, however, notable deviations for individual sources.
For example, both Diamondback and SAND fits to the L1 GJ~1048B yield an inflated radius R = 0.17~R$_\odot$,
and a {\teff} $\approx$ 1700~K which is considerably lower than expected from the source spectral type.
This deviation may be due to an interplay between condensate opacity and temperature, where a reduction in the former is compensated by a reduction in the latter.
Conversely, the L8 Gliese~584C (R = 0.064$\pm$0.001~R$_\odot$ from SAND models) and the T6.5 G~204-39B (R = 0.065$\pm$0.003~R$_\odot$ and 0.069$^{+0.002}_{-0.004}$~R$_\odot$ for Diamondback and Elf Owl models, respectively) have radii somewhat lower than expected for brown dwarfs, suggesting that their inferred {\teff}s may be slightly overestimated.
For Gliese~584C, this may reflect a phase of cloud clearing not captured in the SAND models.

\begin{deluxetable}{llccccccccc}
\tablecaption{MCMC Fitting Results. \label{tab:MCMCresults}} 
\tabletypesize{\tiny}
\tablehead{ 
\colhead{\textbf{Source}} & 
\colhead{\textbf{SpT}} & 
\colhead{$\mathbf{T_{\mathbf{eff}}}$} & 
\colhead{\textbf{log $\mathbf{g}$}} &
\colhead{$[\mathbf{M/H}]$} &
\colhead{\bf C/O\tablenotemark{a}} &
[$\mathbf{\alpha}$/\textbf{\ch{Fe}}] &
\colhead{\textbf{log} $\mathbf{\kappa_{\mathbf{zz}}}$} &
\colhead{$\mathbf{f}_{\mathbf{sed}}$} &
\colhead{\textbf{R}} &
\colhead{$\mathbf{\chi}^\mathbf{2}_\mathbf{r}$} \\ 
\colhead{} & 
\colhead{} & 
\colhead{(K)} & 
\colhead{(cm/s$^2$)} &
\colhead{} &
\colhead{} &
\colhead{} &
\colhead{(cm$^2$/s)} &
\colhead{} &
\colhead{(R$_\odot$)} &
\colhead{}  
}
\startdata
\cline{1-11} 
\multicolumn{11}{c}{\textbf{Diamondback Model Set}} \\
\cline{1-11}
\textbf{LP 465-70B} & L0  & 2045$^{+30}_{-23}$ & 5.35$^{+0.09}_{-0.10}$ & 0.41$^{+0.07}_{-0.10}$ & \nodata & \nodata & \nodata & 1.26$^{+0.20}_{-0.21}$ & 0.111$^{+0.003}_{-0.003}$ & \textbf{12}\tablenotemark{\rm *} \\
\textbf{HD 89744B} & L0  & 2036$^{+18}_{-29}$ & 5.50$^{+0.00}_{-0.05}$ & 0.22$^{+0.06}_{-0.09}$ & \nodata & \nodata & \nodata & 1.00$^{+0.02}_{-0.00}$ & 0.118$^{+0.003}_{-0.002}$ & \textbf{93}\tablenotemark{\rm *} \\
\textbf{GJ 1048B} & L1  & 1696$^{+11}_{-36}$ & 5.48$^{+0.02}_{-0.10}$ & 0.36$^{+0.06}_{-0.13}$ & \nodata & \nodata & \nodata & 2.04$^{+0.31}_{-0.12}$ & 0.173$^{+0.008}_{-0.002}$ & \textbf{132}\tablenotemark{\rm *} \\
\textbf{Gl 618.1B} & L2.5  & 1499$^{+21}_{-15}$ & 5.13$^{+0.09}_{-0.14}$ & 0.42$^{+0.08}_{-0.12}$ & \nodata & \nodata & \nodata & 3.34$^{+0.40}_{-0.87}$ & 0.157$^{+0.004}_{-0.004}$ & \textbf{19}\tablenotemark{\rm *} \\
{G 62-33B} & L2.5  & 1650$^{+211}_{-94}$ & 5.40$^{+0.10}_{-0.26}$ & 0.30$^{+0.13}_{-0.12}$ & \nodata & \nodata & \nodata & 1.39$^{+1.80}_{-0.39}$ & 0.130$^{+0.016}_{-0.024}$ & {19} \\
G 196-3B & L3$\beta$  & 1343$^{+13}_{-13}$ & 4.50$^{+0.05}_{-0.00}$ & 0.50$^{+0.00}_{-0.07}$ & \nodata & \nodata & \nodata & 2.61$^{+0.11}_{-0.38}$ & 0.206$^{+0.003}_{-0.005}$ & 36 \\
{Gliese 584C} & L8  & 1415$^{+12}_{-9}$ & 5.48$^{+0.02}_{-0.02}$ & 0.50$^{+0.00}_{-0.03}$ & \nodata & \nodata & \nodata & 8.19$^{+0.45}_{-0.72}$ & 0.073$^{+0.001}_{-0.001}$ & {7} \\
HN Peg B & T2.5  & 1348$^{+24}_{-23}$ & 4.59$^{+0.12}_{-0.09}$ & 0.10$^{+0.06}_{-0.06}$ & \nodata & \nodata & \nodata & 7.89$^{+0.59}_{-0.44}$ & 0.068$^{+0.002}_{-0.003}$ & 39 \\
\textbf{G 204-39B} & T6.5  & 1087$^{+25}_{-19}$ & 4.66$^{+0.19}_{-0.15}$ & 0.29$^{+0.11}_{-0.06}$ & \nodata & \nodata & \nodata & 9.57$^{+0.32}_{-0.25}$ & 0.065$^{+0.003}_{-0.003}$ & \textbf{78}\tablenotemark{\rm *} \\
HD 3651B & T7.5  & 900$^{+10}_{-0}$ & 4.58$^{+0.10}_{-0.07}$ & 0.08$^{+0.09}_{-0.07}$ & \nodata & \nodata & \nodata & 9.54$^{+0.35}_{-0.32}$ & 0.067$^{+0.000}_{-0.001}$ & 6 \\
Gliese 570D & T7.5  & 900$^{+6}_{-0}$ & 4.63$^{+0.11}_{-0.09}$ & -0.02$^{+0.06}_{-0.04}$ & \nodata & \nodata & \nodata & 9.65$^{+0.25}_{-0.28}$ & 0.065$^{+0.000}_{-0.001}$ & 8 \\
\cline{1-11} 
\multicolumn{11}{c}{\textbf{Elf Owl Model Set}} \\
\cline{1-11}
LP 465-70B & L0  & 2261$^{+36}_{-37}$ & 4.76$^{+0.16}_{-0.15}$ & 0.42$^{+0.08}_{-0.12}$ & 1.47$^{+0.03}_{-0.06}$ & \nodata & 2.9$^{+0.9}_{-0.5}$ & \nodata & 0.090$^{+0.003}_{-0.003}$ & 25 \\
HD 89744B & L0  & 2320$^{+5}_{-17}$ & 5.46$^{+0.03}_{-0.18}$ & 0.50$^{+0.00}_{-0.10}$ & 1.50$^{+0.00}_{-0.02}$ & \nodata & 2.2$^{+0.4}_{-0.2}$ & \nodata & 0.089$^{+0.001}_{-0.000}$ & 277 \\
GJ 1048B & L1  & 2100$^{+0}_{-0}$ & 5.50$^{+0.00}_{-0.07}$ & 0.50$^{+0.00}_{-0.00}$ & 1.50$^{+0.00}_{-0.00}$ & \nodata & 4.0$^{+0.0}_{-0.0}$ & \nodata & 0.108$^{+0.000}_{-0.000}$ & 764 \\
Gl 618.1B & L2.5  & 1900$^{+26}_{-0}$ & 4.50$^{+0.17}_{-0.00}$ & 0.50$^{+0.00}_{-0.02}$ & 1.50$^{+0.00}_{-0.05}$ & \nodata & 4.0$^{+0.0}_{-0.2}$ & \nodata & 0.093$^{+0.000}_{-0.003}$ & 334 \\
G 62-33B & L2.5  & 1918$^{+58}_{-18}$ & 4.60$^{+0.22}_{-0.06}$ & 0.50$^{+0.00}_{-0.06}$ & 1.50$^{+0.00}_{-0.08}$ & \nodata & 4.1$^{+0.4}_{-0.4}$ & \nodata & 0.092$^{+0.002}_{-0.006}$ & 161 \\
G 196-3B & L3$\beta$  & 1800$^{+41}_{-0}$ & 4.50$^{+0.12}_{-0.00}$ & 0.50$^{+0.00}_{-0.01}$ & 1.50$^{+0.00}_{-0.08}$ & \nodata & 4.0$^{+0.0}_{-0.0}$ & \nodata & 0.100$^{+0.000}_{-0.005}$ & 656 \\
Gliese 584C & L8  & 1800$^{+80}_{-22}$ & 4.58$^{+0.22}_{-0.08}$ & 0.50$^{+0.00}_{-0.09}$ & 1.45$^{+0.05}_{-0.13}$ & \nodata & 4.0$^{+0.4}_{-0.5}$ & \nodata & 0.042$^{+0.001}_{-0.004}$ & 102 \\
HN Peg B & T2.5  & 1432$^{+30}_{-22}$ & 4.65$^{+0.15}_{-0.10}$ & 0.45$^{+0.05}_{-0.07}$ & 1.25$^{+0.07}_{-0.06}$ & \nodata & 2.5$^{+0.6}_{-0.4}$ & \nodata & 0.059$^{+0.002}_{-0.003}$ & 40 \\
\textbf{G 204-39B} & T6.5  & 1054$^{+26}_{-18}$ & 4.64$^{+0.25}_{-0.14}$ & 0.36$^{+0.04}_{-0.14}$ & 1.20$^{+0.06}_{-0.11}$ & \nodata & 2.3$^{+0.5}_{-0.3}$ & \nodata & 0.069$^{+0.002}_{-0.004}$ & \textbf{77}\tablenotemark{\rm *} \\
\textbf{HD 3651B} & T7.5  & 808$^{+13}_{-14}$ & 4.50$^{+0.08}_{-0.00}$ & 0.09$^{+0.07}_{-0.06}$ & 1.18$^{+0.08}_{-0.08}$ & \nodata & 3.0$^{+0.5}_{-0.3}$ & \nodata & 0.083$^{+0.003}_{-0.003}$ & \textbf{4}\tablenotemark{\rm *} \\
\textbf{Gliese 570D} & T7.5  & 812$^{+27}_{-21}$ & 4.57$^{+0.09}_{-0.07}$ & 0.02$^{+0.07}_{-0.06}$ & 1.06$^{+0.11}_{-0.11}$ & \nodata & 3.0$^{+0.7}_{-0.5}$ & \nodata & 0.082$^{+0.006}_{-0.007}$ & \textbf{5}\tablenotemark{\rm *} \\
\cline{1-11} 
\multicolumn{11}{c}{\textbf{SAND Model Set}} \\
\cline{1-11}
LP 465-70B & L0  & 1900$^{+2}_{-3}$ & 5.50$^{+0.01}_{-0.01}$ & -0.29$^{+0.11}_{-0.06}$ & \nodata & 0.13$^{+0.02}_{-0.01}$ & \nodata & \nodata & 0.132$^{+0.000}_{-0.001}$ & 27 \\
HD 89744B & L0  & 1900$^{+0}_{-0}$ & 5.50$^{+0.00}_{-0.00}$ & -0.15$^{+0.00}_{-0.00}$ & \nodata & 0.15$^{+0.00}_{-0.00}$ & \nodata & \nodata & 0.138$^{+0.000}_{-0.000}$ & 287 \\
{GJ 1048B} & L1  & 1744$^{+0}_{-8}$ & 5.50$^{+0.03}_{-0.00}$ & -0.05$^{+0.00}_{-0.00}$ & \nodata & 0.02$^{+0.00}_{-0.00}$ & \nodata & \nodata & 0.171$^{+0.002}_{-0.000}$ & {136} \\
\textbf{Gl 618.1B} & L2.5  & 1778$^{+23}_{-10}$ & 5.46$^{+0.02}_{-0.03}$ & 0.14$^{+0.07}_{-0.03}$ & \nodata & -0.01$^{+0.03}_{-0.02}$ & \nodata & \nodata & 0.115$^{+0.001}_{-0.002}$ & \textbf{19}\tablenotemark{\rm *} \\
\textbf{G 62-33B} & L2.5  & 1797$^{+18}_{-9}$ & 5.47$^{+0.02}_{-0.03}$ & 0.13$^{+0.05}_{-0.02}$ & \nodata & 0.00$^{+0.03}_{-0.02}$ & \nodata & \nodata & 0.112$^{+0.001}_{-0.002}$ & \textbf{15}\tablenotemark{\rm *} \\
\textbf{G 196-3B} & L3$\beta$  & 1687$^{+32}_{-11}$ & 4.50$^{+0.00}_{-0.00}$ & -0.08$^{+0.03}_{-0.06}$ & \nodata & 0.03$^{+0.01}_{-0.01}$ & \nodata & \nodata & 0.129$^{+0.002}_{-0.006}$ & \textbf{13}\tablenotemark{\rm *} \\
\textbf{Gliese 584C} & L8  & 1523$^{+14}_{-18}$ & 5.27$^{+0.11}_{-0.09}$ & 0.05$^{+0.08}_{-0.14}$ & \nodata & 0.03$^{+0.14}_{-0.03}$ & \nodata & \nodata & 0.064$^{+0.001}_{-0.001}$ & \textbf{4}\tablenotemark{\rm *} \\
\textbf{HN Peg B} & T2.5  & 1250$^{+17}_{-8}$ & 5.59$^{+0.10}_{-0.09}$ & -0.08$^{+0.11}_{-0.08}$ & \nodata & 0.13$^{+0.03}_{-0.02}$ & \nodata & \nodata & 0.082$^{+0.001}_{-0.005}$ & \textbf{29}\tablenotemark{\rm *} \\
G 204-39B & T6.5  & 935$^{+22}_{-25}$ & 4.70$^{+0.13}_{-0.05}$ & -0.09$^{+0.05}_{-0.04}$ & \nodata & 0.02$^{+0.01}_{-0.01}$ & \nodata & \nodata & 0.099$^{+0.006}_{-0.005}$ & 117 \\
HD 3651B & T7.5  & 769$^{+24}_{-39}$ & 4.52$^{+0.10}_{-0.02}$ & 0.10$^{+0.01}_{-0.00}$ & \nodata & 0.02$^{+0.01}_{-0.03}$ & \nodata & \nodata & 0.101$^{+0.013}_{-0.010}$ & 13 \\
Gliese 570D & T7.5  & 700$^{+37}_{-0}$ & 4.50$^{+0.04}_{-0.00}$ & 0.22$^{+0.04}_{-0.12}$ & \nodata & -0.02$^{+0.04}_{-0.01}$ & \nodata & \nodata & 0.114$^{+0.001}_{-0.013}$ & 48 \\
\enddata
\tablenotetext{a}{C/O ratio relative to solar abundance, with (C/O)$_\odot$ = 0.458 \citep{2009LanB...4B..712L}.}
\tablenotetext{\rm *}{Best fit model or models (if $\Delta\chi^2_r \leq 1$) for this source.}
\end{deluxetable}

\subsection{Random Forest Retrieval}

\begin{figure}
    \centering
    \includegraphics[width=\textwidth]{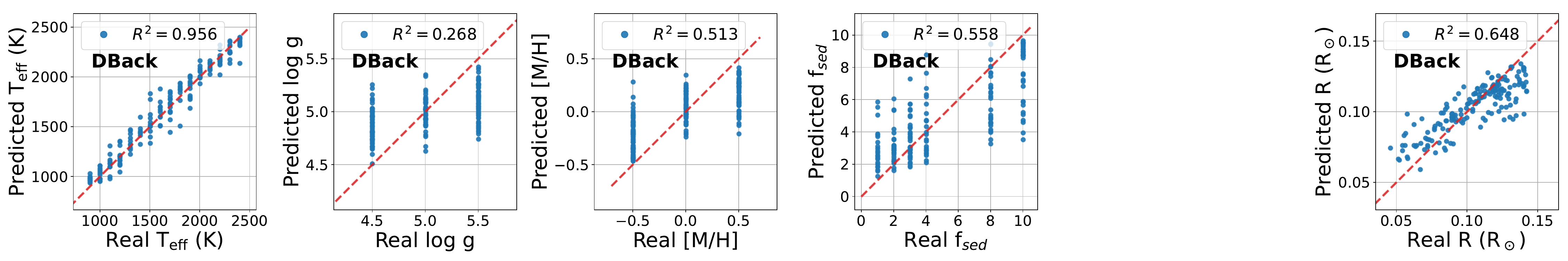}
    \includegraphics[width=\textwidth]{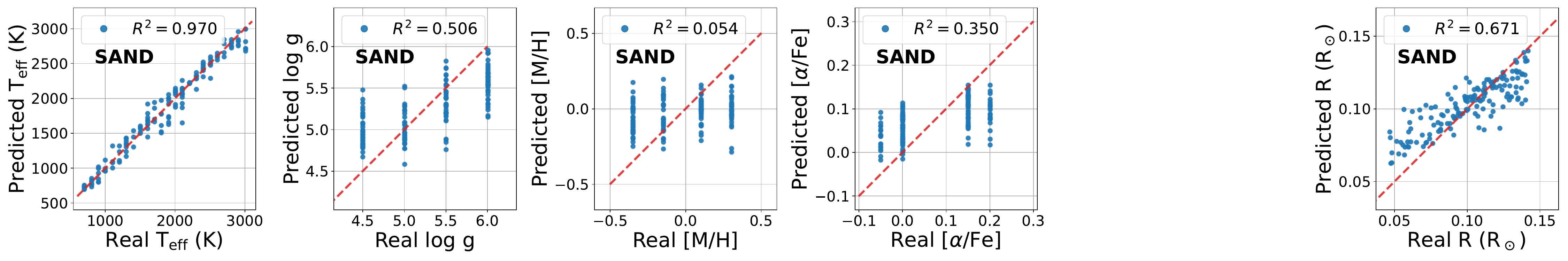}
    \includegraphics[width=\textwidth]{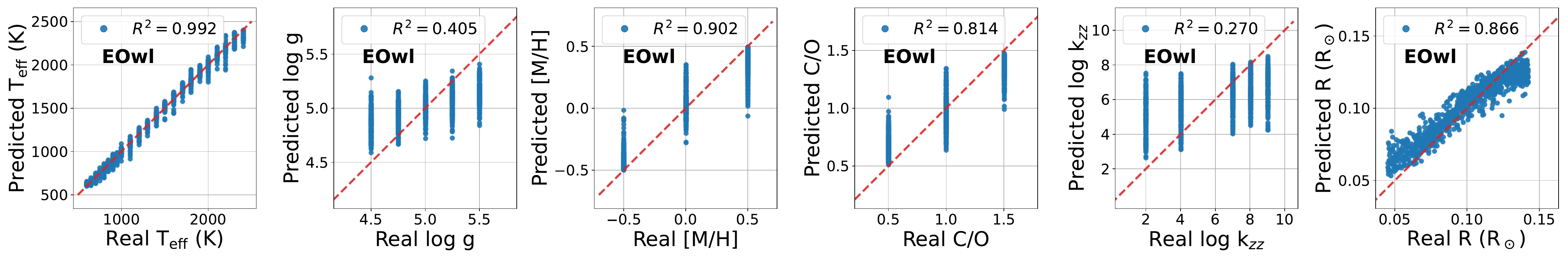}
    \caption{Real versus predicted (RvP) comparison for random forest model training and testing on the evaluated model grids. Each row represents one grid with its corresponding model parameters. The red dashed line in each panel indicates perfect agreement ($R^2$ = 1).} 
    \label{fig:RvP}
\end{figure}

The performance of the random forest retrieval models can be characterized both by the quality of fits and the training metrics of real versus predicted (RvP) parameters and feature importance. Figure~\ref{fig:RvP} shows the RvP plots for each of the model parameters and for each model. Very high self-consistency is found for effective temperature, with $R^2 > 0.95$ for all of the grids. For the other parameters, $R^2$ varies widely across models, generally showing worse performance when fewer parameter values are available. For example, for metallicity $R^2$ ranges from 0.05 for the SAND models to 0.90 for Elf Owl.

\begin{figure}
    \centering
    \includegraphics[width=\textwidth]{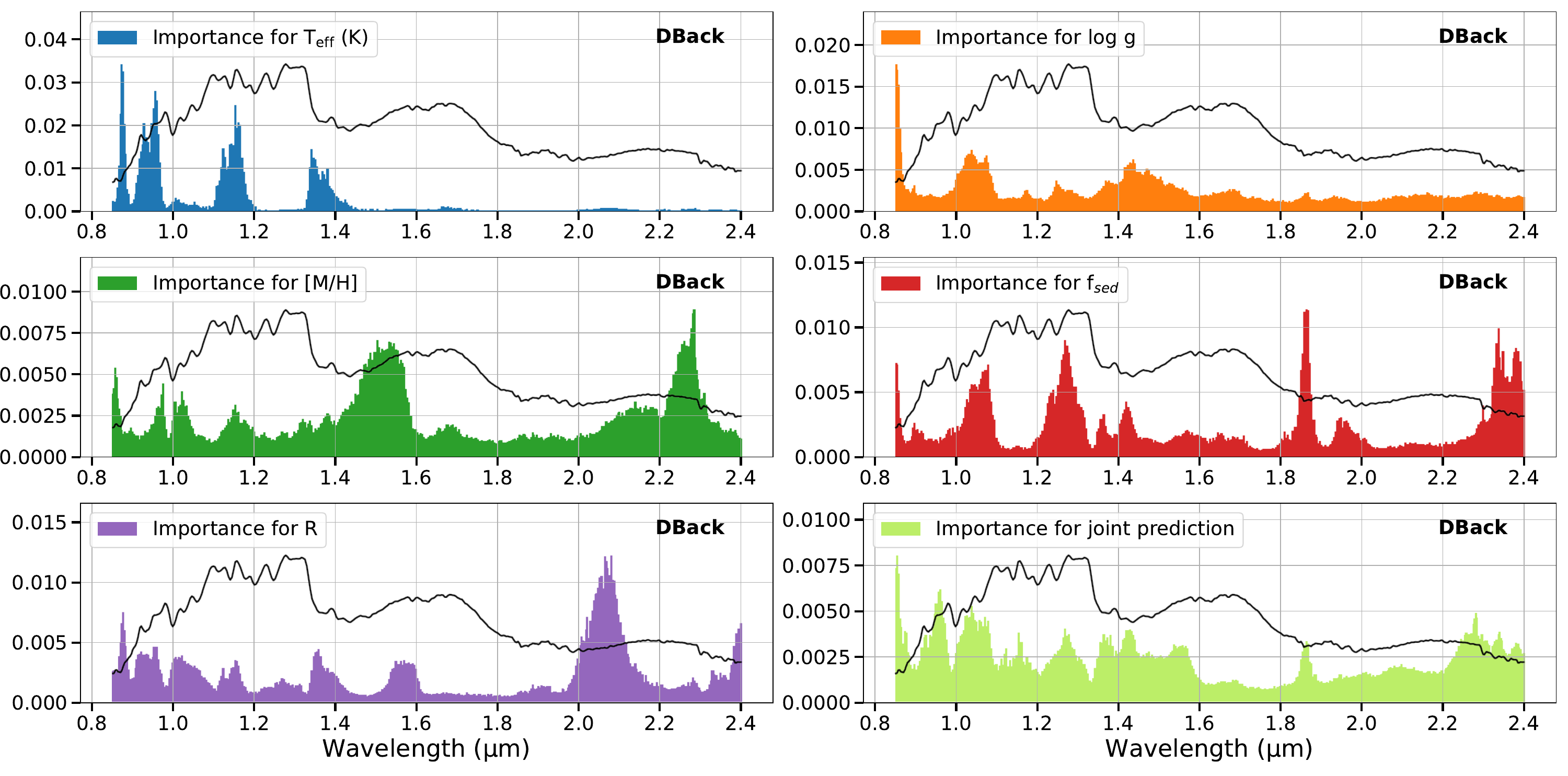}
    \caption{Feature importance plots for the random forest retrieval model trained on the Diamondback grid for temperature T$\mathrm{eff}$ (blue), surface gravity $\log{g}$ (orange), metallicity [$M/H$] (green), cloud sedimentation efficiency $f_\mathrm{sed}$ (red), radius R (calculated from the calibration factor $f$; purple), and joint retrieval (lime). For comparison, the scaled observed spectrum of HD~89744B (solid, black line) has been added to each plot.}
    \label{fig:Dback_feature_HD89744B}
\end{figure}

\begin{figure}
    \centering
    \includegraphics[width=\textwidth]{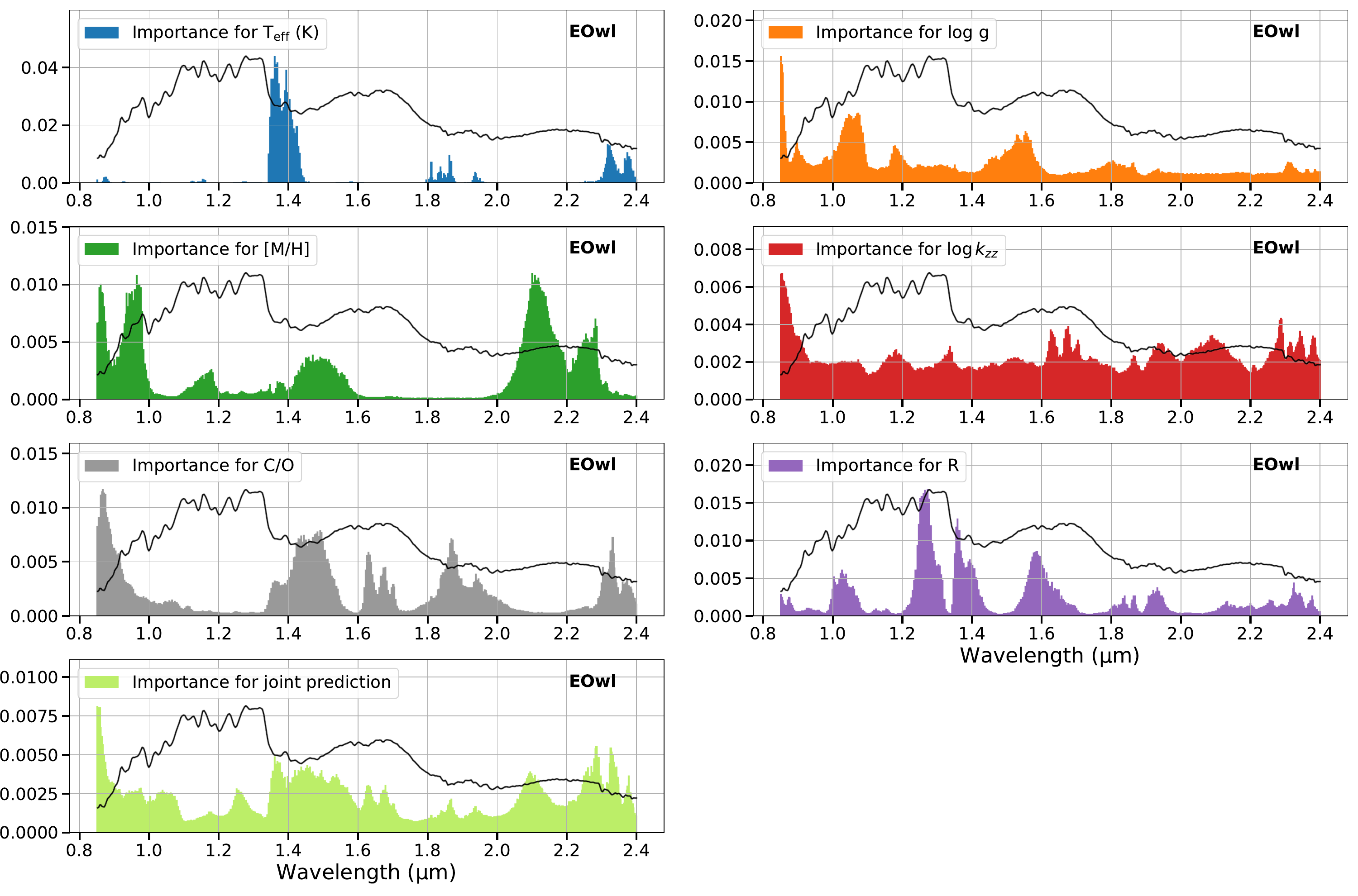}
    \caption{Same as Figure~\ref{fig:Dback_feature_HD89744B} for the  Elf Owl model grid, for temperature T$\mathrm{eff}$ (blue), surface gravity $\log{g}$ (orange), metallicity [$M/H$] (green), vertical eddy diffusion coefficient $\log k_{\rm zz}$ (red), carbon-to-oxygen ratio C/O (gray), radius R (purple), and joint retrieval (lime). }
    \label{fig:Elfowl_feature_HD89744B}
\end{figure}

\begin{figure}
    \centering
    \includegraphics[width=\textwidth]{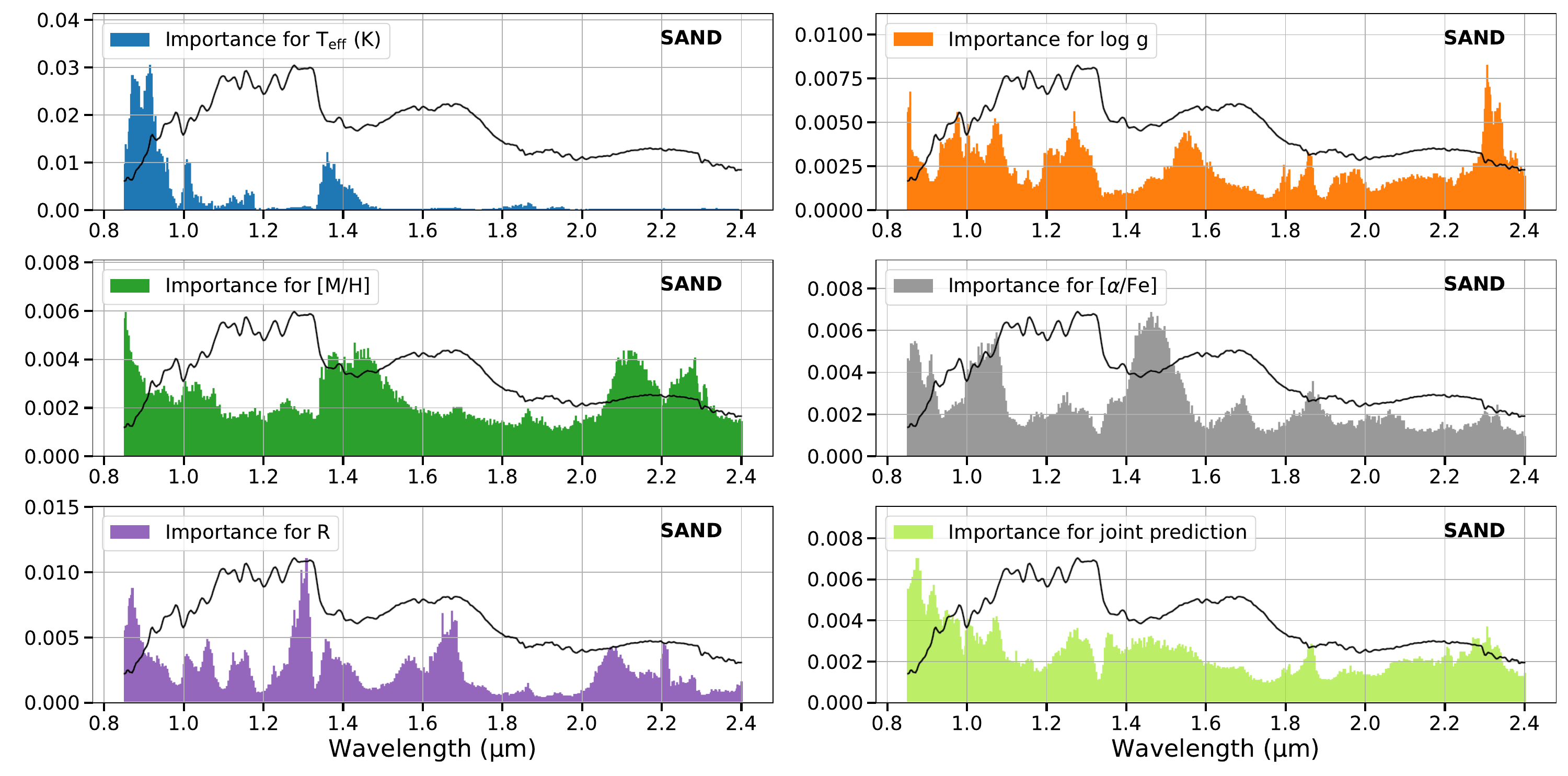}
    \caption{Same as Figure~\ref{fig:Dback_feature_HD89744B} for SAND model grid,
    for temperature T$\mathrm{eff}$ (blue), surface gravity $\log{g}$ (orange), metallicity [$M/H$] (green), alpha element enrichment [$\alpha$/\ch{Fe}] (gray), radius R (purple), and joint retrieval (lime).}
    \label{fig:Sand_feature_HD89744B}
\end{figure}

Figures~\ref{fig:Dback_feature_HD89744B} to \ref{fig:Sand_feature_HD89744B} display the feature importance diagrams for individual model parameters for the three atmosphere models, compared to the scaled spectrum of HD~89744B to guide assessment of the relevant spectral features. For {\teff}, the Diamondback and SAND models show a consistent structure in feature importance, aligning with the strong molecular absorption bands of \ch{H2O} at 0.92, 1.1 and 1.4~$\mu$m, and the red optical spectral slope. The Elf Owl models differ slightly, with the strongest feature corresponding to \ch{H2O} at 1.4~$\mu$m but additional features aligned with the 1.8~$\mu$m \ch{H2O} band and the 2.3~$\mu$m \ch{CO} band. For surface gravity, all grids consistently indicate that the wavelength range below $1.2\,\mu$m is crucial, consistent with the findings of \citet{2023ApJ...954...22L}. The lowest wavelengths are particularly sensitive to the presence and strength of the K~I line wings centered at 0.77~$\mu$m, which are known to be key indicators of an object's surface gravity \citep{2005A&A...440.1195A}. An additional wavelength range of note in the SAND model grid is the $2.2-2.4\,\mu$m region, which likely traces collision-induced \ch{H2} absorption known to be sensitive to  atmospheric pressure and thus surface gravity in the T dwarfs \citep{1969ApJ...156..989L,2002ApJ...564..421B}. Feature importance patterns for [$M/H$] across all model grids, as well as for C/O in the Elf Owl models and alpha enhancement [$\alpha$/\ch{Fe}] in the SAND models, show similar structures aligned with the strong molecular absorption bands of \ch{H2O}, \ch{CO}, and \ch{CH4}. Peaks are also found at wavelengths below $1.2\,\mu$m, again likely due to the influence of alkali line features. 
Similarly, the parameter {\logk} in the Elf Owl models has the greatest feature importance shortward of 0.9~$\mu$m and in the \ch{CO} band, indicating comparable abundance effects from vertical mixing.
The parameter $f_{sed}$ in the Diamondback models show several discrete peaks that overlap with the 1.05~$\mu$m and 1.25~$\mu$m flux peaks, where condensate grain scattering is most prominent;
and the 1.8~$\mu$m \ch{H2O} and 2.3~$\mu$m \ch{CO} bands which are relatively enhanced in the dustiest L dwarfs \citep{2023ApJ...943L..16S}.
Finally, for the radius scaling parameter $f$, all model grids behave differently. 
The most significant feature in the Diamondback model grid is centered at 2.1~$\mu$m, a pseudocontinuum region of the L dwarfs and a flux peak for the T dwarfs.
The Elf Owl model grid shows numerous discrete features associated with flux peaks at $1.05\,\mu$m, $1.25\,\mu$m, and $1.58\,\mu$m, and absorption features from \ch{H2O} at $1.4\,\mu$m and \ch{CO} at $2.3\,\mu$m.
The SAND model grid shows different features associated with the $<$0.9~$\mu$m red optical slope, the 2.1~$\mu$m flux peak, and the 1.6~$\mu$m \ch{CH4} absorption band.
These distinctions in feature importance indicate that a direct comparison between different model grids should be treated with caution.

\begin{figure}
\centering
\includegraphics[width=0.32\textwidth]{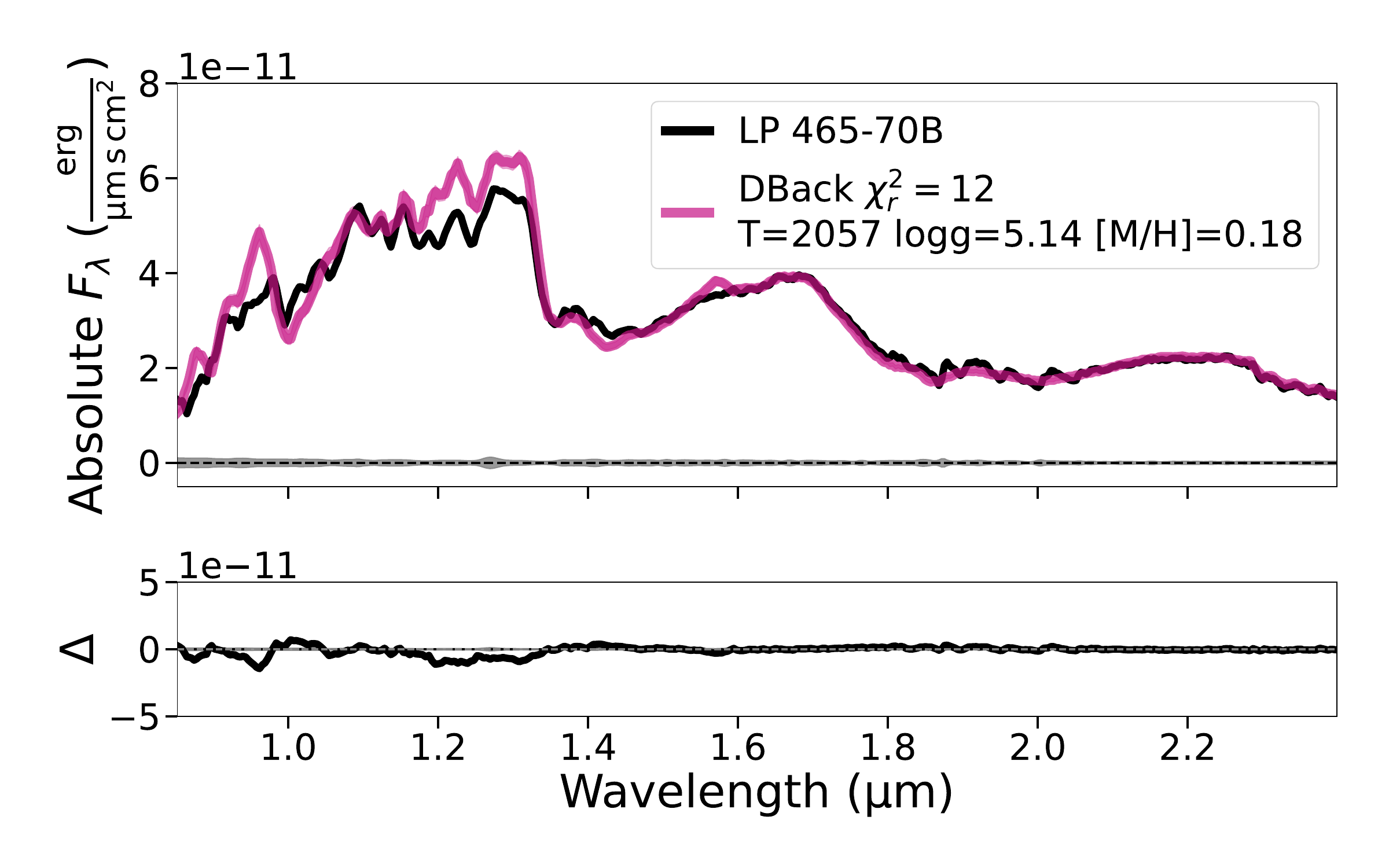}
\includegraphics[width=0.32\textwidth]{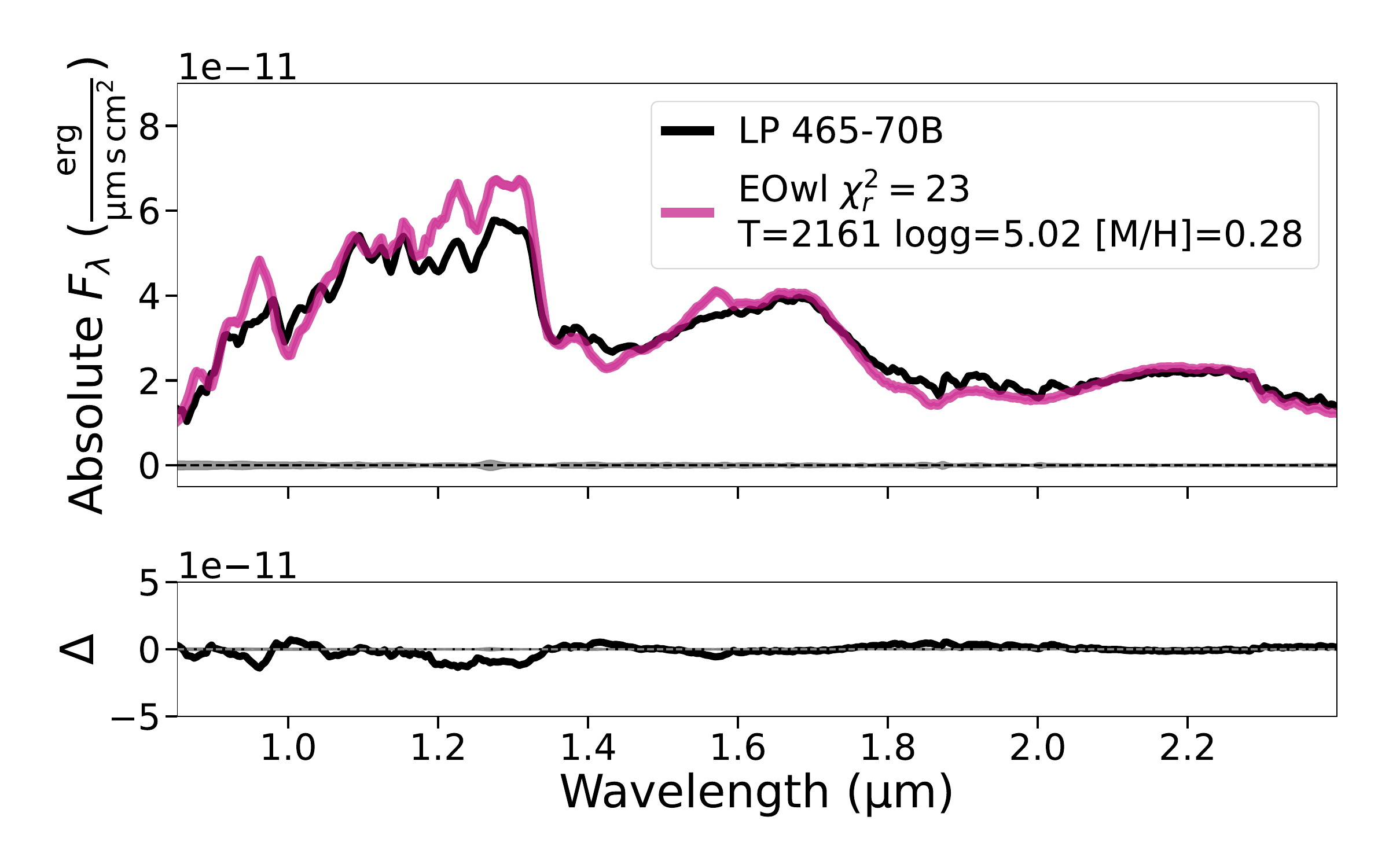}
\includegraphics[width=0.32\textwidth]{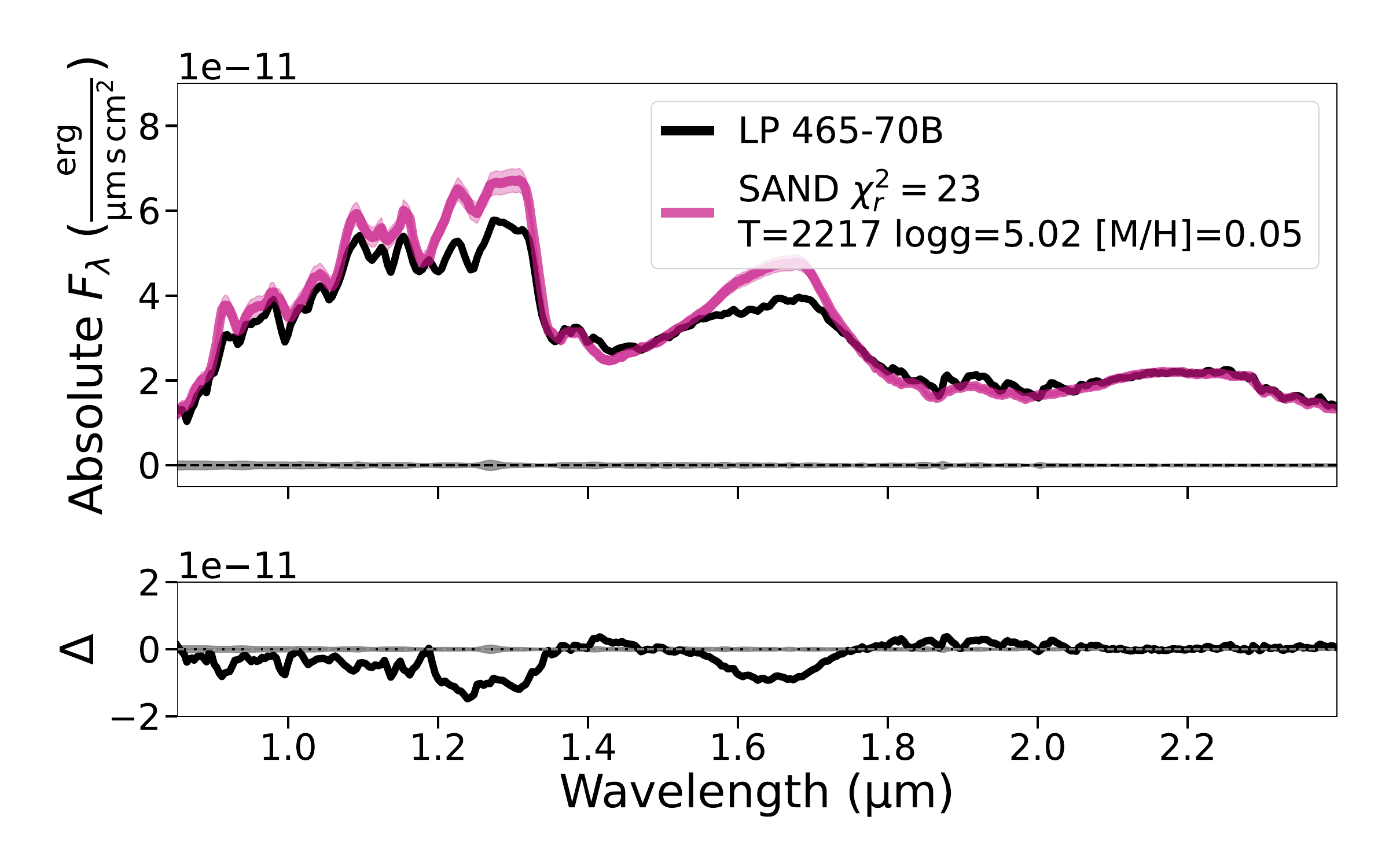} \\
\includegraphics[width=0.32\textwidth]{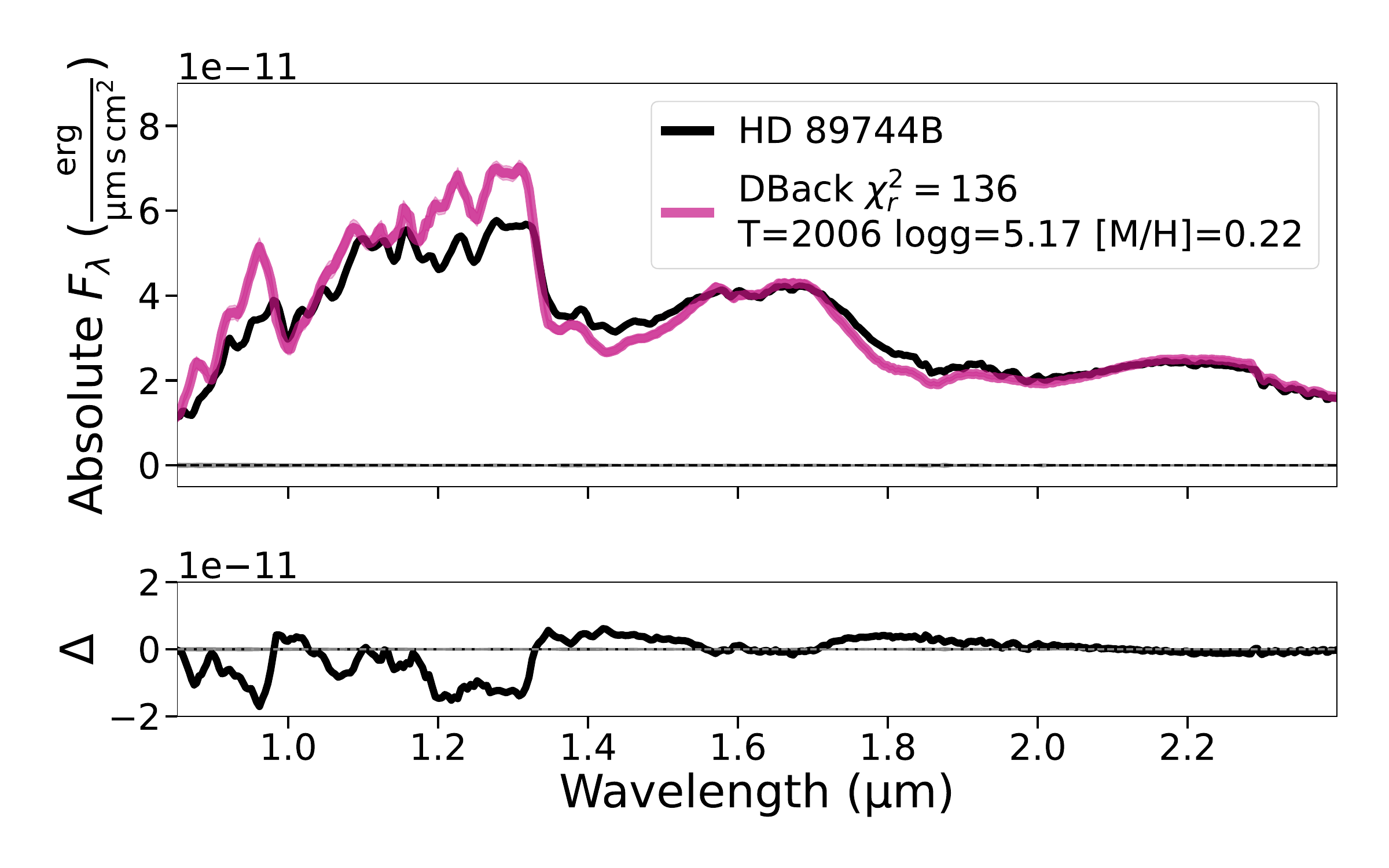} 
\includegraphics[width=0.32\textwidth]{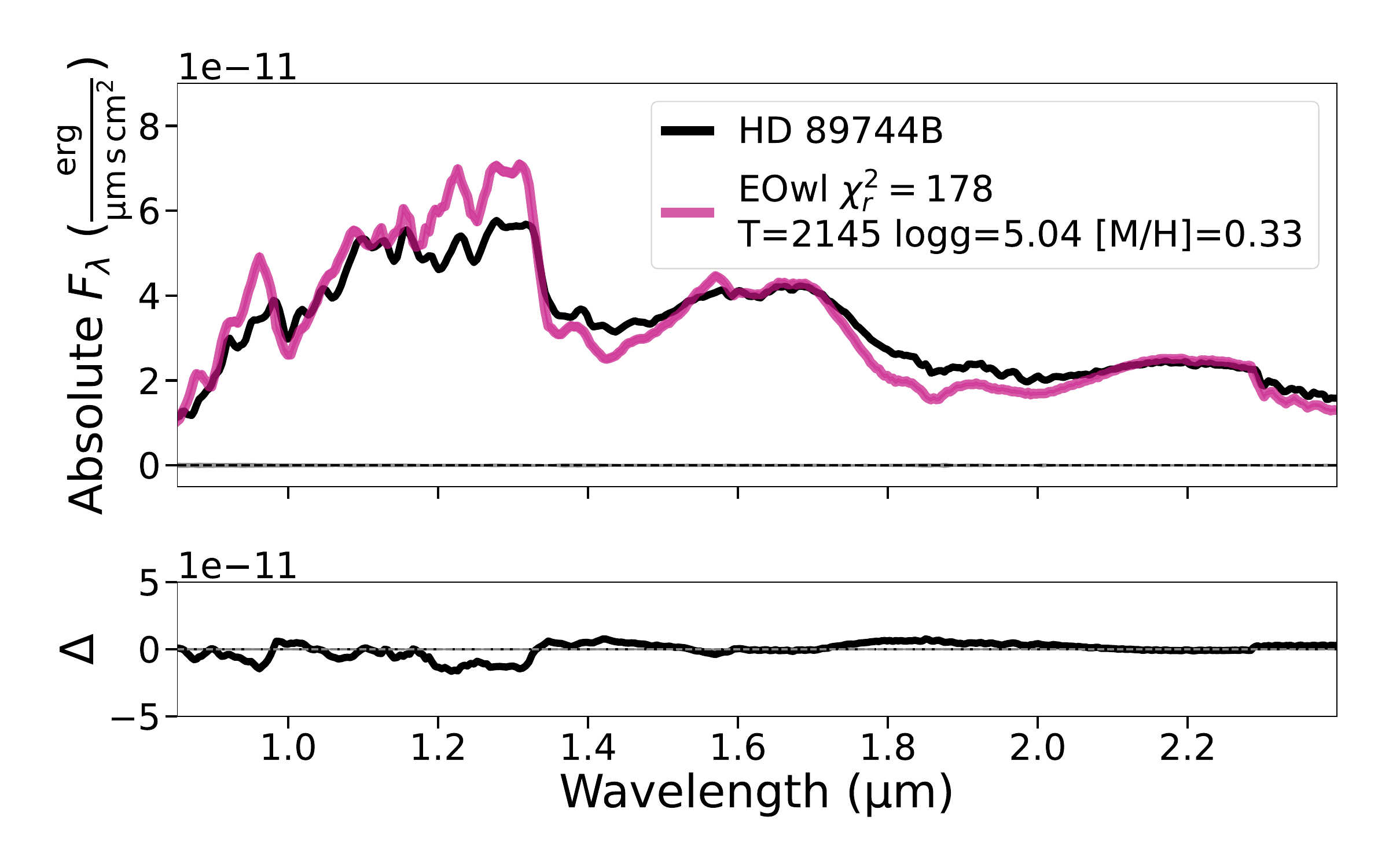} 
\includegraphics[width=0.32\textwidth]{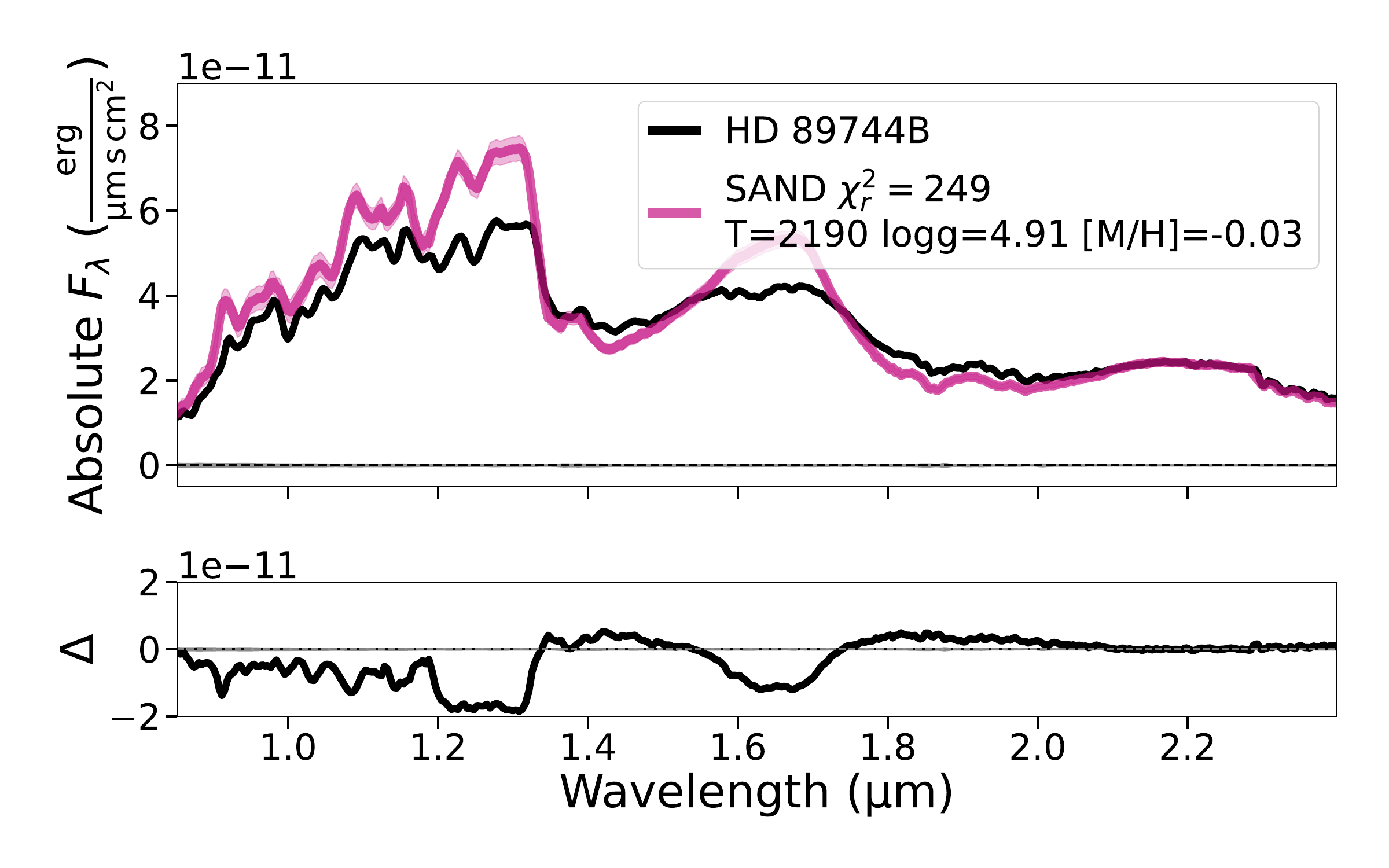} \\
\includegraphics[width=0.32\textwidth]{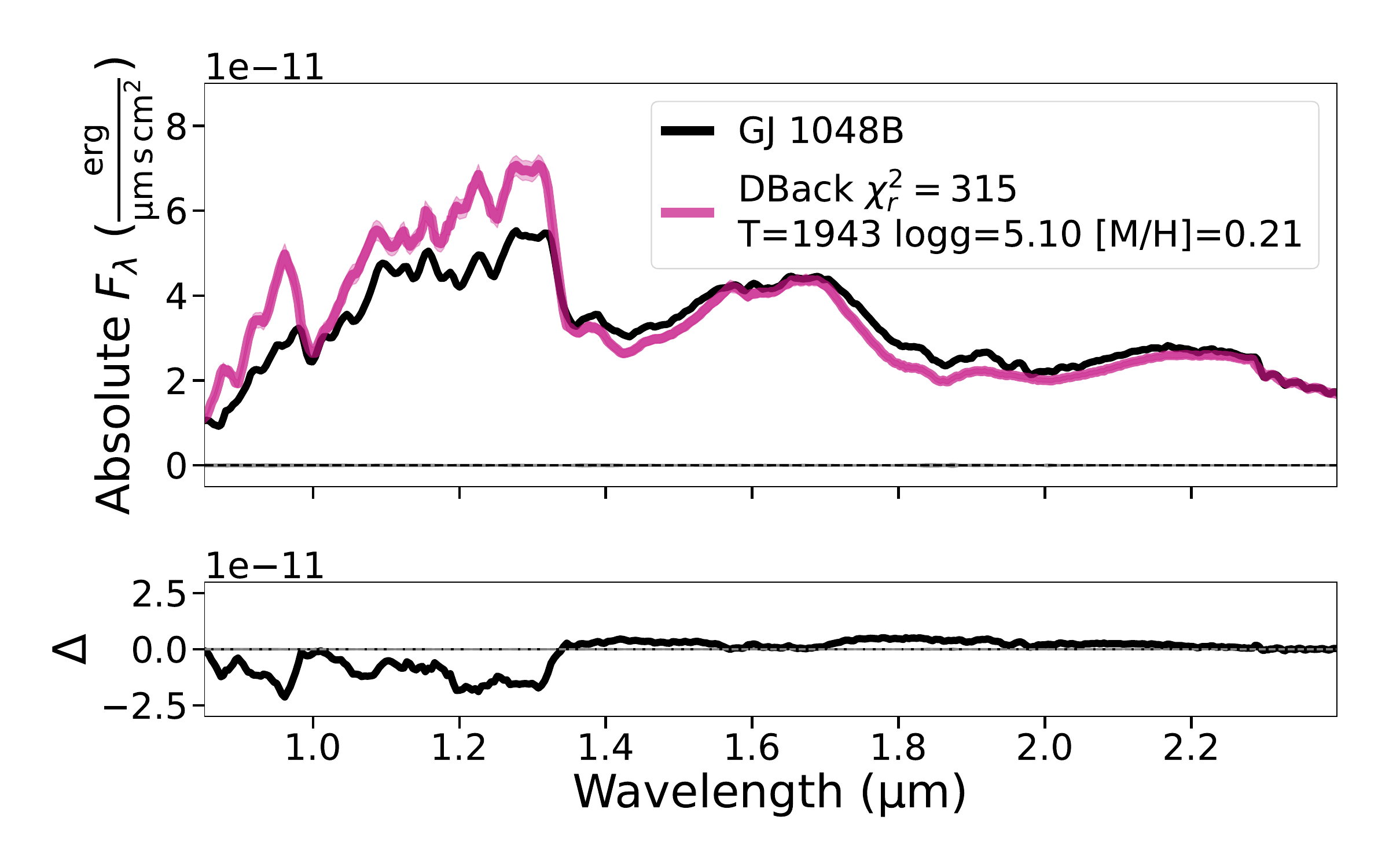} 
\includegraphics[width=0.32\textwidth]{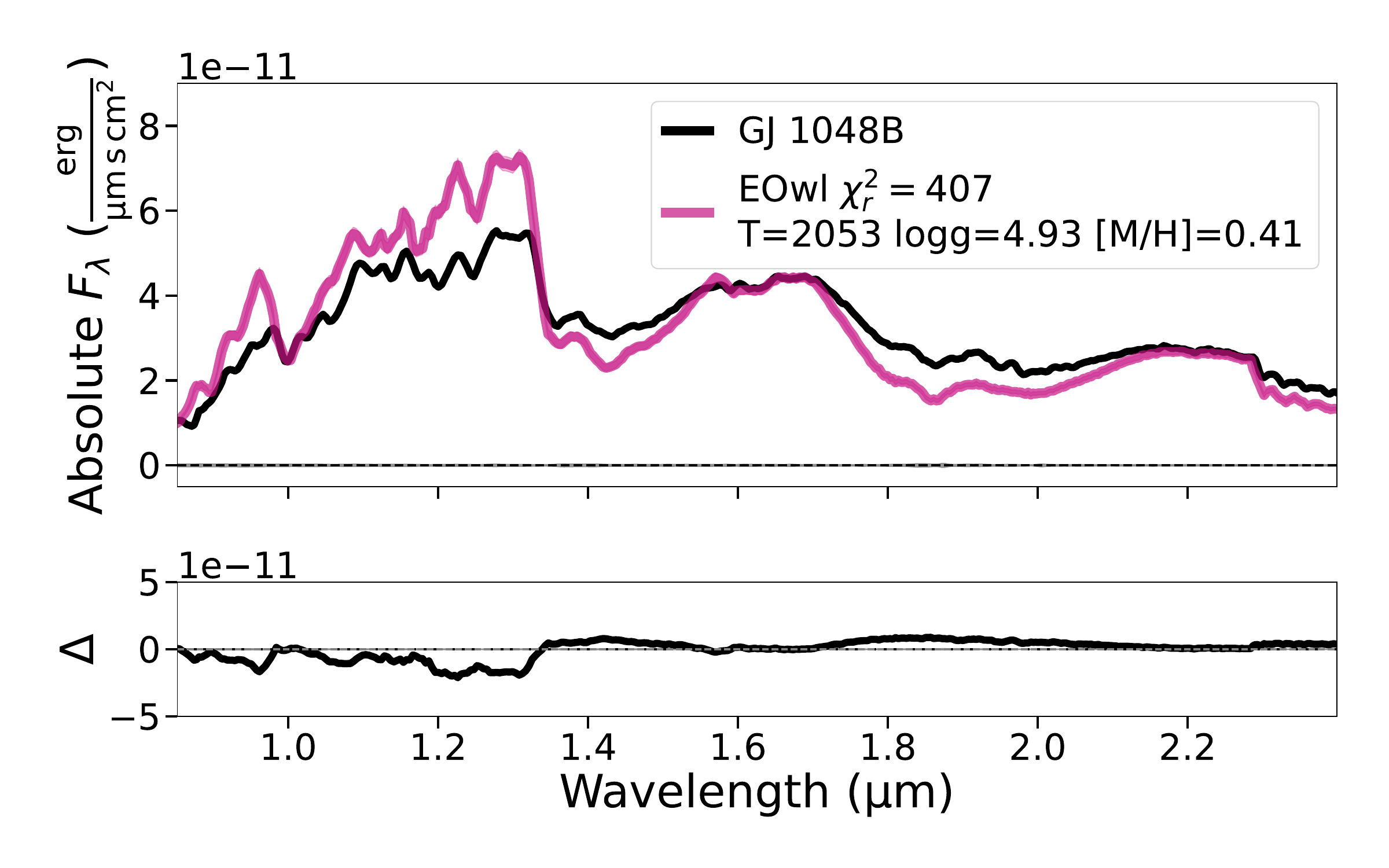} 
\includegraphics[width=0.32\textwidth]{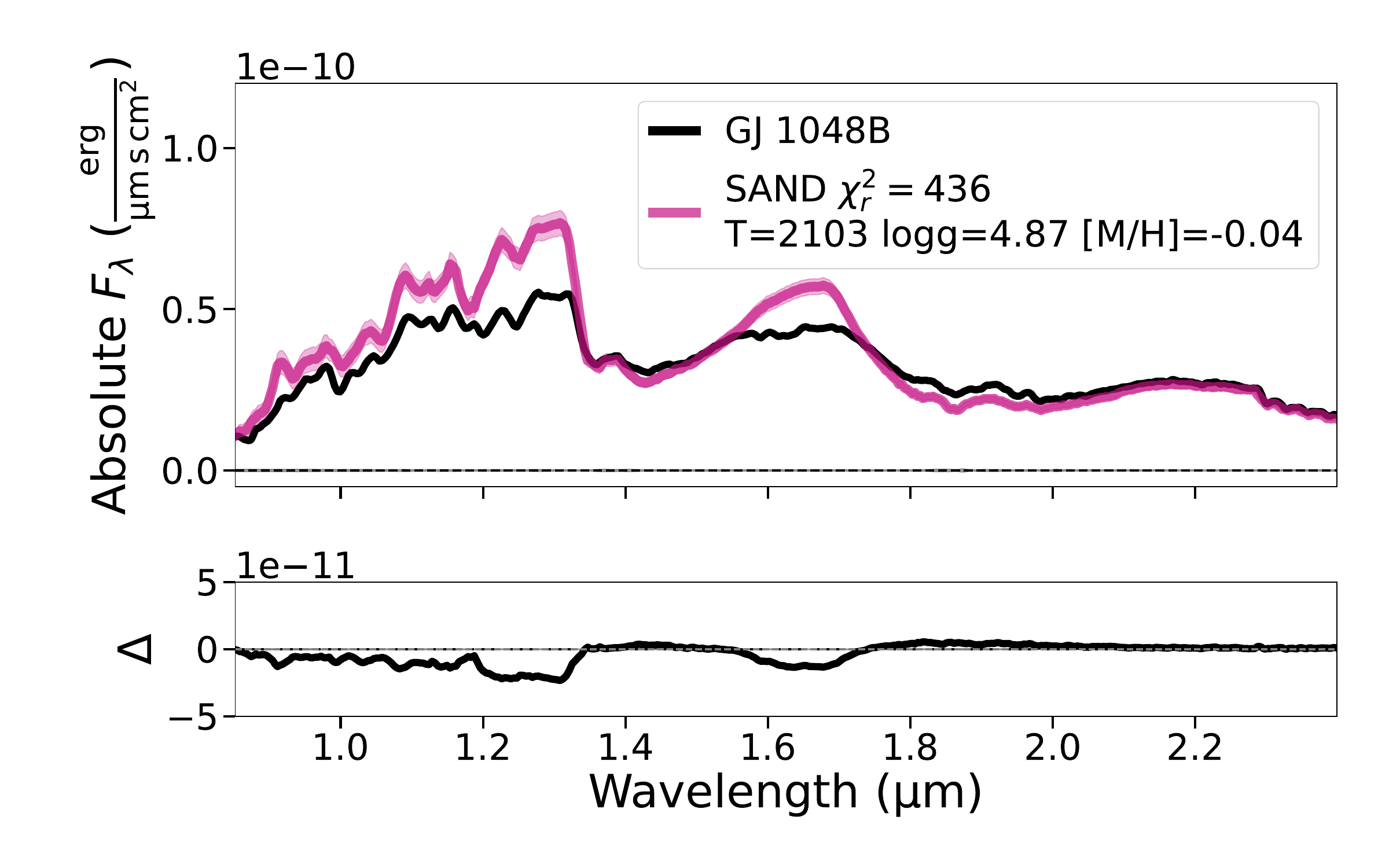} \\
\includegraphics[width=0.32\textwidth]{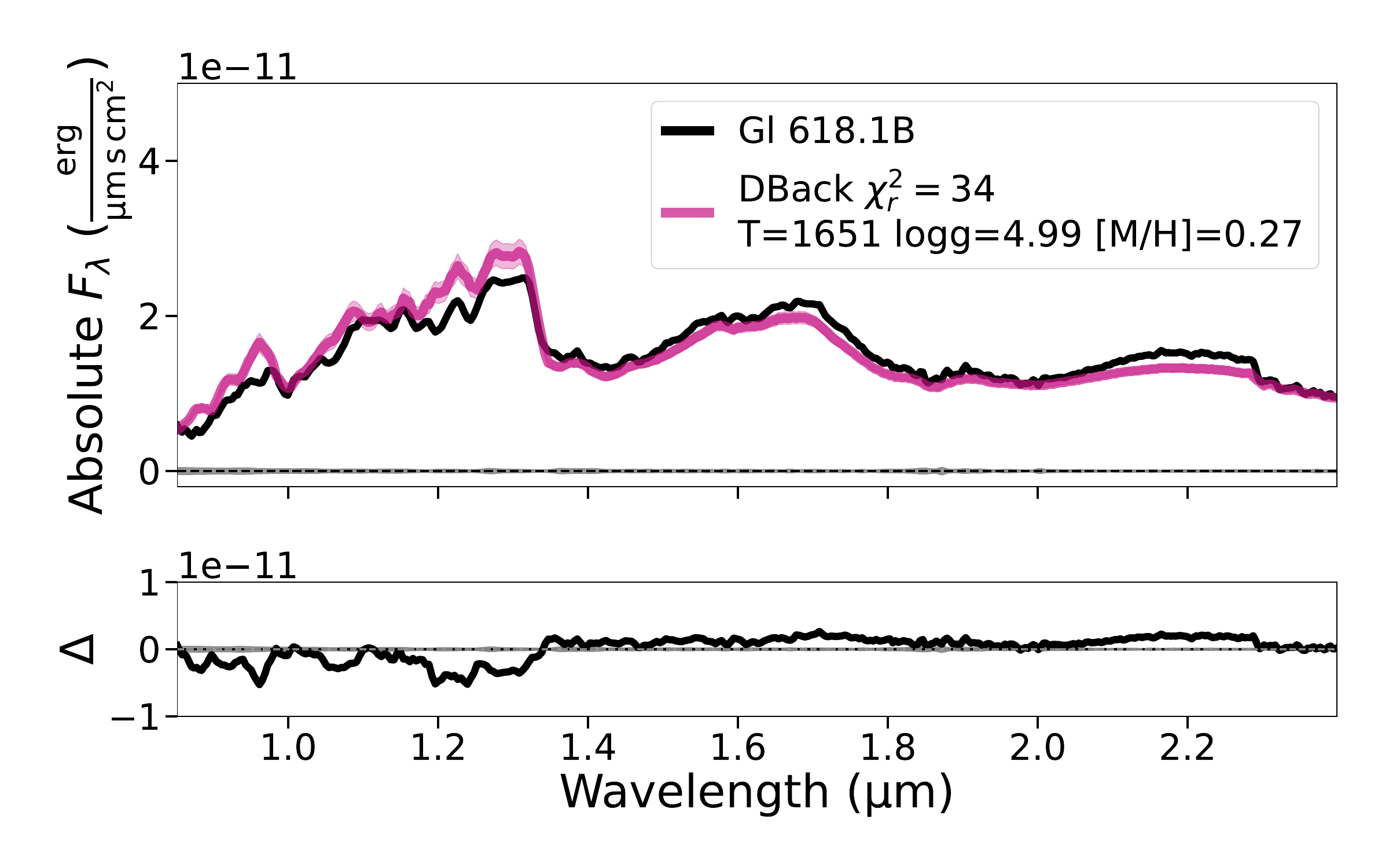} 
\includegraphics[width=0.32\textwidth]{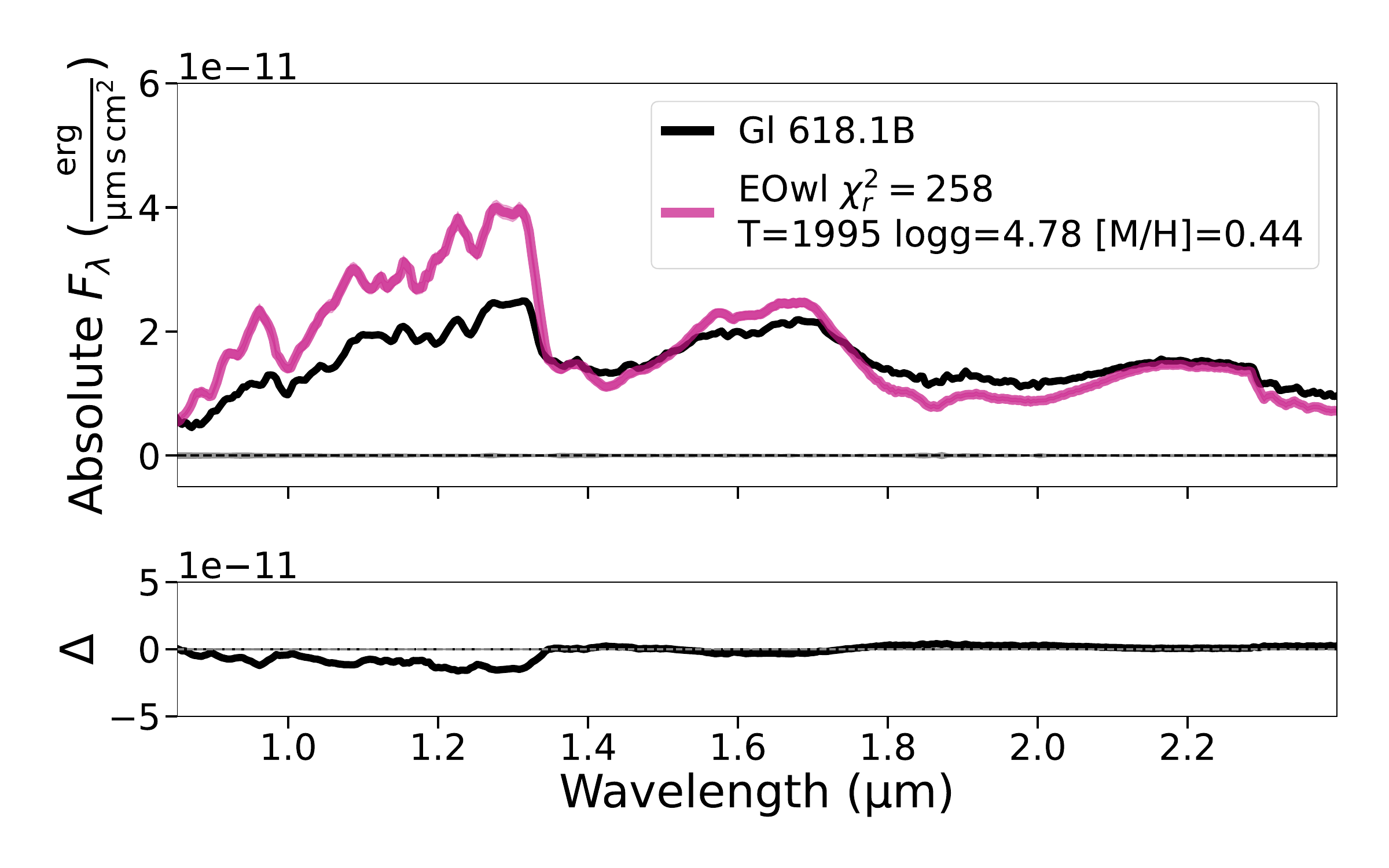} 
\includegraphics[width=0.32\textwidth]{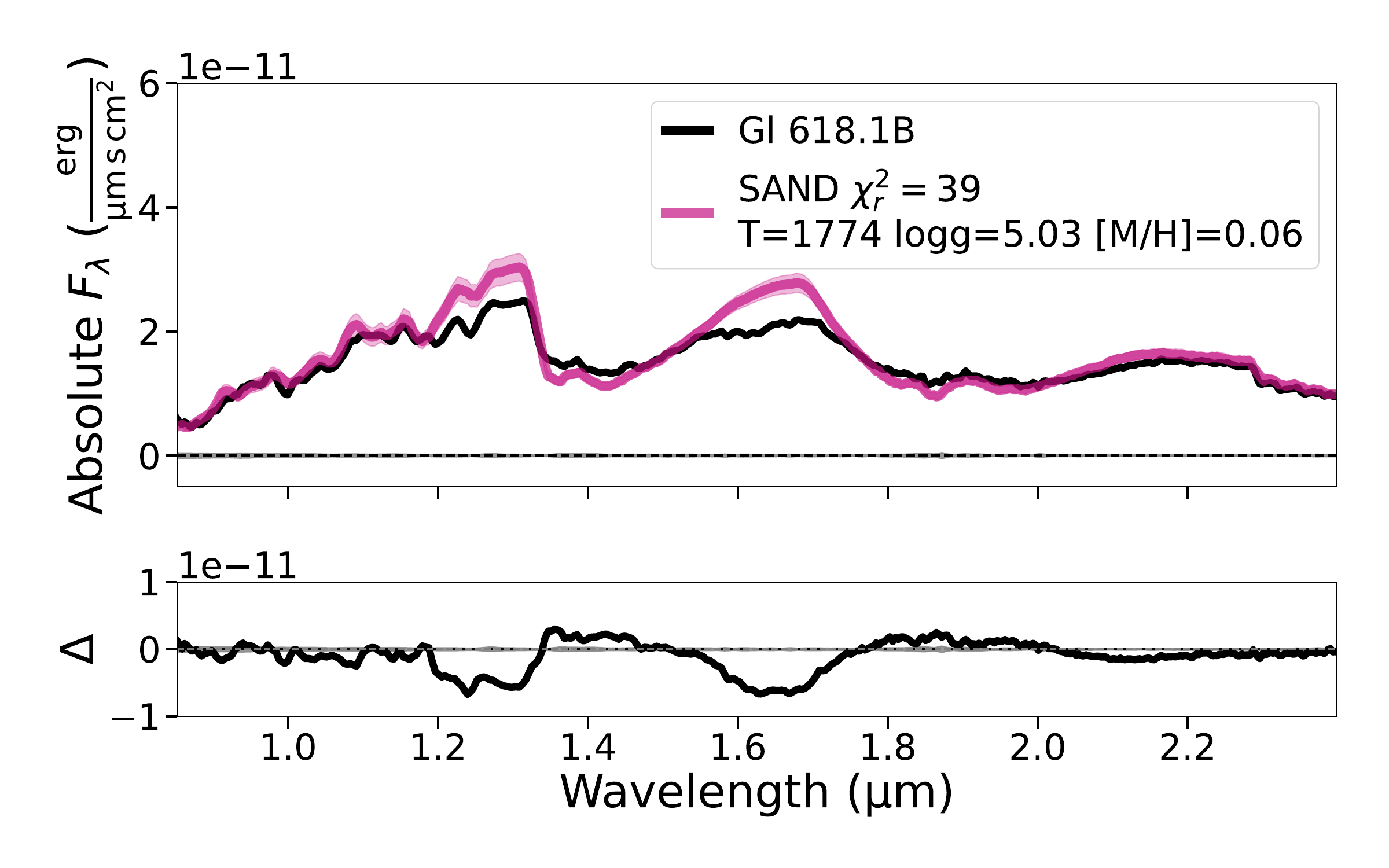}  \\
\includegraphics[width=0.32\textwidth]{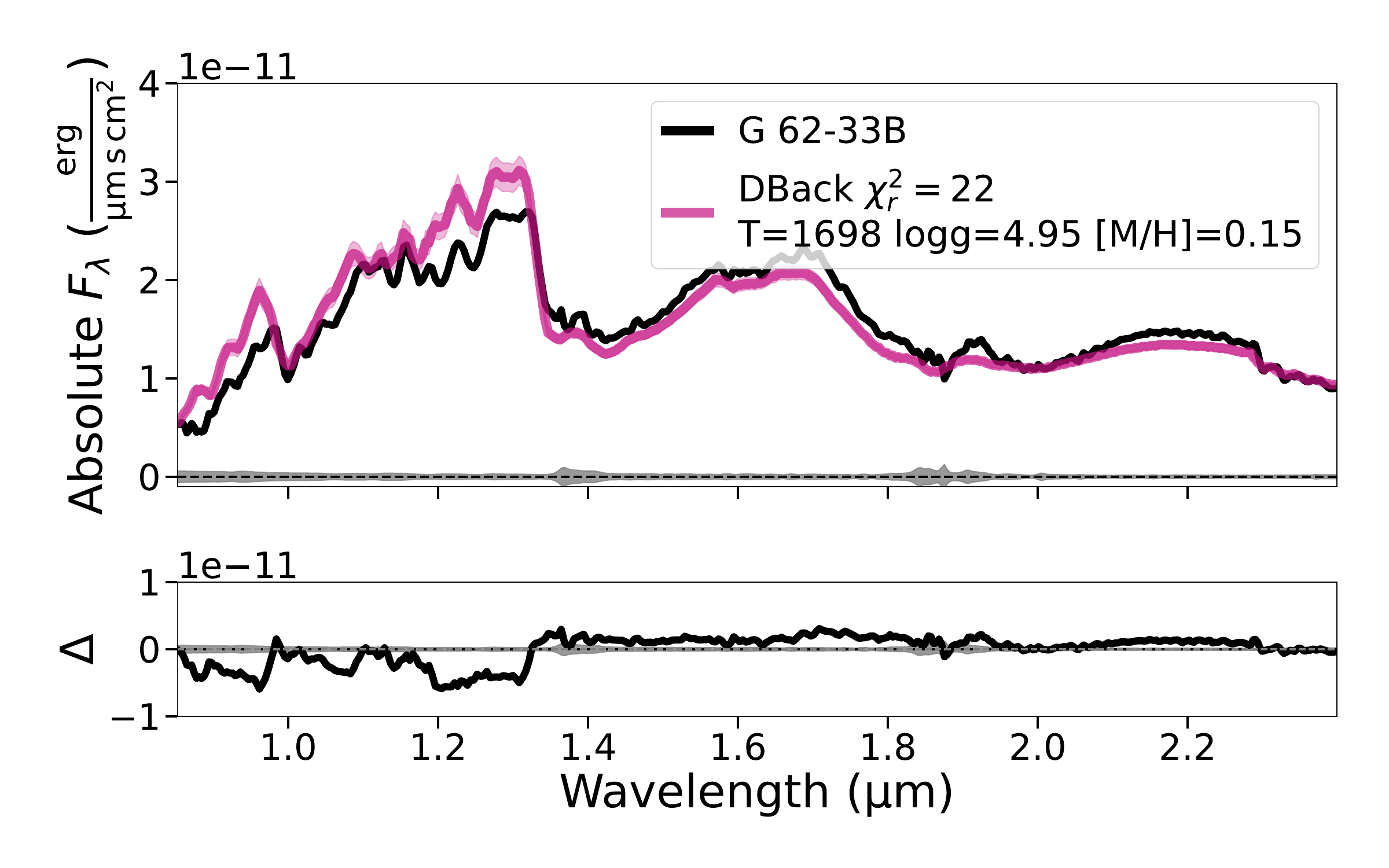}
\includegraphics[width=0.32\textwidth]{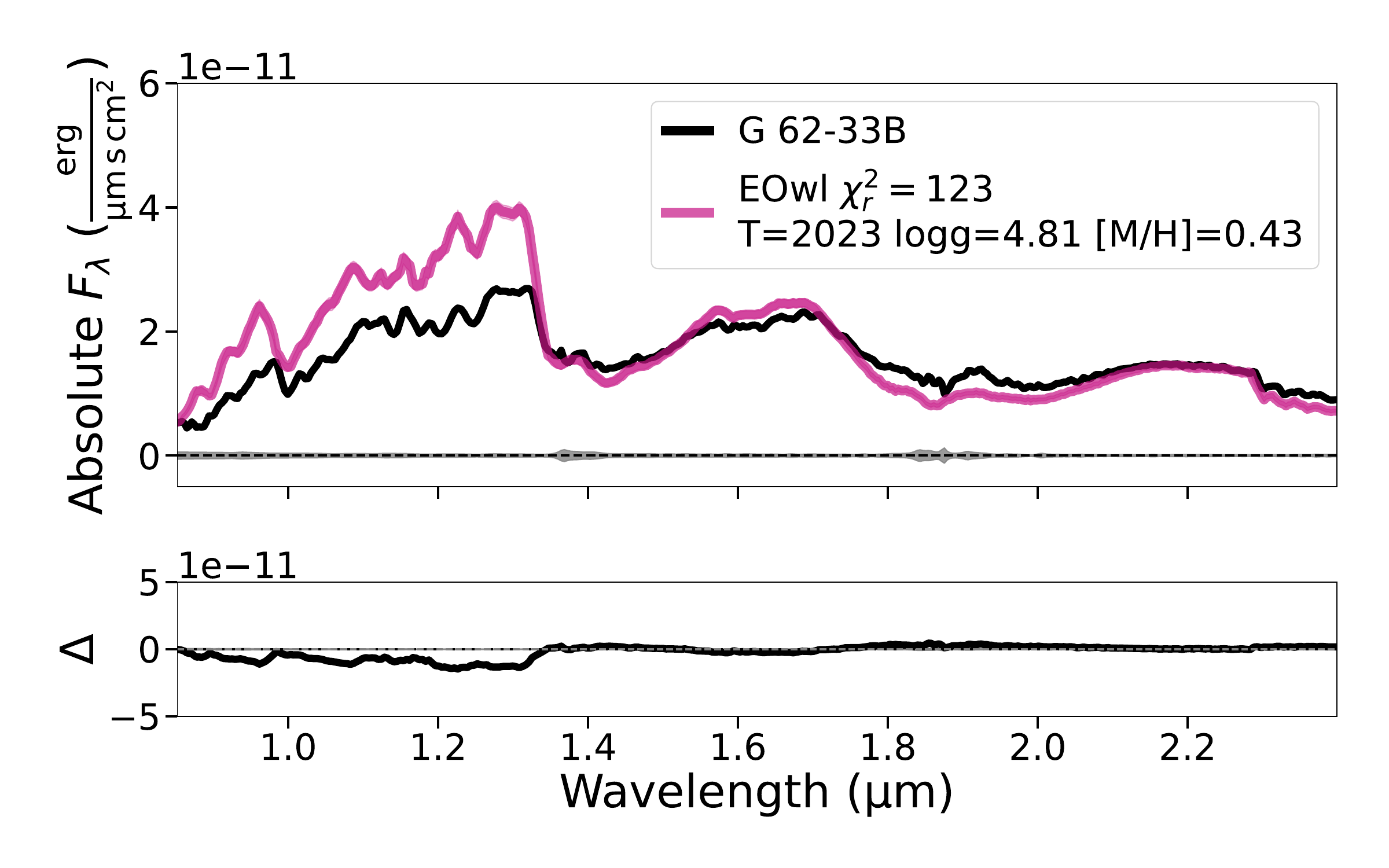}
\includegraphics[width=0.32\textwidth]{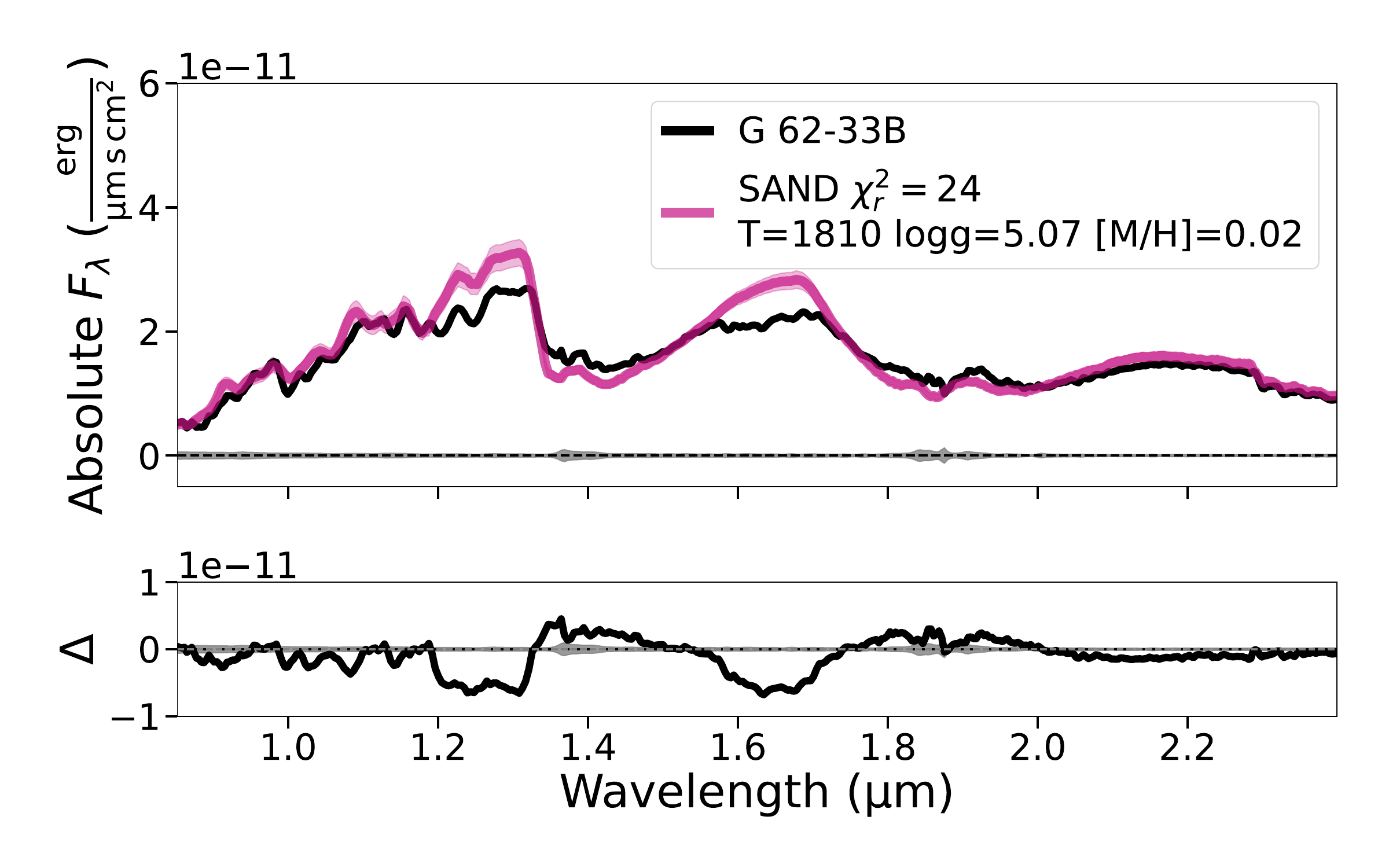} \\
\includegraphics[width=0.32\textwidth]{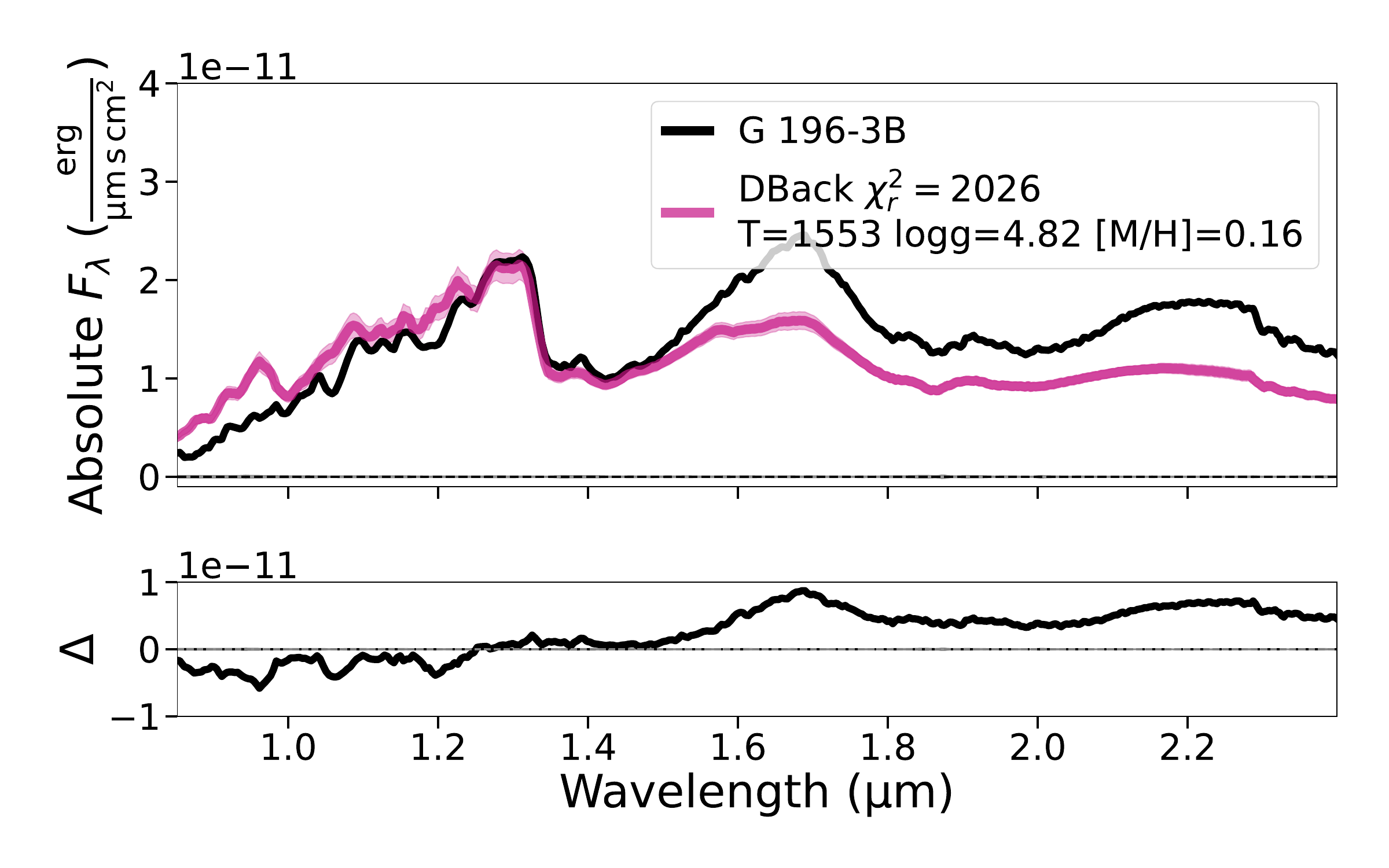} 
\includegraphics[width=0.32\textwidth]{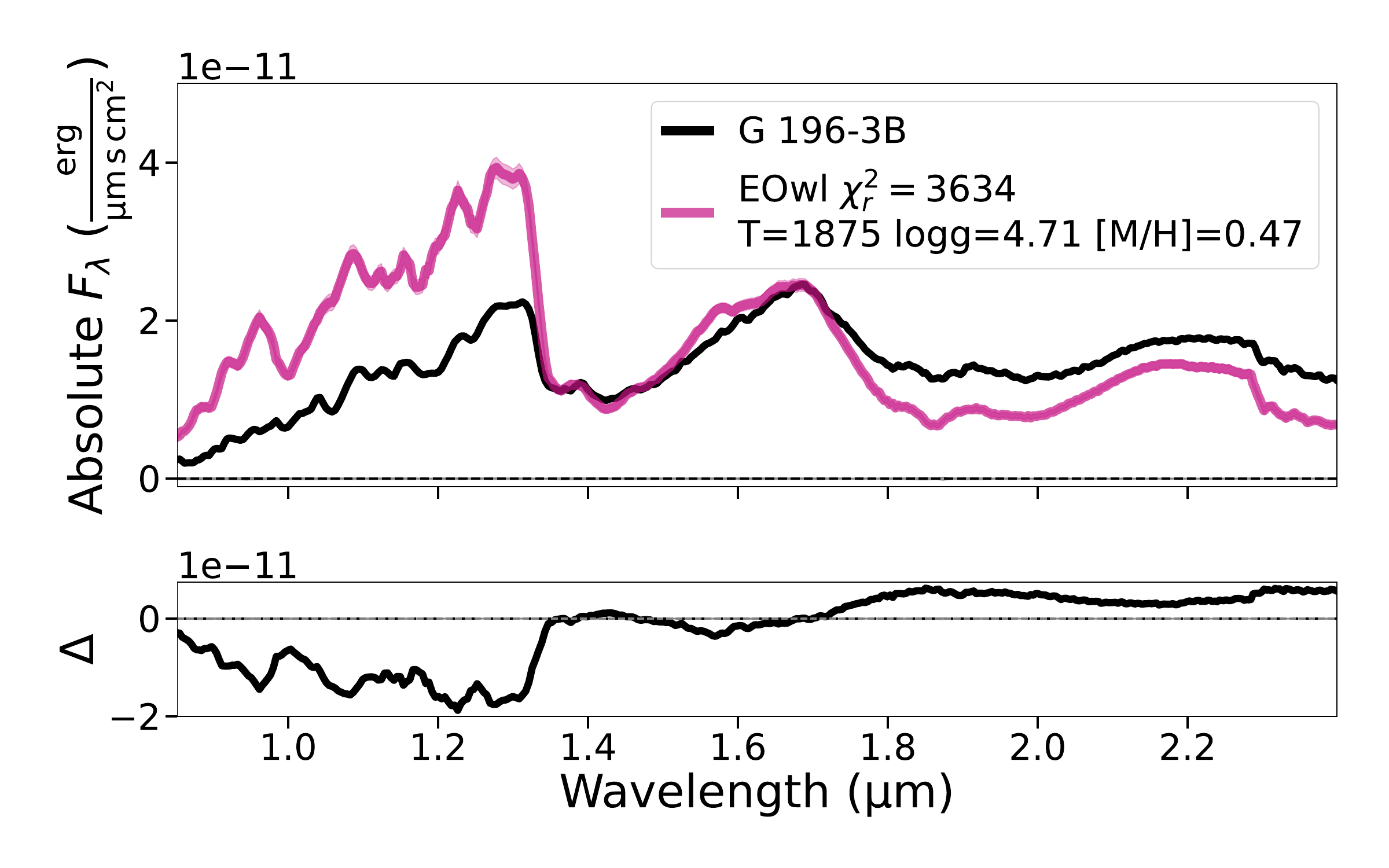} 
\includegraphics[width=0.32\textwidth]{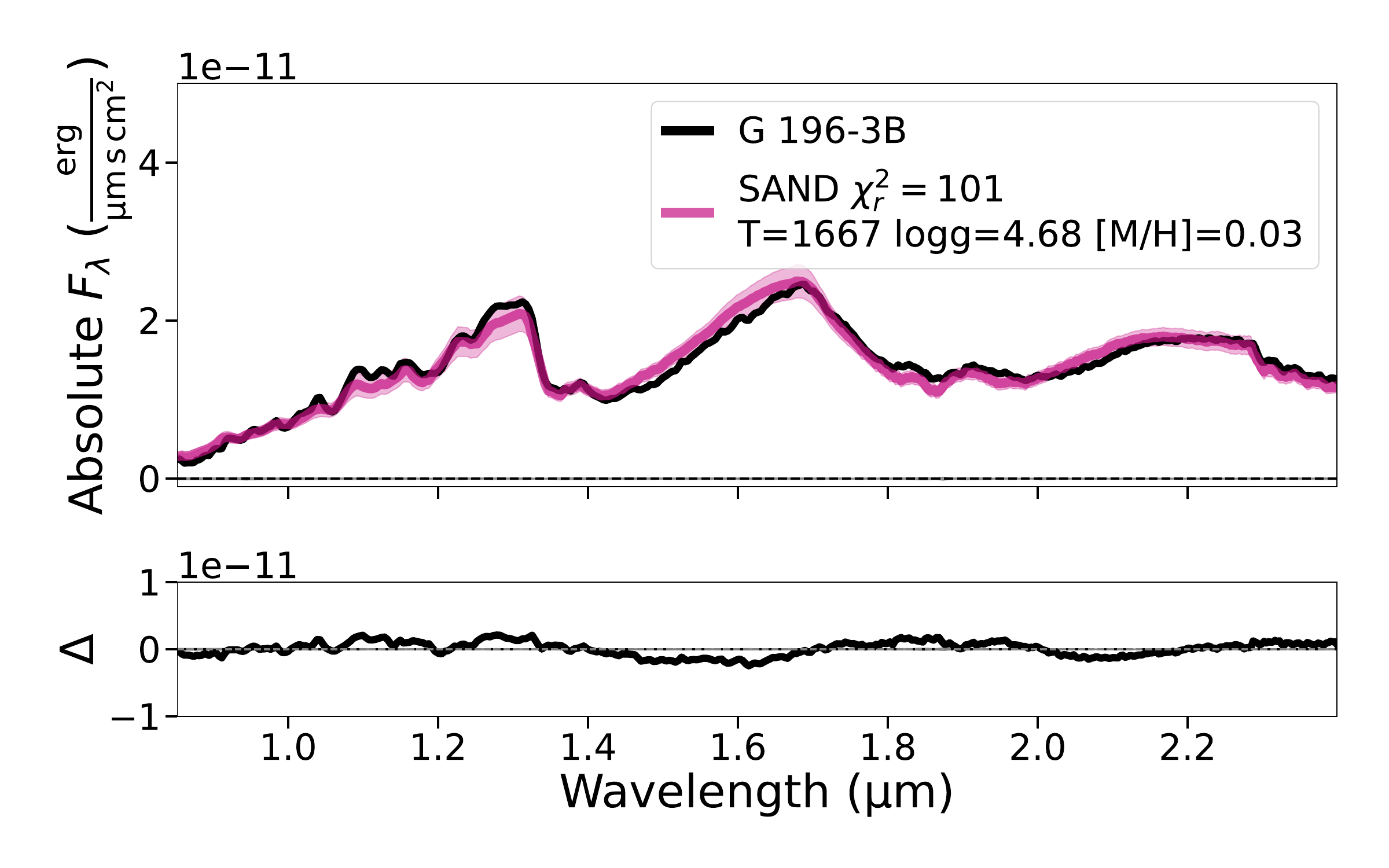}  \\
\caption{Same as Figure~\ref{fig:mcmc-comp1} but for random forest retrieval spectral fits for Diamondback (DBack, left), Elf Owl (EOwl, middle), and the SAND (right) model grids. The bottom panel in each plot compares the difference ($\Delta$ = model - data) to the $\pm1\sigma$ uncertainty spectrum (gray bands). Magenta curves display the median fits while shaded areas display the 1$\sigma$ uncertainties based on posterior draws.
\label{fig:rfr-comp1}}
\end{figure}

\begin{figure}
\centering
\includegraphics[width=0.32\textwidth]{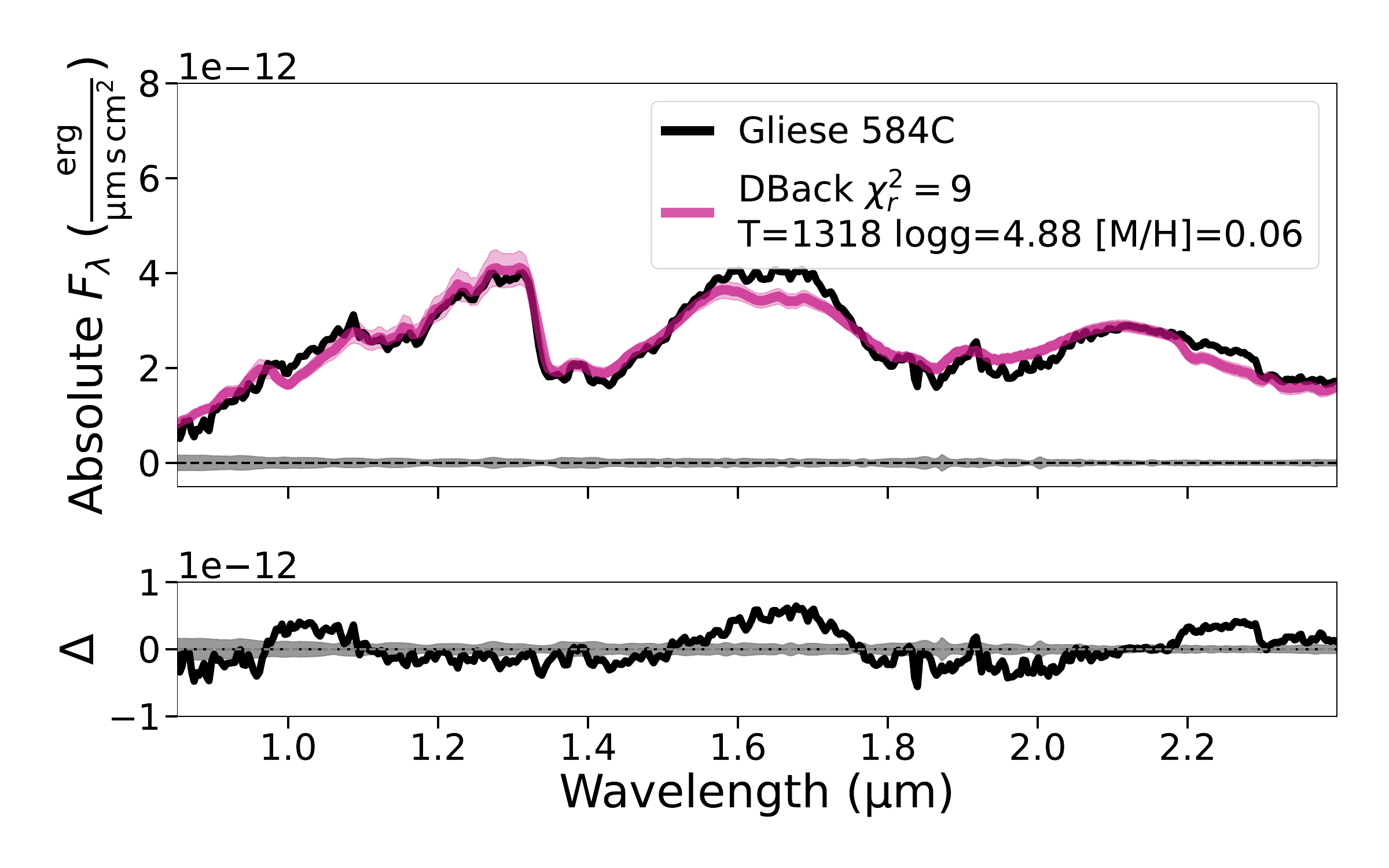}
\includegraphics[width=0.32\textwidth]{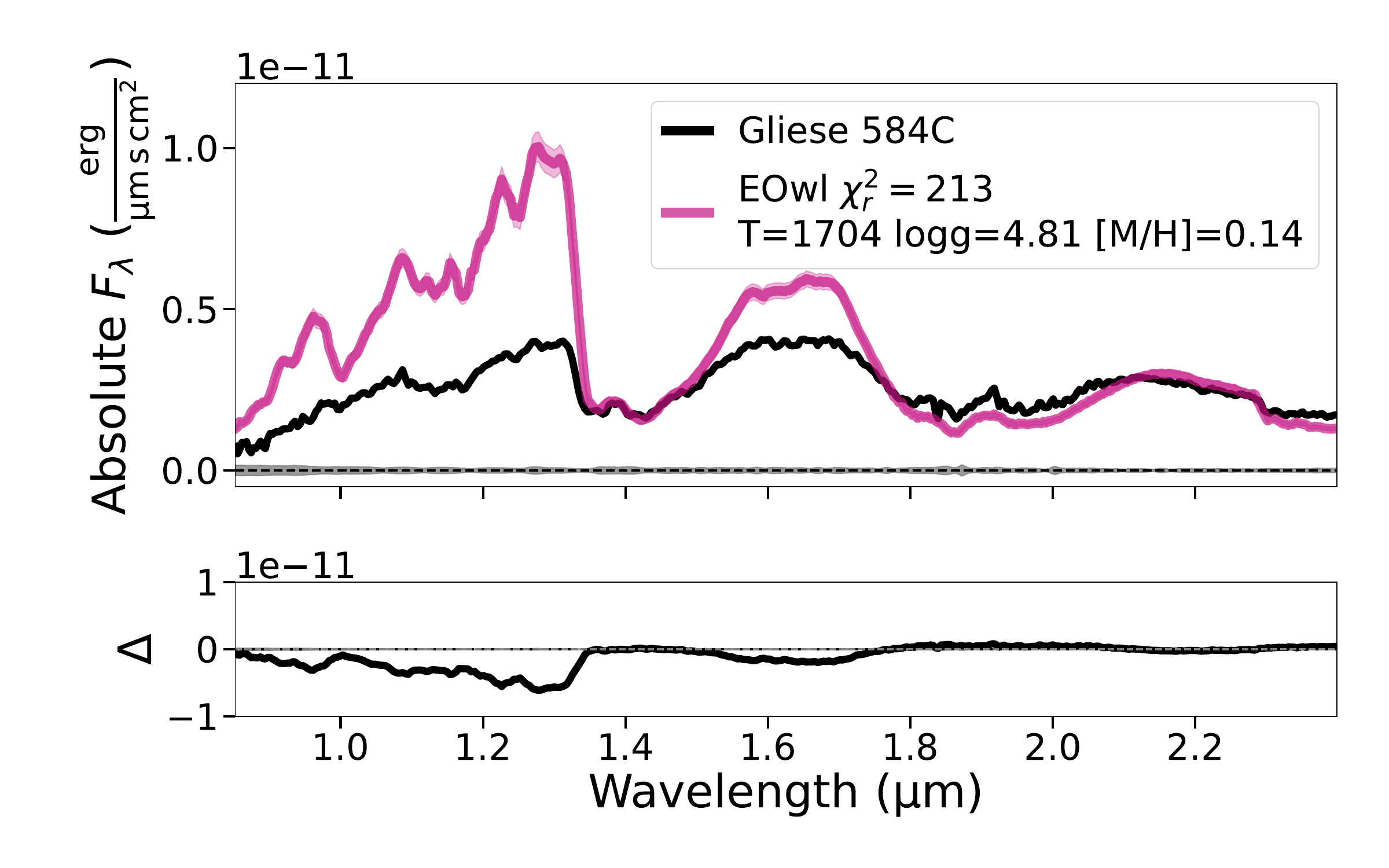}
\includegraphics[width=0.32\textwidth]{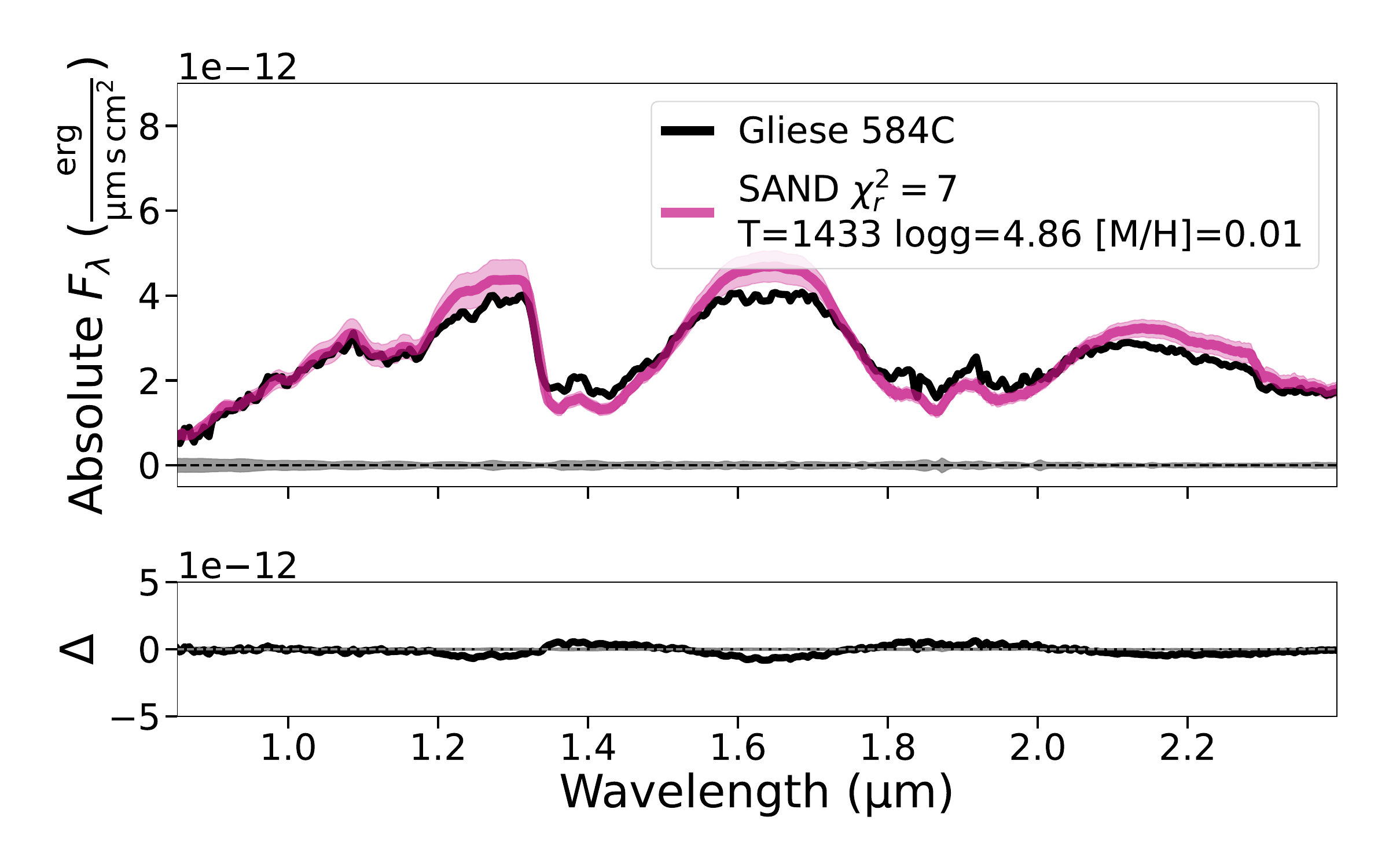} \\
\includegraphics[width=0.32\textwidth]{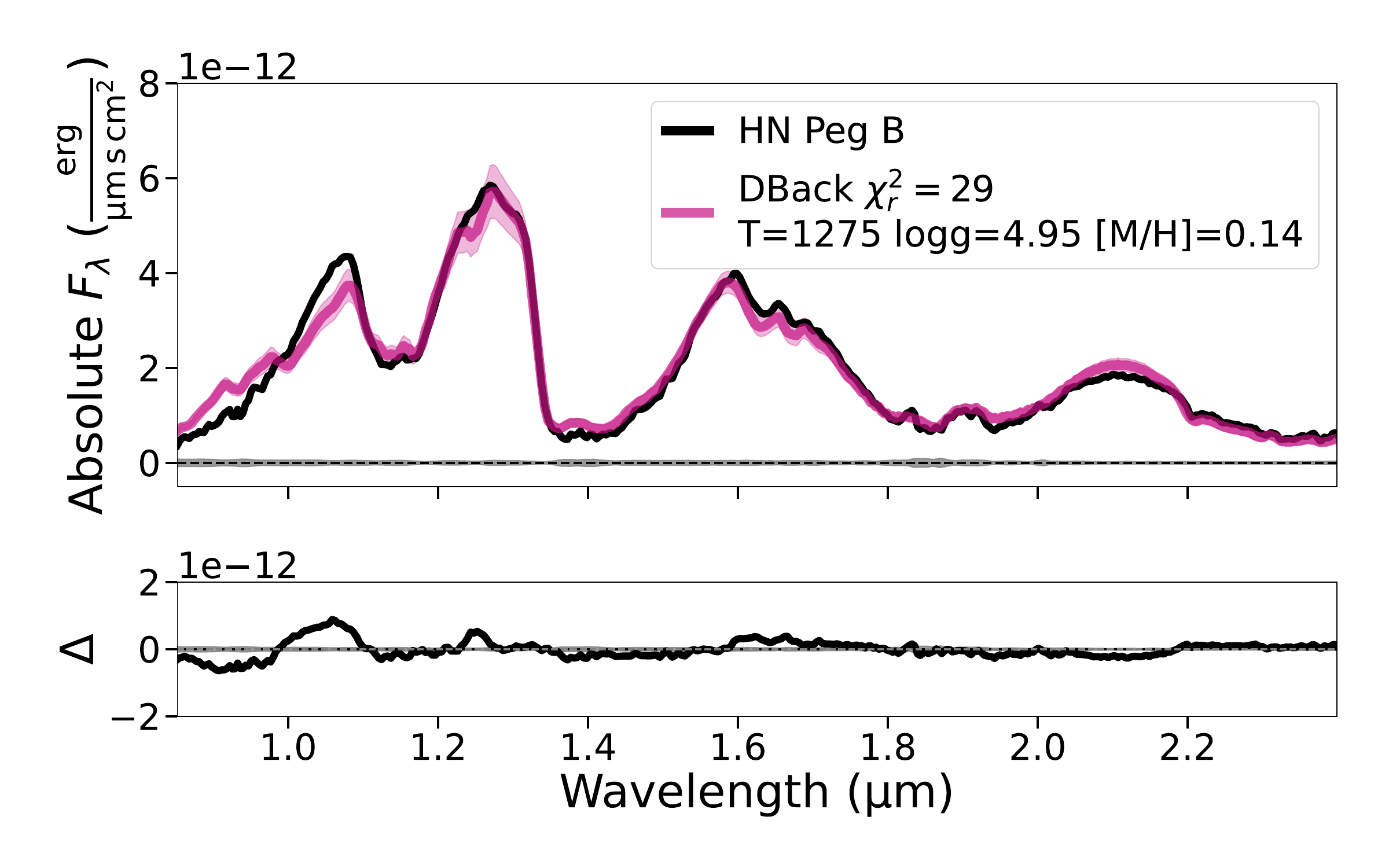} 
\includegraphics[width=0.32\textwidth]{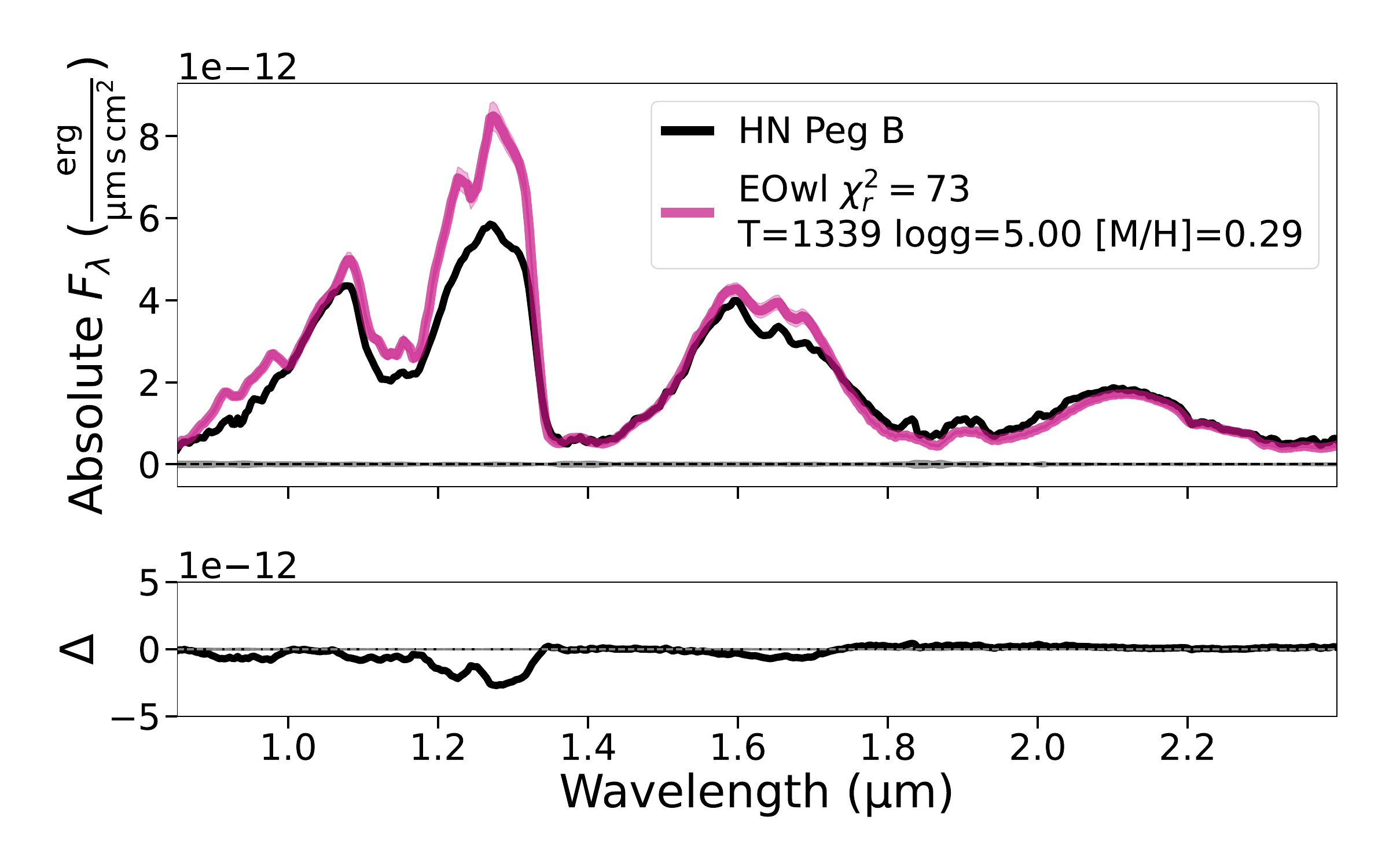} 
\includegraphics[width=0.32\textwidth]{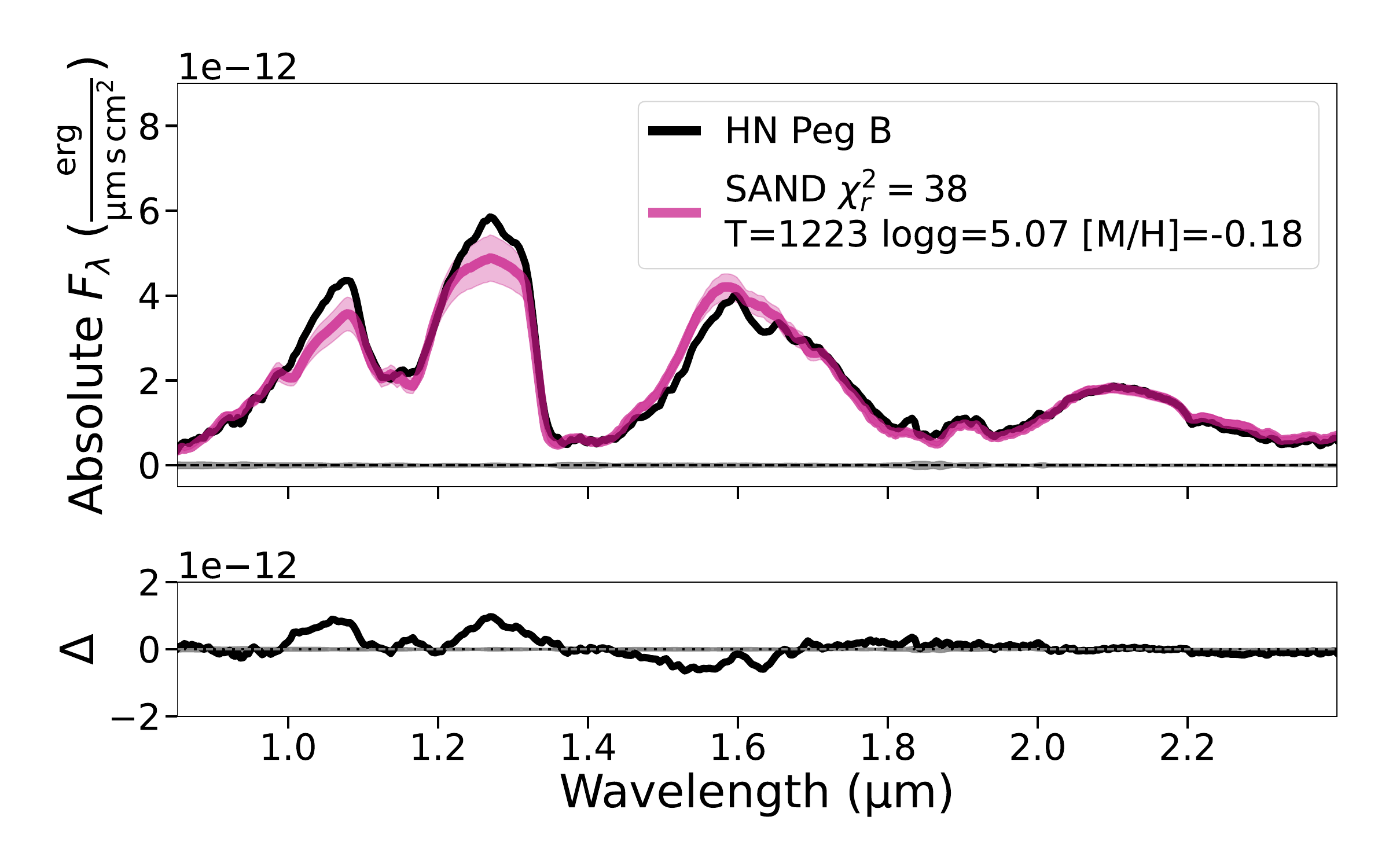} \\
\includegraphics[width=0.32\textwidth]{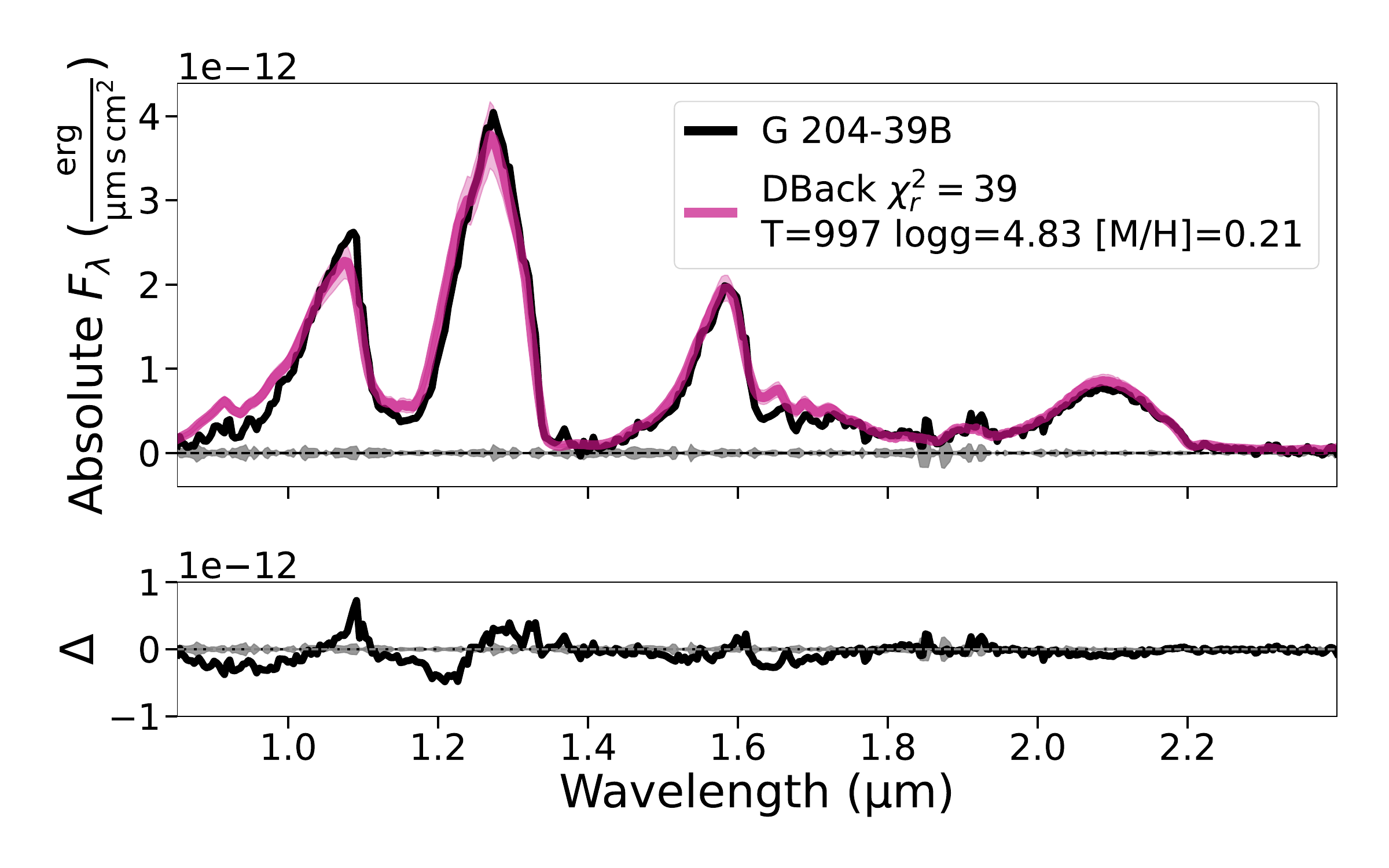} 
\includegraphics[width=0.32\textwidth]{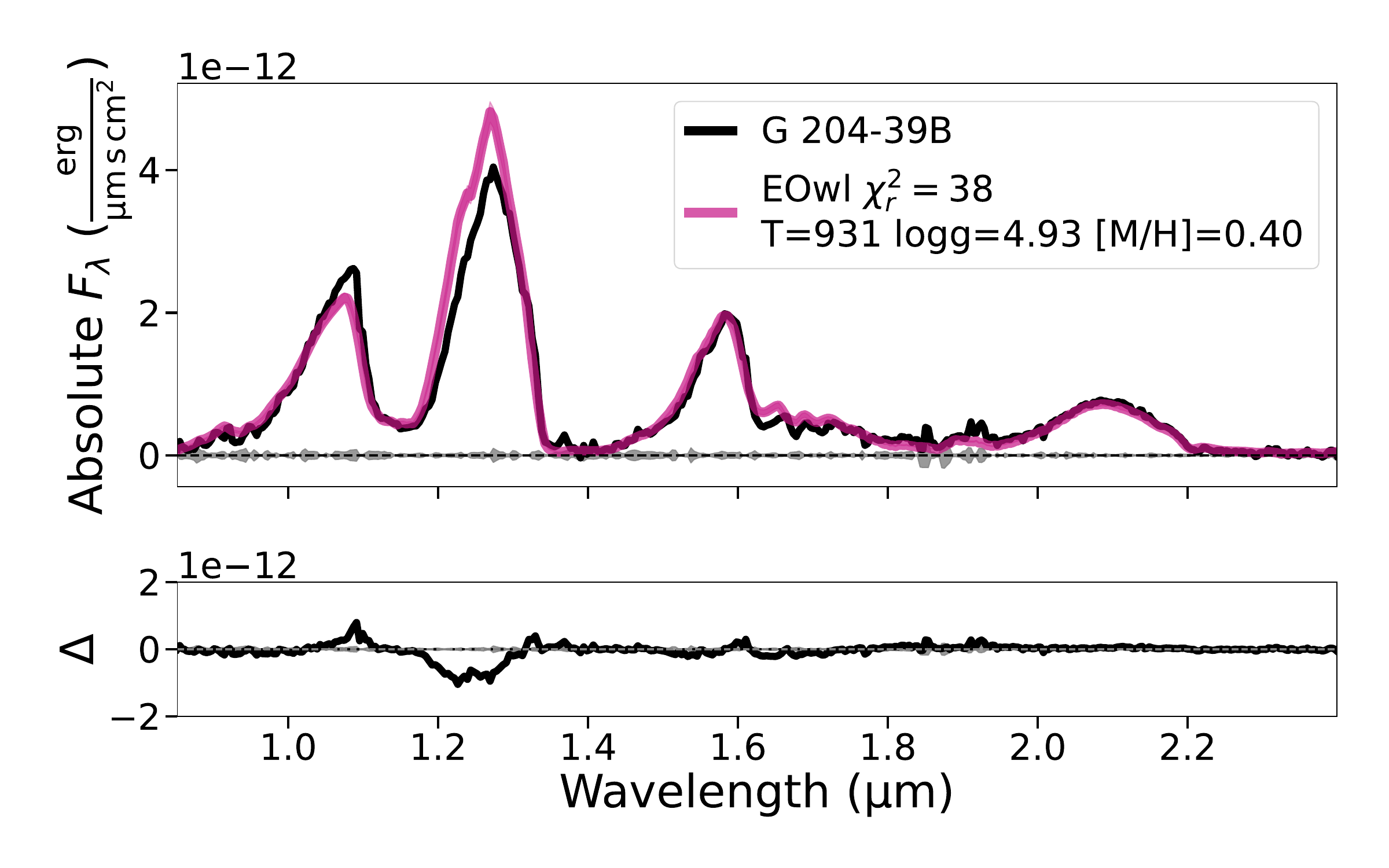} 
\includegraphics[width=0.32\textwidth]{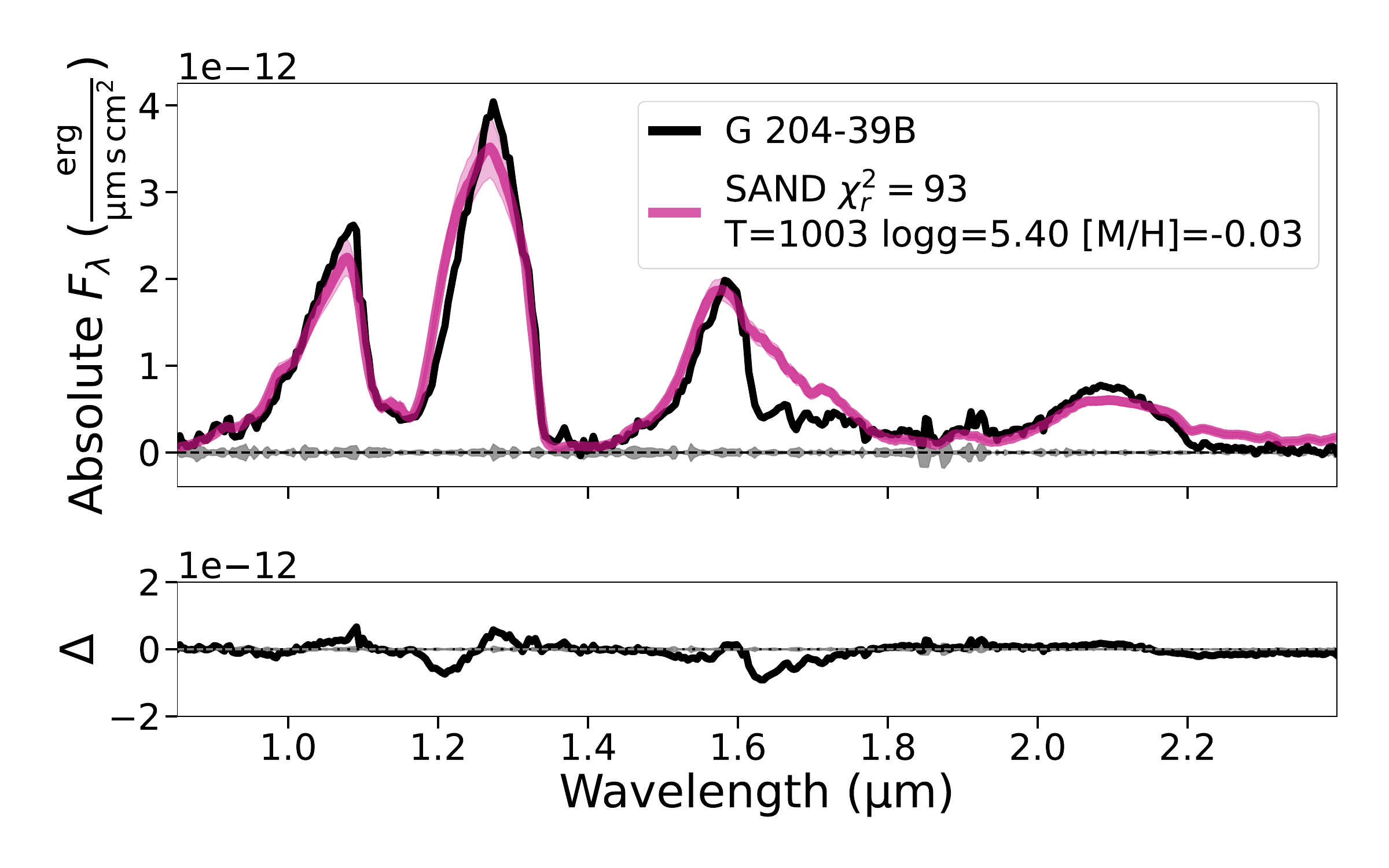} \\
\includegraphics[width=0.32\textwidth]{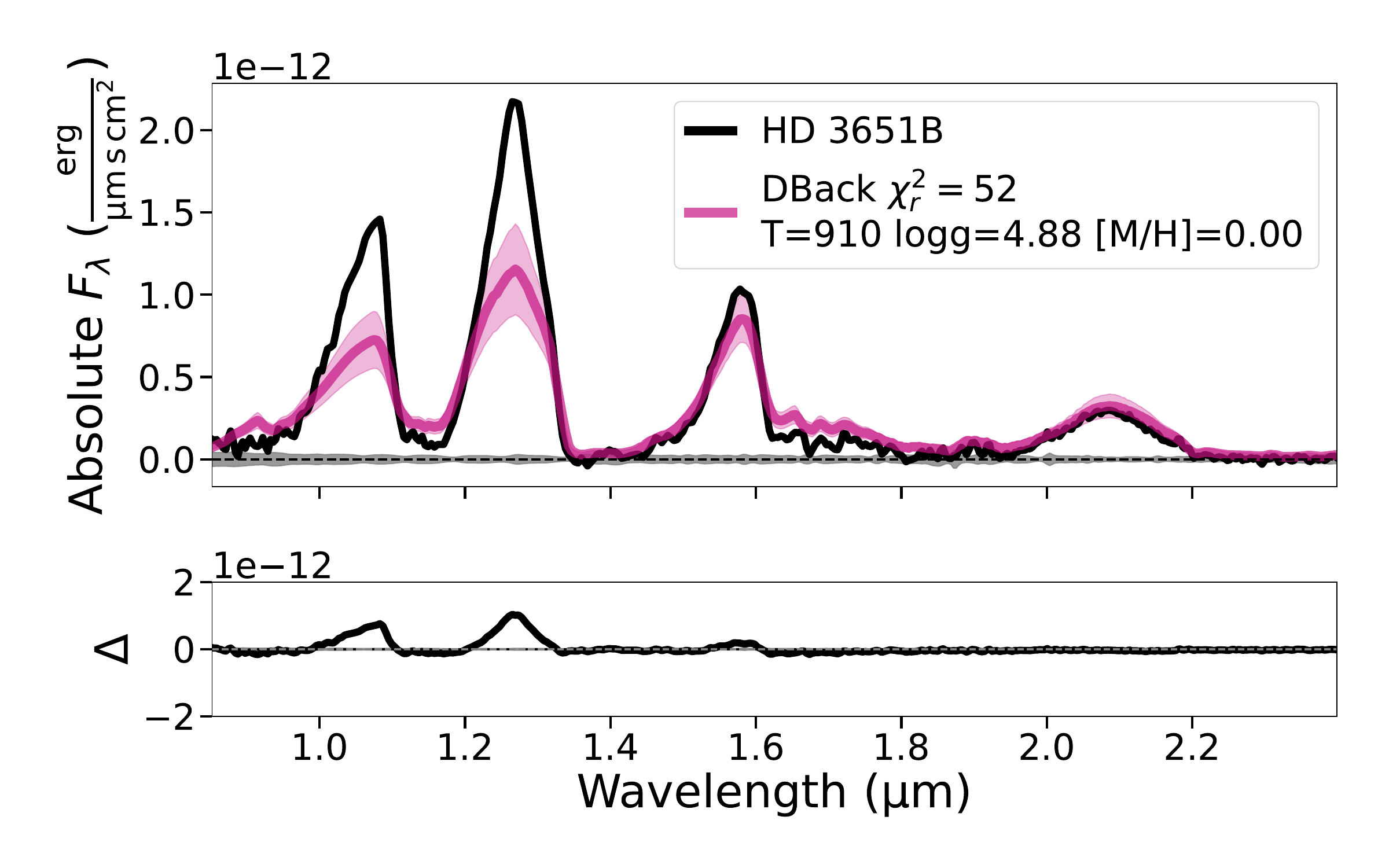} 
\includegraphics[width=0.32\textwidth]{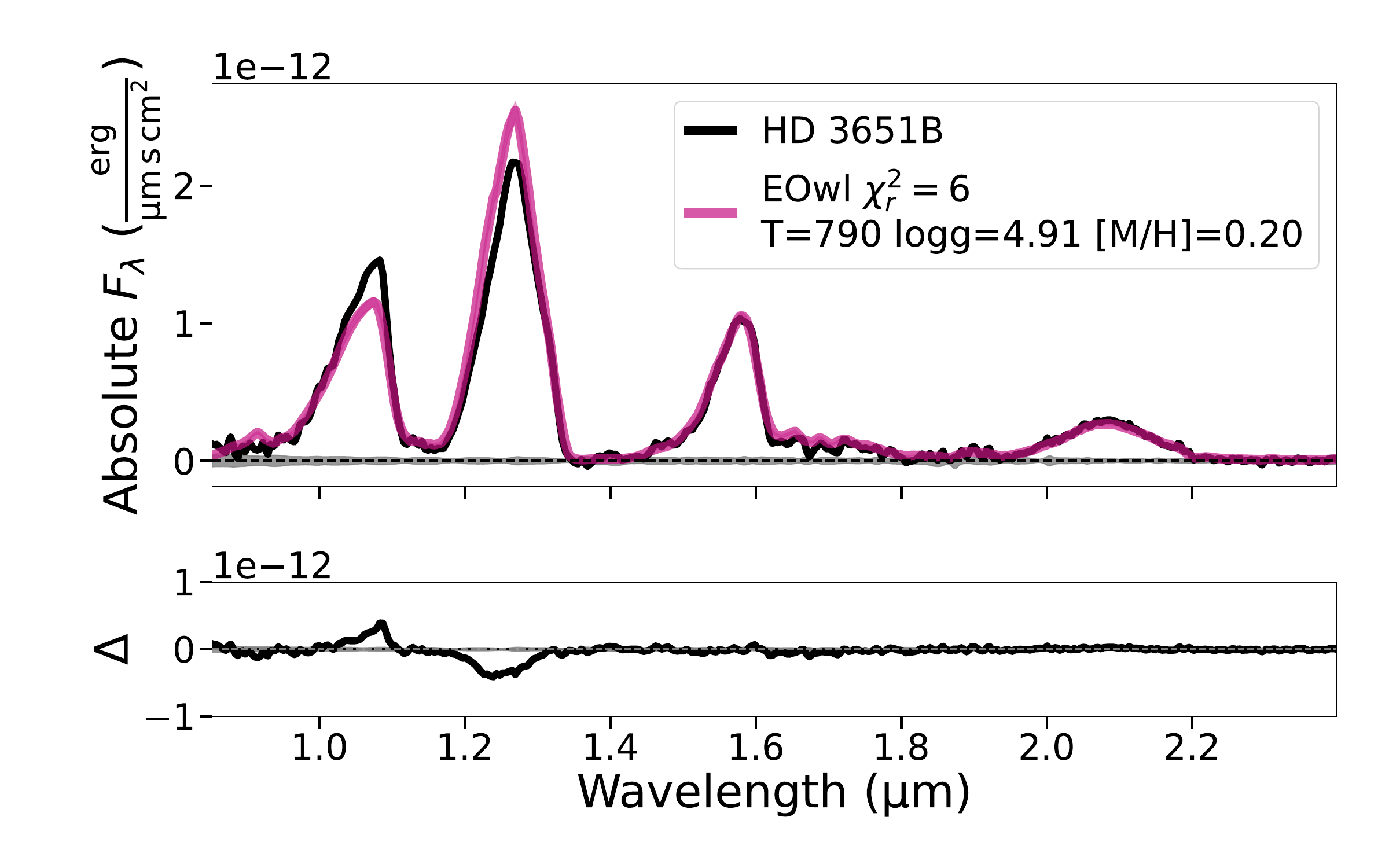} 
\includegraphics[width=0.32\textwidth]{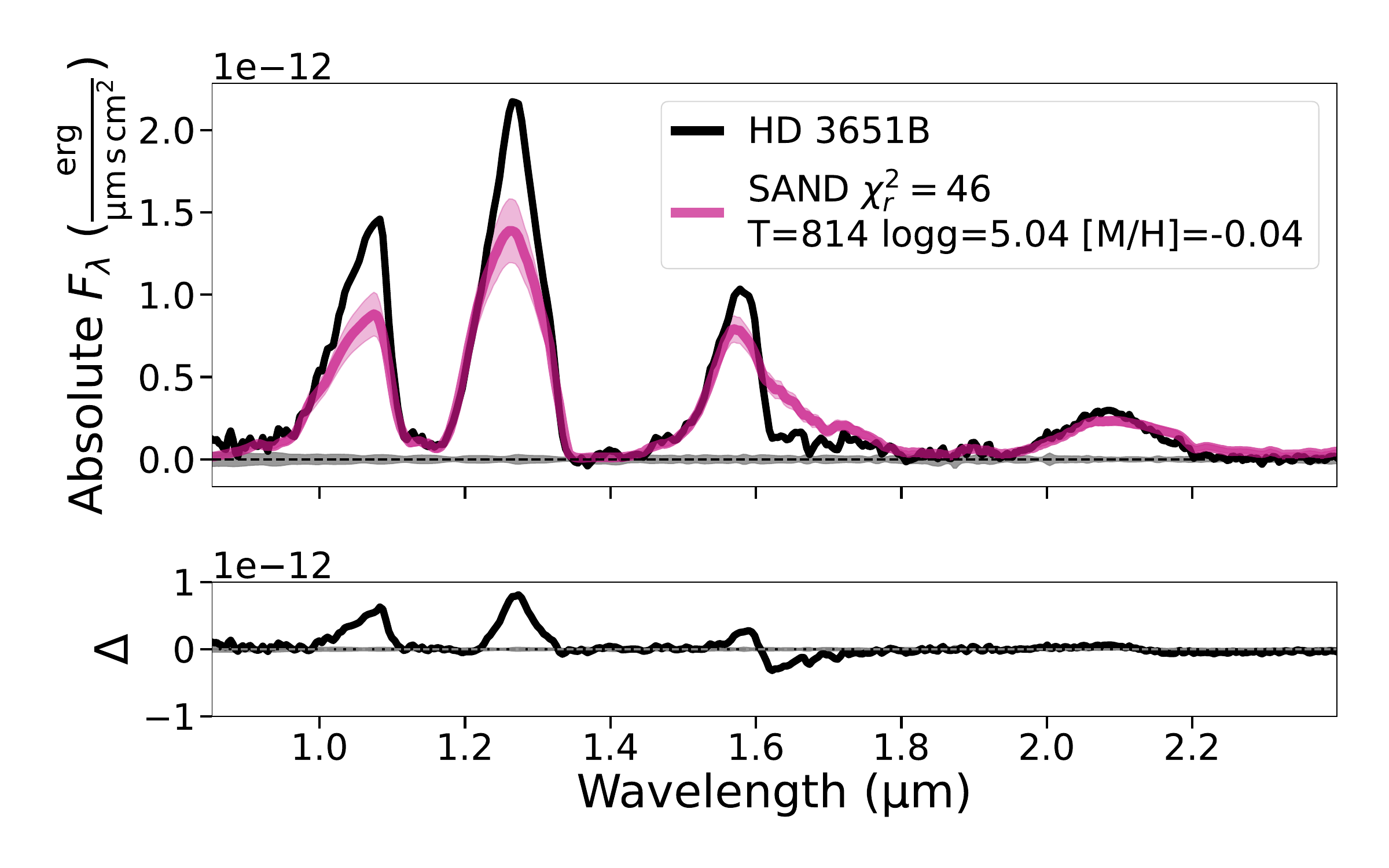} \\ 
\includegraphics[width=0.32\textwidth]{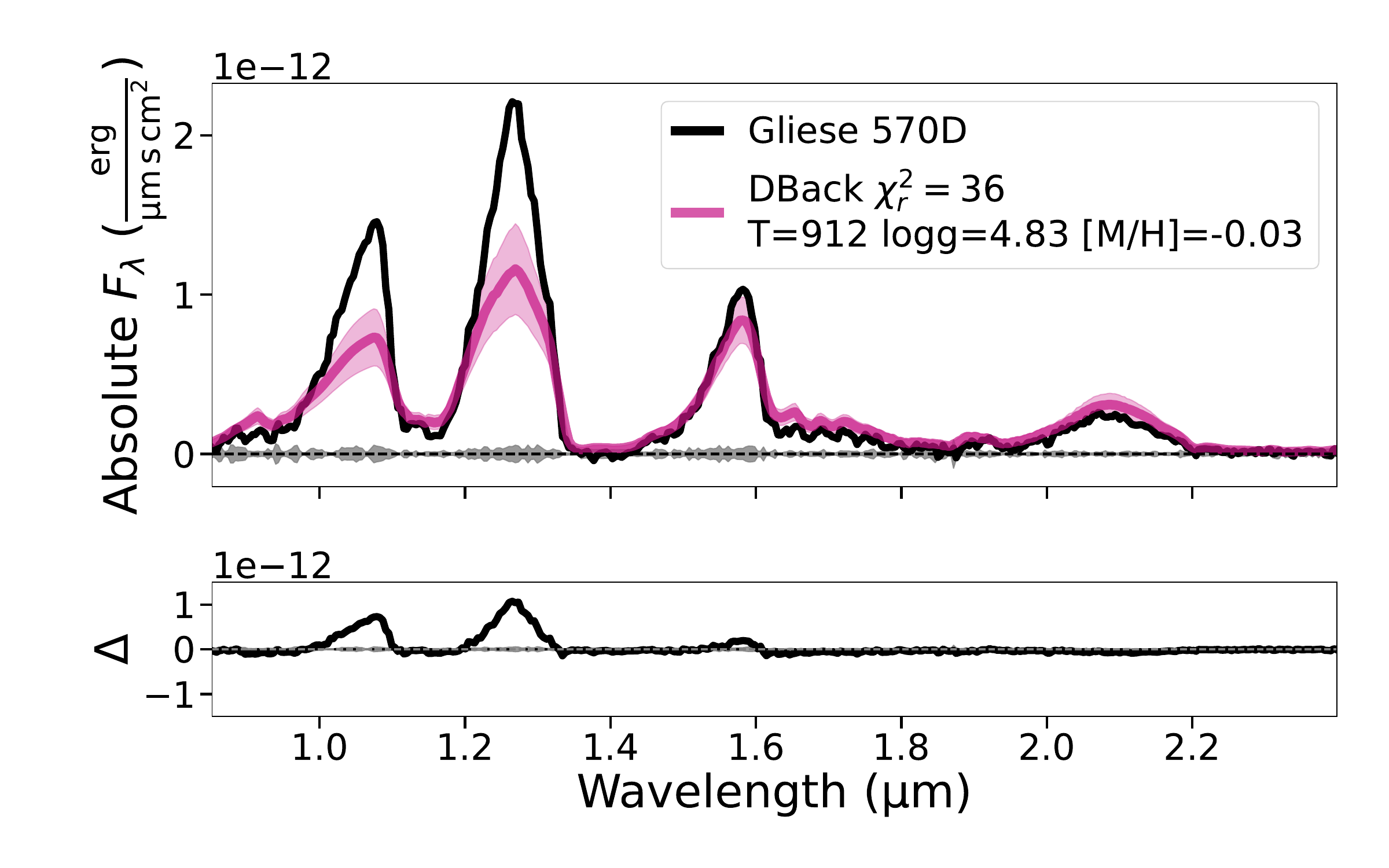}
\includegraphics[width=0.32\textwidth]{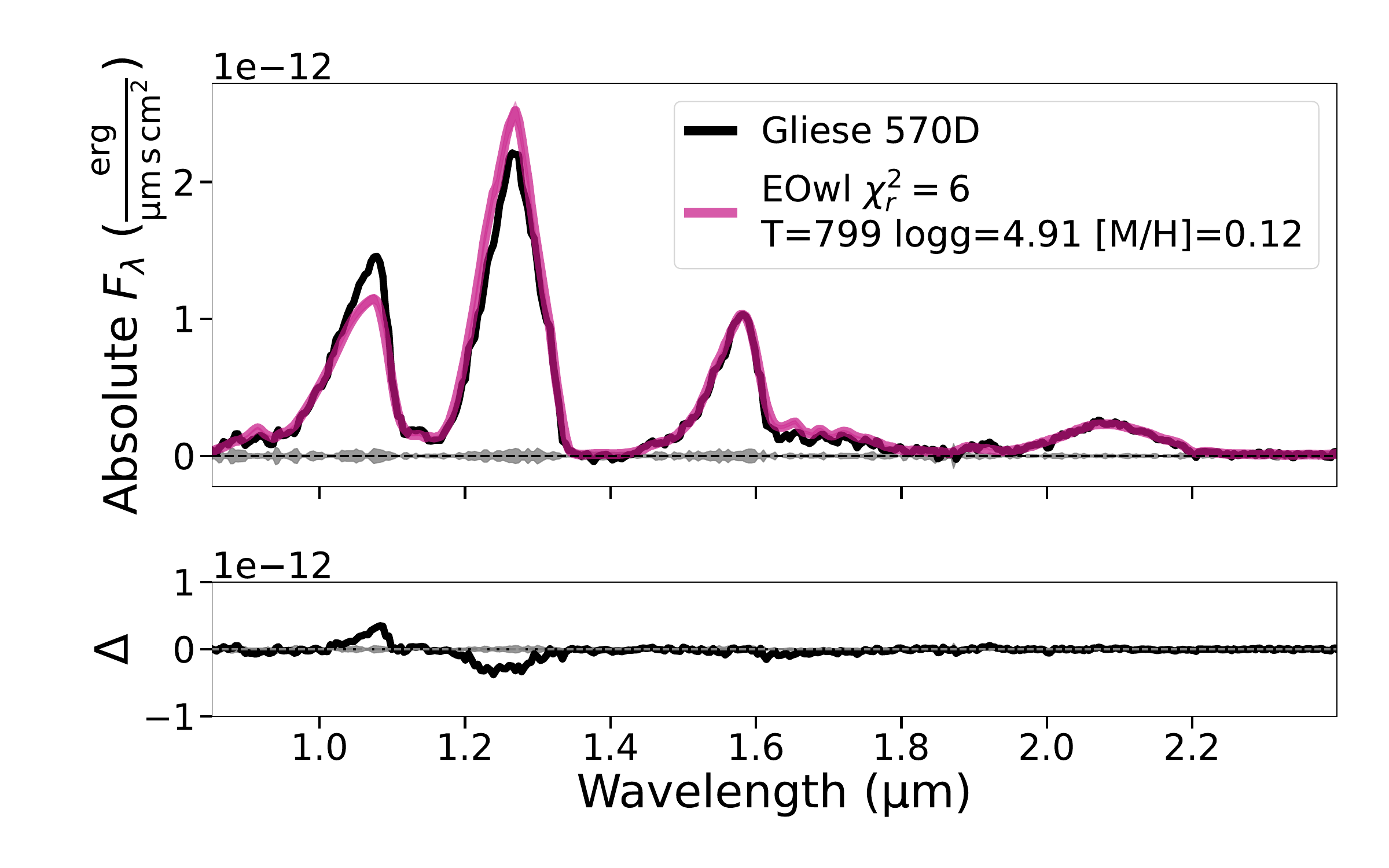}
\includegraphics[width=0.32\textwidth]{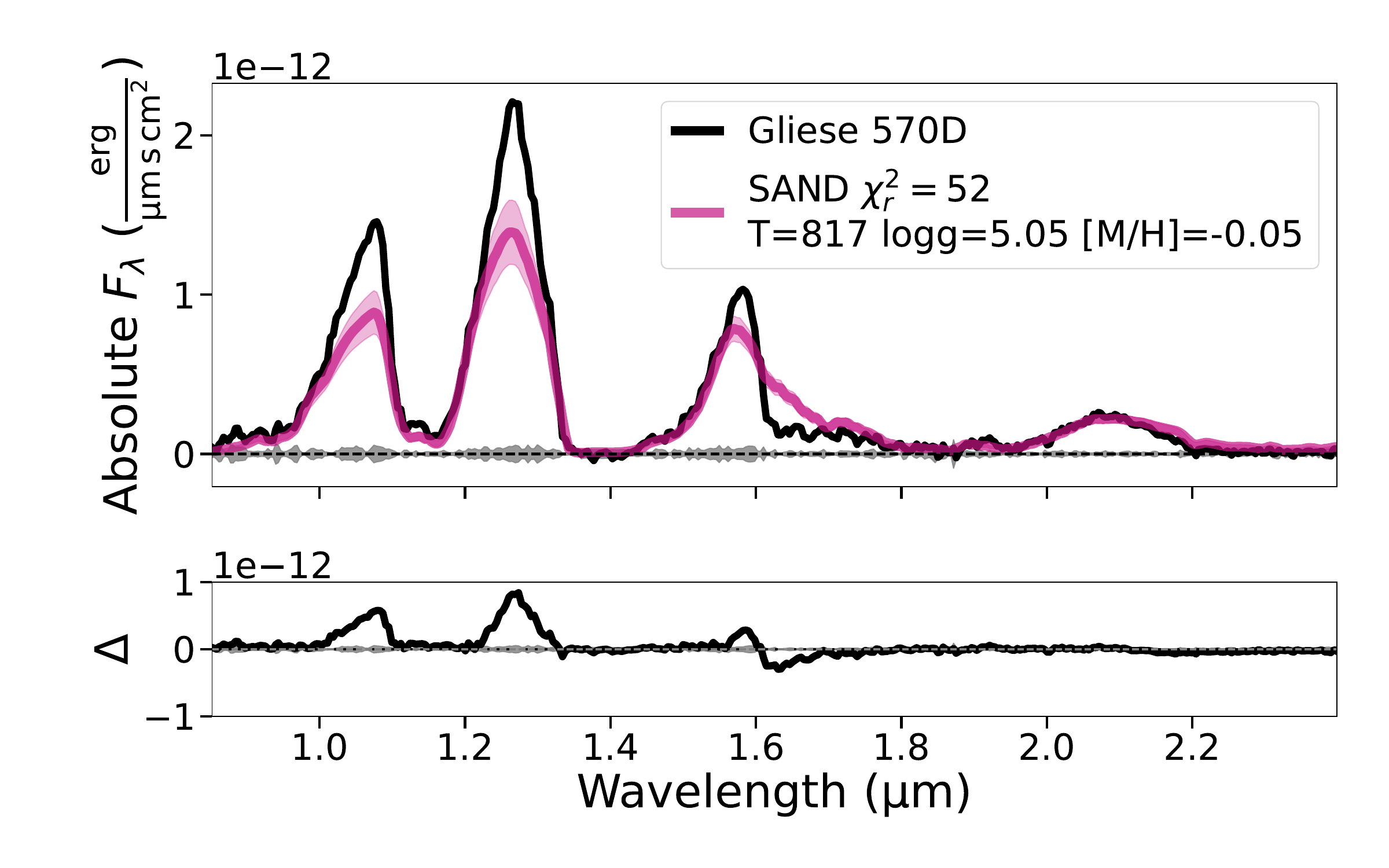} \\
\caption{Figure~\ref{fig:rfr-comp1} continued. \label{fig:rfr-comp2}}
\end{figure}

Figures~\ref{fig:rfr-comp1} and~\ref{fig:rfr-comp2} display the best-fit models for each of the spectra based on the Diamondback, Elf Owl, and SAND models, while Table~\ref{tab:RFRresults} summarizes the corresponding posterior parameters. As observed in prior work on random forest retrievals \citep{2023ApJ...954...22L}, the median retrieved spectra show significant differences between the three grids, with both over- and underestimates of spectral fluxes compared to the measured spectra of up to an order of magnitude.
Nevertheless, each source can be reasonably well fit by one of the model grids, following the same pattern as the MCMC fits:
L dwarfs and early-T dwarfs are best fit by the Diamondback models, mid- and late-type L dwarfs (most notably G~196-3B) by the SAND models, and mid- and late-type T dwarfs by the Elf Owl models. 
Detailed fits of molecular features shortward of 1.3~$\mu$m in the early L dwarfs are again problematic for all three models,
with the L1 GJ~1048B showing the worst fits ($\chi_r^2$ $>$ 100),
while the SAND models are unable to properly reproduce the 1.6~$\mu$m \ch{CH4} band.
We also see for these fits that the Diamondback and Elf Owl models diverge for the latest T dwarfs, with the latter providing clearly superior fits.
Like the MCMC fits, we see the anticipated trends among best-fit parameters from the RFR modeling,
with effective temperatures decreasing toward later spectral type, lower surface gravities for the young
brown dwarfs G~196-3B and HN~Peg~B, and $f_{sed}$ increasing across the L/T transition for the Diamondback models.
Furthermore, while the best-fit radius for GJ~1048B from the Diamondback fits, R = 0.130$^{+0.008}_{-0.006}$~R$_\odot$, is higher than expected from evolutionary models, it is not as extreme as the MCMC fit.

\begin{deluxetable}{llccccccccc}
\tablecaption{Random Forest Retrieval Results. \label{tab:RFRresults}} 
\tabletypesize{\tiny}
\tablehead{ 
\colhead{\textbf{Source}} & 
\colhead{\textbf{SpT}} & 
\colhead{$\mathbf{T_{\mathbf{eff}}}$} & 
\colhead{\textbf{log $\mathbf{g}$}} &
\colhead{$[\mathbf{M/H}]$} &
\colhead{\bf C/O\tablenotemark{a}} &
[$\mathbf{\alpha}$/\textbf{\ch{Fe}}] &
\colhead{\textbf{log} $\mathbf{\kappa_{\mathbf{zz}}}$} &
\colhead{$\mathbf{f}_{\mathbf{sed}}$} &
\colhead{\textbf{R}} &
\colhead{$\mathbf{\chi}^\mathbf{2}_\mathbf{r}$} \\ 
\colhead{} & 
\colhead{} & 
\colhead{(K)} & 
\colhead{(cm/s$^2$)} &
\colhead{} &
\colhead{} &
\colhead{} & 
\colhead{(cm$^2$/s)} &
\colhead{} &
\colhead{(R$_\odot$)} &
\colhead{}  
}
\startdata
\cline{1-11} 
\multicolumn{11}{c}{\textbf{Diamondback Model Set}} \\
\cline{1-11}
\textbf{LP 465-70B} & L0 & 2057$^{+243}_{-157}$ & 5.14$^{+0.36}_{-0.64}$ & 0.18$^{+0.33}_{-0.18}$ & \nodata & \nodata & \nodata & 2.9$^{+1.1}_{-1.9}$  & 0.112$^{+0.020}_{-0.023}$ & \textbf{12}\tablenotemark{\rm *} \\ 
\textbf{HD 89744B} & L0  & 2006$^{+294}_{-106}$ & 5.17$^{+0.33}_{-0.17}$ & 0.22$^{+0.28}_{-0.22}$ & \nodata & \nodata & \nodata &2.7$^{+1.3}_{-1.7}$  & 0.122$^{+0.015}_{-0.027}$ & \textbf{136}\tablenotemark{\rm *} \\ 
\textbf{GJ 1048B} & L1 & 1943$^{+57}_{-143}$ & 5.10$^{+0.40}_{-0.35}$ & 0.21$^{+0.29}_{-0.21}$ & \nodata & \nodata & \nodata & 2.4$^{+0.6}_{-1.4}$  & 0.130$^{+0.008}_{-0.006}$ & \textbf{315}\tablenotemark{\rm *} \\ 
\textbf{Gl 618.1B} & L2.5 & 1651$^{+49}_{-109}$ & 4.99$^{+0.51}_{-0.49}$ & 0.27$^{+0.23}_{-0.27}$ & \nodata & \nodata & \nodata & 2.2$^{+0.8}_{-1.2}$  & 0.124$^{+0.015}_{-0.010}$ & \textbf{34}\tablenotemark{\rm *} \\ 
\textbf{G 62-33B} & L2.5  & 1698$^{+102}_{-98}$ & 4.95$^{+0.56}_{-0.45}$ & 0.15$^{+0.35}_{-0.65}$ & \nodata & \nodata & \nodata & 2.6$^{+0.4}_{-1.6}$  & 0.120$^{+0.020}_{-0.014}$ & \textbf{22}\tablenotemark{\rm *} \\ 
G 196-3B & L3$\beta$ & 1553$^{+47}_{-53}$ & 4.82$^{+0.68}_{-0.32}$ & 0.16$^{+0.34}_{-0.46}$ & \nodata & \nodata & \nodata & 2.4$^{+0.6}_{-1.4}$  & 0.125$^{+0.012}_{-0.010}$ & 2026 \\ 
\textbf{Gliese 584C} & L8 & 1318$^{+130}_{-143}$ & 4.88$^{+0.12}_{-0.38}$ & 0.06$^{+0.44}_{-0.56}$ & \nodata & \nodata & \nodata & 2.9$^{+1.1}_{-1.9}$ & 0.090$^{+0.028}_{-0.024}$ & 9 \\ 
\textbf{HN Peg B} & T2.5 & 1275$^{+225}_{-175}$ & 4.95$^{+0.55}_{-0.45}$ & 0.14$^{+0.36}_{-0.64}$ & \nodata & \nodata & \nodata & 7.0$^{+3.0}_{-3.0}$ & 0.080$^{+0.028}_{-0.025}$ & \textbf{29}\tablenotemark{\rm *} \\ 
\textbf{G 204-39B} & T6.5 & 997$^{+103}_{-97}$ &  4.83$^{+0.17}_{-0.33}$ & 0.21$^{+0.29}_{-0.21}$ & \nodata & \nodata & \nodata & 8.6$^{+1.4}_{-0.6}$  & 0.082$^{+0.024}_{-0.016}$ & \textbf{39}\tablenotemark{\rm *} \\ 
HD 3651B & T7 & 910$^{+10}_{-10}$ & 4.88$^{+0.48}_{-0.38}$ & 0.00$^{+0.50}_{-0.50}$ & \nodata & \nodata & \nodata & 5.9$^{+2.1}_{-2.9}$ &  0.060$^{+0.010}_{-0.012}$ & 52 \\ 
Gliese 570D & T8 & 912$^{+12}_{-12}$ & 4.83$^{+0.67}_{-0.33}$ & -0.03$^{+0.19}_{-0.47}$ & \nodata & \nodata & \nodata & 6.0$^{+2.0}_{-3.0}$ &  0.059$^{+0.011}_{-0.012}$ & 36 \\ 
\cline{1-11} 
\multicolumn{11}{c}{\textbf{Elf Owl Model Set}} \\
\cline{1-11}
LP 465-70B & L0 & 2161$^{+239}_{-161}$ & 5.02$^{+0.48}_{-0.27}$ & 0.28$^{+0.22}_{-0.28}$ & 1.30$^{+0.20}_{-0.30}$ & \nodata & 5.8$^{+3.1}_{-3.8}$ & \nodata & 0.100$^{+0.015}_{-0.018}$ & 23 \\ 
HD 89744B & L0  & 2145$^{+155}_{-145}$ & 5.04$^{+0.46}_{-0.29}$ & 0.33$^{+0.17}_{-0.33}$ & 1.45$^{+0.05}_{-0.05}$ & \nodata & 5.4$^{+2.6}_{-3.4}$ & \nodata & 0.105$^{+0.021}_{-0.019}$ & 178 \\ 
GJ 1048B & L1 & 2053$^{+147}_{-153}$ & 4.93$^{+0.32}_{-0.43}$ & 0.41$^{+0.09}_{-0.41}$ & 1.47$^{+0.03}_{-0.03}$ & \nodata & 4.3$^{+3.7}_{-2.3}$ & \nodata & 0.114$^{+0.015}_{-0.014}$ & 407 \\ 
Gl 618.1B & L2.5 & 1995$^{+205}_{-195}$ & 4.78$^{+0.22}_{-0.28}$ & 0.44$^{+0.06}_{-0.06}$ & 1.41$^{+0.09}_{-0.41}$ & \nodata & 4.7$^{+3.3}_{-2.7}$ & \nodata & 0.090$^{+0.020}_{-0.020}$ & 258 \\ 
G 62-33B & L2.5  & 2023$^{+277}_{-223}$ & 4.81$^{+0.20}_{-0.31}$ & 0.43$^{+0.07}_{-0.07}$ & 1.44$^{+0.06}_{-0.06}$ & \nodata & 4.7$^{+3.3}_{-2.7}$ & \nodata & 0.088$^{+0.017}_{-0.020}$ & 123 \\ 
G 196-3B & L3$\beta$ & 1875$^{+125}_{-175}$ & 4.71$^{+0.29}_{-0.21}$ & 0.47$^{+0.03}_{-0.03}$ & 1.36$^{+0.14}_{-0.36}$ & \nodata & 4.9$^{+3.1}_{-2.9}$ & \nodata & 0.098$^{+0.017}_{-0.019}$ & 3634 \\ 
Gliese 584C & L8 & 1704$^{+96}_{-104}$ & 4.81$^{+0.19}_{-0.31}$ & 0.14$^{+0.36}_{-0.14}$ & 1.15$^{+0.35}_{-0.65}$ & \nodata & 6.0$^{+3.0}_{-4.0}$ & \nodata & 0.057$^{+0.007}_{-0.008}$ & 213 \\ 
HN Peg B & T2.5 & 1339$^{+161}_{-139}$ & 5.00$^{+0.50}_{-0.50}$ & 0.29$^{+0.21}_{-0.29}$ & 1.27$^{+0.23}_{-0.27}$ & \nodata & 5.8$^{+2.2}_{-3.8}$ & \nodata & 0.073$^{+0.015}_{-0.015}$ & 73 \\ 
\textbf{G 204-39B} & T6.5 & 931$^{+69}_{-81}$ & 4.93$^{+0.43}_{-0.43}$ & 0.40$^{+0.10}_{-0.40}$ & 1.36$^{+0.14}_{-0.36}$ & \nodata & 5.2$^{+2.8}_{-3.2}$ & \nodata & 0.093$^{+0.024}_{-0.025}$ & \textbf{38}\tablenotemark{\rm *} \\ 
\textbf{HD 3651B} & T7 & 790$^{+110}_{-90}$ & 4.91$^{+0.34}_{-0.41}$ & 0.20$^{+0.30}_{-0.20}$ & 1.21$^{+0.29}_{-0.21}$ & \nodata & 4.5$^{+3.5}_{-2.5}$ & \nodata & 0.092$^{+0.032}_{-0.026}$ & \textbf{6}\tablenotemark{\rm *} \\ 
\textbf{Gliese 570D} & T8 & 799$^{+101}_{-99}$ & 4.91$^{+0.34}_{-0.41}$ & 0.12$^{+0.39}_{-0.12}$ & 1.16$^{+0.34}_{-0.66}$ & \nodata & 5.2$^{+2.8}_{-3.2}$ & \nodata & 0.090$^{+0.028}_{-0.025}$ & \textbf{6}\tablenotemark{\rm *} \\ 
\cline{1-11} 
\multicolumn{11}{c}{\textbf{SAND Model Set}} \\
\cline{1-11}
LP 465-70B & L0 & 2217$^{+157}_{-317}$ & 5.02$^{+0.48}_{-0.52}$ & 0.05$^{+0.25}_{-0.20}$ & \nodata & 0.03$^{+0.12}_{-0.03}$ & \nodata & \nodata & 0.099$^{+0.029}_{-0.018}$ & 23 \\ 
HD 89744B & L0  & 2190$^{+210}_{-289}$ & 4.91$^{+0.59}_{-0.41}$ & -0.03$^{+0.33}_{-0.32}$ & \nodata & 0.02$^{+0.13}_{-0.07}$ & \nodata & \nodata & 0.106$^{+0.027}_{-0.019}$ & 249 \\ 
GJ 1048B & L1 & 2103$^{+197}_{-203}$ & 4.87$^{+0.63}_{-0.37}$ & -0.04$^{+0.34}_{-0.32}$ & \nodata & 0.01$^{+0.01}_{-0.06}$ & \nodata & \nodata & 0.116$^{+0.019}_{-0.027}$ & 436 \\ 
\textbf{Gl 618.1B} & L2.5 & 1774$^{+126}_{-174}$ & 5.03$^{+0.47}_{-0.53}$ & 0.06$^{+0.24}_{-0.21}$ & \nodata & 0.06$^{+0.09}_{-0.11}$ & \nodata & \nodata & 0.115$^{+0.023}_{-0.019}$ & 39 \\ 
\textbf{G 62-33B} & L2.5  & 1810$^{+190}_{-110}$ & 5.07$^{+0.43}_{-0.57}$ & 0.02$^{+0.28}_{-0.37}$ & \nodata & 0.06$^{+0.09}_{-0.11}$ & \nodata & \nodata & 0.112$^{+0.026}_{-0.022}$ & 24 \\ 
\textbf{G 196-3B} & L3$\beta$ & 1667$^{+33}_{-67}$ & 4.68$^{+0.32}_{-0.18}$ & 0.03$^{+0.09}_{-0.18}$ & \nodata & 0.00$^{+0.00}_{-0.05}$ & \nodata & \nodata & 0.129$^{+0.012}_{-0.013}$ & \textbf{101}\tablenotemark{\rm *} \\ 
\textbf{Gliese 584C} & L8 & 1433$^{+167}_{-233}$ & 4.86$^{+0.65}_{-0.36}$ & 0.01$^{+0.29}_{-0.36}$ & \nodata & 0.12$^{+0.08}_{-0.12}$ & \nodata & \nodata & 0.076$^{+0.019}_{-0.017}$ & \textbf{7}\tablenotemark{\rm *} \\ 
HN Peg B & T2.5 & 1223$^{+177}_{-123}$ & 5.07$^{+0.43}_{-0.57}$ & -0.18$^{+0.05}_{-0.17}$ & \nodata & 0.04$^{+0.11}_{-0.09}$ & \nodata & \nodata & 0.089$^{+0.030}_{-0.024}$ & 38 \\ 
G 204-39B & T6.5 & 1003$^{+197}_{-103}$ & 5.40$^{+0.60}_{-0.40}$ & -0.03$^{+0.33}_{-0.32}$ & \nodata & 0.06$^{+0.09}_{-0.11}$ & \nodata & \nodata & 0.088$^{+0.026}_{-0.035}$ & 93 \\ 
HD 3651B & T7 & 814$^{+86}_{-114}$ & 5.04$^{+0.47}_{-0.54}$ & -0.04$^{+0.14}_{-0.32}$ & \nodata & 0.06$^{+0.14}_{-0.07}$ & \nodata & \nodata & 0.086$^{+0.028}_{-0.028}$ & 46 \\ 
Gliese 570D & T8 & 817$^{+83}_{-117}$ & 5.05$^{+0.45}_{-0.55}$ & -0.05$^{+0.15}_{-0.30}$ & \nodata & 0.06$^{+0.14}_{-0.06}$ & \nodata & \nodata & 0.085$^{+0.026}_{-0.027}$ & 52 \\ 
\enddata
\tablenotetext{a}{C/O ratio relative to solar abundance, with (C/O)$_\odot$ = 0.458 \citep{2009LanB...4B..712L}.}
\tablenotetext{\rm *}{Best fit model or models (if $\Delta\chi^2_r \leq 1$) for this source.}
\end{deluxetable}

\subsection{Comparison of MCMC and RFR Approaches}

\subsubsection{Comparison of Fit Quality}

The MCMC and RFR approaches generally yield comparable results in terms of preferred model grid and inferred parameters, albeit with some informative differences.
First, the quality of the best fits as quantified by $\chi_r^2$ is consistently better for the MCMC approach.
In several cases the difference is substantial; for example, SAND fits to the L3$\beta$ G~196-3B yield $\chi^2_r$ = 101 for RFR versus $\chi^2_r$ = 13 for MCMC, the latter being an obviously better fit (cf.\ Figures~\ref{fig:mcmc-comp1} and~\ref{fig:rfr-comp1}). On the other hand, Elf Owl fits for T6.5 G~204-39B are significantly better for the RFR approach ($\chi^2_r$ = 38) than the MCMC approach ($\chi^2_r$ = 77).
There are also cases where the best-fit model set differs between approaches; for example, the best MCMC fit for HN~Peg~B comes from the SAND models, while the best RFR fit comes from the Diamondback model fit, both having identical $\chi^2_r$ = 29.
As such, while both approaches yield broadly comparable fits, the MCMC approach provides an average better qualities of fit.

\subsubsection{Comparison of Fit Parameters}

\begin{figure}
    \centering
    \includegraphics[width=\textwidth]{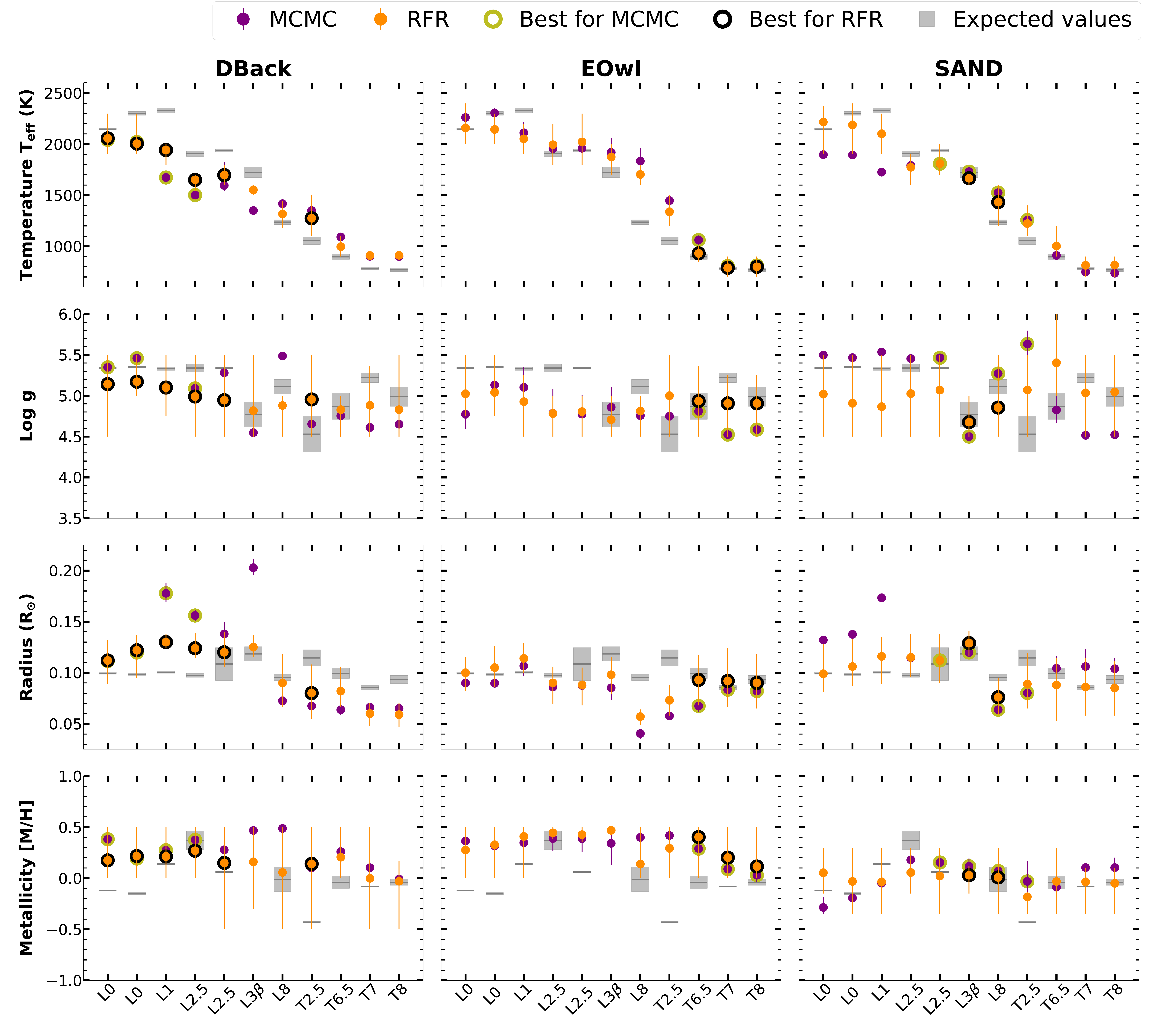}
    \caption{Model fit parameters and 1$\sigma$ uncertainties for our full L and T dwarf sequence, comparing the MCMC (purple points) and RFR (orange points) approaches, and the Diamondback (DBack, left), Elf Owl (EOwl, middle), and SAND (right) model grids. 
    Parameters show from top to bottom are: effective temperature ({\teff} in units of K), surface gravity ({\logg} in log units of cm/s$^2$), radius ($R$ in units of $R_\odot$), and metallicity ([$M/H$] in log units relative to solar).
    For each method, the models with the best $\chi^2$-values for each object are highlighted with circles (green for MCMC, black for RFR).
    Gray bars indicate the expected values from evolutionary models ({\teff}, {\logg}, radius) or primaries (metallicity) as listed in Table~\ref{tab:sample}. 
    \label{fig:Posteriors_comparison}}
\end{figure}

Figure~\ref{fig:Posteriors_comparison} compares the fit parameters and 1$\sigma$ uncertainties for the MCMC and RFR approaches for each of our sources, with the best-fit model for each approach highlighted.
There is a clear difference in the inferred parameter uncertainties, with RFR uncertainties being significantly larger. 
For example, typical uncertainties in {\teff} are $\sim$5--10\% for RFR and $\sim$1\% for MCMC, while for {\logg} uncertainties are $\sim$0.5~dex for RFR and $\sim$0.1~dex for MCMC.
The lower uncertainties for the MCMC parameters may partly reflect the better fidelity in reproducing the observed spectra, 
but may also indicate underestimation of parameter uncertainties in the approach applied here. 
Within their mutual uncertainties, the parameter values are largely consistent for the best-fit models,
with a few notable exception.
The aforementioned low {\teff} and inflated radius from MCMC Diamondback model fits to the L1 GJ~1048B are significantly distinct from the RFR SAND model fits for this source, which has a more reasonable (but more uncertiain) {\teff} = 2100$\pm$200~K and radius R = 0.116$^{+0.019}_{-0.027}$~R$_\odot$.
In contrast, RFR Elf Model fits  
yield a marginally larger and more realistic radius 
(R = 0.093$^{+0.024}_{-0.025}$~R$_\odot$)
for the T6.5 G~204-39B as compared to the MCMC fits,
We also find that the surface gravities from RFR Elf Owl model to the late T dwarfs HD~3651B and Gliese~570D are higher than MCMC values for these sources and more in line with those expected for older brown dwarfs; while the surface gravity for the best-fit RFR Diamondback model fits for HN~Peg~B is lower than the MCMC SAND model fits for this source and more in line with this system's young age.
Even among these variances, the best-fit MCMC and RFR model parameters are generally consistent with each other, albeit within the relatively large uncertainties of the RFR values.

\subsubsection{Comparison to Evolutionary Model Predictions}

Figure~\ref{fig:Posteriors_comparison} compares fit parameters to the expected evolutionary model-based estimates of {\teff}, {\logg}, and radius from \citet[see Table~\ref{tab:sample}]{2023ApJ...959...63S}.
While there is broad agreement in the spectra type trends among these values, there are several deviations that merit scrutiny.
Across the early L dwarf sequence, best-fit {\teff}s and radii from the Diamondback models are consistently lower and larger, respectively, compared to the expected values, more so for the MCMC fits. This trend reverses for the T dwarfs, although RFR Elf Owl model fits for the mid- and late-T dwarfs are consistent with expected temperatures and radii. The SAND models show particularly good agreement with expected {\teff}s and radii across the full sample, even when providing the lowest quality fit to the spectral data, with a notable match to the inflated radius expected for the young L3$\beta$ G~196-3B. With regard to surface gravities, Diamondback and Elf Owl fits are consistently lower than expectations for the L dwarfs, in line with inflated radii, and are essentially constant for the T dwarfs. 
The SAND models show differing behaviors between MCMC and RFR approaches across the sample, the latter generally underpredicting {\logg} for L dwarfs and overpredicting {\logg} for T dwarfs.
Overall, the early L dwarfs show the largest deviations between fit parameters and evolutionary model-based estimates, whereas parameters for the mid- and late-L dwarfs and T dwarfs are generally consistent with these estimates.

\subsubsection{Comparison to System Parameters}

As benchmark companions, our sample has independent constraints on the atmosphere parameters of metallicity and surface gravity based on the elemental composition and ages of the host primaries.
For metallicity, Figure~\ref{fig:Posteriors_comparison} shows that there are mixed outcomes comparing between fit metallicities and primary iron abundances, with 
about half of the fits yielding metallicities that are too high.
Elf Owl models consistently yield super-solar metallicities for the L dwarfs and early-/mid-type T dwarfs for both approaches, often reaching the limits of the parameter grid. 
This positive metallicity skew and similarly inflated C/O ratios likely compensate for missing condensate opacity in the Elf Owl model grid, as the two late T dwarfs HD~3651B and Gliese~570D show no such skew.
In contrast, the SAND models yield metallicities that are largely consistent with the iron abundances of their primaries, even when the model does not provide the best fit overall. Metallicities from the Diamondback models are too uncertain to make a reliable comparison. The better performance of the SAND models for metallicity inference is somewhat surprising given its low RvP value for this parameter (Figure~\ref{fig:RvP}).

For surface gravities, while these are skewed low for the L dwarfs they are generally in line with the ages of the host systems across our sample. High surface gravities for old systems such as LP~465-70B (Diamondback {\logg} = 5.35$^{+0.09}_{-0.10}$ for MCMC, {\logg} = 5.14$^{+0.36}_{-0.64}$ for RFR) are consistent with ages in the $\tau \gtrsim$ 2~Gyr range, while low surface gravities for young systems such as G~196-3B (SAND {\logg} $\leq$ 4.5 for MCMC, {\logg} = 4.68$^{+0.32}_{-0.18}$ for RFR) are consistent with ages in the $\tau \lesssim$ 1~Gyr range, based on evolutionary models for the effective temperatures of these sources \citet{2001RvMP...73..719B,2003A&A...402..701B}.
There are again exceptions. The T2.5 dwarf HN~Peg~B, part of the 240$\pm$30~Myr HN Peg system \citep{2007ApJ...669.1167B}, has a best-fit MCMC SAND model with {\logg} = 5.59$^{+0.10}_{-0.09}$, consistent with an age $\gtrsim$ 5~Gyr.\footnote{The RFR Diamondback models yield {\logg} values that are too uncertain for a robust age test.} 
Similarly, the best-fit MCMC Elf Owl model for the late-T dwarfs HD~3651B and Gliese~570D yield {\logg} = 4.5--4.6 implying ages $\lesssim$ 500~Myr, considerably younger than their primaries.

\subsection{Case studies\label{sec:casestudies}}

Here we examine in detail fits to three representative objects that encompass our sample:
the early L dwarf LP~465-70B, the L/T transition object Gliese~584C, and the late T dwarf Gliese~570D. 
All three objects show robust fits to one of the model sets with the parameters listed in Tables~\ref{tab:MCMCresults} and \ref{tab:RFRresults}. 
In addition to the spectral fits shown above, we display joint posterior parameter distributions for both MCMC and RFR approaches for these three sources in Figures~\ref{fig:mcmc-corner-LP465-70B} through~\ref{fig:Elfowl_posteriors_Gliese570D} of the Appendix.

\subsubsection{LP 465-70B and Diamondback Models}

The spectrum of the L0 LP~465-70B is best fit by the Diamondback models, reflecting the important role of condensates in shaping the infrared spectrum of this source.
Both MCMC and RFR fits yield the same $\chi_r^2$ = 12, albeit with different deviations from the data.
The MCMC fit accurately reproduces the SED but fails to match detailed absorption features shortward of 1.2~$\mu$m, particularly around the 0.99~$\mu$m \ch{FeH} band;
the RFR fit overpredicts the 1.1--1.3~$\mu$m continuum and shows similar feature divergences shortward of 1.0~$\mu$m.
All of the fit parameters are formally consistent between the models, with only a 12~K difference in {\teff} and nearly identical radii.
The condensation parameter $f_{sed}$ is inferred to be low for both fits (1.26$^{+0.20}_{-0.21}$ for MCMC, 2.9$^{+1.1}_{-1.9}$ for RFR), indicating significant condensate opacity in the atmosphere of this L dwarf.
Figures~\ref{fig:mcmc-corner-LP465-70B} and \ref{fig:Dback_posteriors_LP465-70B} demonstrate that these parameters are well-constrained, with single peaks in the marginalized distributions that are set in part by the parameter limits for {\logg}, [M/H], and $f_{sed}$.
We find only one (expected) negative correlation between {\teff} and radius. 
The inferred temperatures are $\sim$100~K cooler and radii 13\% larger than the evolutionary model estimates of \citet{2023ApJ...959...63S}, while the MCMC surface gravity is consistent with the older age of this system ($\tau \geq$ 4.5~Gyr). 
Both fits also yield a high metallicity ([M/H] = +0.41$^{+0.07}_{-0.10}$ for MCMC), in contradiction to the subsolar iron abundance inferred for the primary ([Fe/H] = $-$0.120$\pm$0.020; \citealt{2022AJ....163..152S}). 
This offset reflects the positive metallicity bias seen among the Diamondback and Elf Owl model fits of all of the L dwarfs in our sample.

\subsubsection{Gliese 584C and SAND Models}

The spectrum of the L8 Gliese~584C is well reproduced by the SAND models, which provide an intermediary between the condensate-infused Diamondback models (a close second-best fit with a large value of $f_{sed}$) 
and the condensate-free Elf Owl models (which poorly fit).
Both MCMC and RFR analyses produce best fits with similar $\chi_r^2$ values.
The MCMC fit is slightly superior ($\chi_r^2$ = 4 for MCMC versus $\chi_r^2$ = 7 for RFR) with only a modestly reduced 1.25~$\mu$m flux peak, while the RFR fit shows considerable variance around the three main flux peaks at 1.25~$\mu$m, 1.6~$\mu$m, and 2.1~$\mu$m which inflates the final parameter uncertainties.
Both analyses nevertheless yield equivalent {\teff}s and effectively solar metallicities, the latter
consistent with iron abundance measurements of the G dwarf primaries ([Fe/H] = $-$0.01$\pm$0.12; \citealt{2024ApJ...960..105Z}). Both [M/H] and [$\alpha$/H] are poorly constrained in these fits, and we find a strong negative correlation between these parameters which may inflate their uncertainties. We also see a complex correlation between {\logg} and [M/H] in the RFR fits that drives up the uncertainties for these parameters.
There is a significant difference in the inferred surface gravities between the analyses, being higher and more consistent with the age of the system for the MCMC fits. In contrast, the inferred MCMC radius of R = 0.064$\pm$0.001~R$_\odot$ is too small to be physical, while
the slightly larger RFR radius of R = 0.076$^{+0.019}_{-0.017}$~R$_\odot$ is also low but formally consistent with the expected model-dependent radius.
The small radii inferred from these fits reflect the $\sim$200~K higher effective temperature inferred in this analysis as compared to the {\teff} = 1237$\pm$24~K model estimate from \citet{2023ApJ...959...63S}, although prior spectral model fits \citep{2004ApJ...607..499N,2020ApJ...891..171Z,2023ApJ...959...63S} find {\teff}s in the 1300--1500~K range. 
Our ability to accurately reproduce the spectrum with Gliese~584C with the SAND models indicates progress in modeling the complex L/T transition \citep{2002ApJ...571L.151B,2005AN....326..627H,2016ApJ...817L..19T}, but the large variances and uncertainties in temperature, surface gravity, and metallicity suggest that additional spectral data are needed to tightly constrain physical properties.

\subsubsection{Gliese 570D and Elf Owl Models}

The spectrum of the T8 dwarf Gliese~570D is best fit by the Elf Owl models, with the Diamondback models being a close second and the SAND models well behind due to poor reproduction of the 1.6~$\mu$m CH$_4$ band.
Again, both MCMC and RFR analyses yield comparable best fits ($\chi_r^2$ = 5 and 6, respectively),
with the former showing $\sim$10\% deviations at the 1.05~$\mu$m and 1.55~$\mu$m flux peaks
while the latter shows slightly larger deviations at 1.05~$\mu$m and 1.25~$\mu$m.
Temperatures and metallicities are well-constrained and consistent between the MCMC and RFR analyses, and are also consistent with previously inferred values for the brown dwarf and the K4 dwarf primary \citep{2001ApJ...556..373G,2006ApJ...639.1095B,2015ApJ...807..183L,2021ApJ...921...95Z,2023ApJ...954...22L}.
The radii are also equal to within 10\% and match model-based expectations. 
While all of the fit parameters are consistent within the RFR uncertainties, we do find a notable offset in surface gravity between the analysis methods, with the lower MCMC {\logg} = 4.57$^{+0.09}_{-0.07}$ being significantly below the model-based estimate {\logg} = 4.99$\pm$0.12.
This low surface gravity implies an evolutionary model age of $\tau$ = 450--650~Myr \citep{2001RvMP...73..719B,2003A&A...402..701B} that is significantly younger
that the inferred age of the primary ($\tau$ = 3.3$\pm$1.9~Gyr; \citealt{2021ApJ...921...95Z}), and 
reflects the general underestimate of surface gravities seen in
Elf Owl model fits of T dwarfs \citep{2024ApJ...973...60B}.
This offset may reflect modeling errors in pressure-dependent chemistry or the strength of collision-induced H$_2$ absorption \citep{1969ApJ...156..989L}.
The Elf Owl models do account for non-equilibrium chemistry through vertical mixing, which is quantified by the {\logk} diffusion coefficient and found to be relatively weak in the MCMC fits, ({\logk} = 3.0$^{+0.7}_{-0.5}$).
We do note a slight positive correlation between {\logk} and the C/O ratio, the latter consistent with solar in both analyses. This trend is consistent with increased \ch{CO} and decreased \ch{CH4} abundances in a vertically mixed or oxygen-enriched atmosphere \citep{2006ApJ...647..552S,2007ApJ...655.1079L}.

\section{Discussion} \label{sec:discussion}

The central focus of this study is a critical analysis of two different approaches to modeling low-temperature spectra using pre-computed forward atmosphere models. Here we evaluate our outcomes through the lenses of accuracy, precision, and efficiency.

\subsection{Accuracy: Quality of Spectral Fits}

One of the key outcomes of this study is that the choice of model set is more important in reproducing an observed spectrum than the model fitting approach. For our L and T dwarf sample, both MCMC and RFR models were able to converge to a best-fit model that largely reproduced the overall spectral energy distribution and detailed features, with an on average lower value of $\chi^2$ for the MCMC fits. Even cases with persistently high $\chi^2$ values (e.g., HD~89744B, GJ~1048B, and G~204-39B) show a reasonably visual agreement, suggesting that these values are driven by the high signal-to-noise of the spectra or uncorrected calibration issues.
Much greater variance in fit quality is seen between the model sets themselves, with typically one or two of the Diamondback, Elf Owl, or SAND models providing a clearly superior fit depending on the spectral type of the source, in some cases by a factor of 30 difference in $\chi_r^2$ (e.g, G~196-3B). These differences reflect choices in the underlying gas chemistry (including non-equilibrium effects), treatments of condensation, line broadening profile (relevant for the 0.77~$\mu$m K~I doublet), and adopted molecular opacities, as well as potentially unmodeled physical processes.
As such, the methodology of fitting appears to be less critical for ensuring an accurate fit than evaluating a broad selection of models. 

\subsection{Accuracy: Parameter Values}

Another measure of accuracy considered in this study was the agreement between model fit parameters and the properties inferred from the primaries to our benchmark companions.
Here, our results are mixed and likely driven by shortcomings in the models rather than the model fitting process.
Our best MCMC and RFR fits based on Diamondback and Elf Owl models consistently yielded significantly supersolar metallicities, even for systems with solar or subsolar iron abundance primaries.
This metallicity bias is absent for the late T dwarfs, and for G~196-3B and Gliese~584C which are best fit by the SAND models. 
Our analysis suggests that metallicity may not be robustly measurable with these sets of models and/or with low-resolution near-infrared spectra, although the range of metallicities sampled (primary iron abundances $-$0.43 $\leq$ [Fe/H] $\leq$ +0.37) is arguably narrow.
The other benchmark metric, age or surface gravity, shows better agreement across the sample albeit with larger variance between the MCMC and RFR approaches. For the early L dwarfs and late T dwarfs, the MCMC method yields significantly higher and lower surface gravities, respectively, than the RFR method, the latter diverging from expected values. However, these differences may be driven in part by the relatively large uncertainties in the RFR {\logg} values (0.3-0.4~dex, about 20\% of the parameter range), while the disagreement between fit and expected {\logg} for the late T dwarfs can be ascribed to a known gravity bias in the Elf Owl models. 

A separate accuracy assessment comes through comparison of fit parameter to evolutionary model expectations for {\teff} and radius.
Here we see significant departures for the early L dwarfs, with low temperatures and inflated radii based on best-fit Diamondback models. 
The much better reproduction of {\teff} and radius for the T dwarfs suggests this to be a condensation effect, although we cannot rule out biases in the evolutionary models caused by condensation as well. 
Overall, we find mixed outcomes in reproducing independent physical parameters that indicate continued issues in the spectral models, particularly for the L dwarfs. The large variance between MCMC and RFR parameters doesn't indicate that one method is preferred over the other.

\subsection{Precision: Parameter Uncertainties}

The largest difference found between the MCMC and RFR approaches was the precision of the inferred parameters, considerably larger for the latter approach.
This difference can be ascribed to the use of coarse grid sampling in this study. 
The MCMC approach, based on linear interpolation, can in principle reach any parameter value assuming the interpolation is valid. 
The RFR approach, based on decision trees and ensemble learning, constructs predictive models from the provided model grid which may lead to less accurate parameter estimates when the grid is not sampled at a sufficiently fine scale. 
A key measure of this is the real-versus-predicted statistic $R$ (Figure~\ref{fig:RvP})
which tracks predictive accuracy. 
While high self-consistency was found for \teff$\,$($R^2 > 0.95$ for all grids), predictability was very low for other parameters which were more coarsely sampled.
Recent advancements, such as those outlined in \cite{2023ApJ...954...22L}, suggest that incorporating interpolation between grid points for a more refined parameter mapping could significantly improve the precision of RFR and reduce parameter uncertainties. 
In addition, while model grids are usually constructed through uniform sampling of the parameters (either linearly or log-linearly), \cite{2022ApJ...934...31F} showed that this method is sub-optimal for parameter predictions, and that random or Latin hypercube sampling performed better for predicting parameters in higher-dimensional models. 
This study also emphasized that traditional retrievals with interpolation on a linear grid can produce biased posterior distributions, especially for parameters with nonlinear effects on the spectrum such as C/O ratio, cloud properties, and metallicity. 
As such, the small parameter uncertainties from the MCMC approach may mask systematic errors and biases in the grid sampling or interpolation, while the large parameter uncertainties for the RFR approach represents a fundamental limit for coarse model grids.
Furthermore, while the present study is based on high-S/N spectra, it is important to recognize that the effectiveness of machine learning methodologies like RFR can be adversely affected by noise, particularly in instances of low S/N observations. The work conducted by \cite{Reis2019AJ....157...16R} offers a comprehensive analysis regarding the impact of various levels and types of injected noise in a RFR training set and the precision of the resultant RFR outcomes.

\subsection{Efficiency: Computational Effort}

The computational efficiency of a fitting method is an important factor for practical applications, particularly when dealing with large spectral datasets such as those provided by Sloan Digital Sky Survey \citep{2000AJ....120.1579Y} or expected from Euclid \citep{2010SPIE.7731E..1HL} and SPHEREx \citep{2020SPIE11443E..0IC}. 
MCMC is a relatively inefficient method of sampling a posterior parameter space, and therefore requires significant computational resources. 
The major time sink for RFR, on the other hand, is in the training stage; the subsequent prediction-based model determines parameters relatively quickly.
Our study found that MCMC fits took, on average, 30 to 50 times more computational time per source than RFR fitting (i.e., minutes versus seconds; Table~\ref{tab:time}). 
Thereby, RFR computations were performed on a MacBook Pro (2019) with a 2.4 GHz 8-core Intel Core i9 processor and 32 GB of RAM, while MCMC calculations were done on a MacBook Pro (2021) with a 3.2 GHz 8-core M1 Max processor with 64 GB of RAM.
For large-scale or time-sensitive studies, the efficiency of RFR makes it a more practical approach, given the offsets in quality-of-fit and parameter precision.  
On the other hand, the better performance of MCMC makes it valuable for single fits requiring higher accuracy despite its higher computational cost.

\begin{deluxetable}{lccc}
\tabletypesize{\footnotesize}
\tablecaption{Average computational time of MCMC fitting (performed on a MacBook Pro (2021) with a 3.2 GHz 8-core M1 Max processor and 64 GB of RAM) and random forest retrievals (computed on a MacBook Pro (2019) with a 2.4 GHz 8-core Intel Core i9 processor and 32 GB of RAM).\label{tab:time}} 
\tablehead{ 
\colhead{\textbf{Model}} & 
\colhead{\textbf{DBack}} & 
\colhead{\textbf{EOwl}} & 
\colhead{\textbf{SAND}} \\ 
\colhead{} & 
\colhead{[mins:sec]} & 
\colhead{[mins:sec]} & 
\colhead{[mins:sec]}
} 
\startdata
\textit{MCMC Fitting} & & & \\
Total & 4:31 & 22:31 & 2:11\\
\cline{1-4} 
\textit{Random Forest Retrievals} & & & \\
Training & 0:05 & 0:30 & 0:04\\
Testing & $<$0:01 & 0:02 & $<$0:01\\
Total & $\sim$0:05 & 0:32 & $\sim$0:04\\
\hline
\enddata
\end{deluxetable}

\subsection{Combining MCMC and RFR Approaches}

Based on the comparative strengths and weaknesses of these two approaches, and the existing variance in the quality of fits among present-day atmosphere models for low-temperature spectra, we recommend a three-phase hybrid approach for fitting pre-computed models to source spectra. 
First, a preparation phase in which data and/or models are interpolated onto a common wavelength scale, and a set of RFR models are generated with each trained on a different atmosphere model set.
Second, an RFR fitting phase, with spectra fit to each of the model sets to identify the optimal set and initial parameters for each spectrum.
Third, an MCMC fitting phase in which the parameters are fine-tuned and precise estimates of uncertainty are inferred. 
This general framework allows for efficient evaluation of the optimal atmosphere model for a given source, or multiple models depending on error tolerance. 
It also has numerous avenues for optimization; for example, using subgrids of models tuned to the expected parameters of a sample, applying alternative supervised learning models or Bayesian fitting approaches, incorporating additional model calculations to improve parameter sensitivity or interpolation error, iterative inclusion of additional fitting parameters such as element abundance variance (e.g., \citealt{2016ApJ...828...95V,2024ApJ...973...60B}) 
once an initial optimal parameter set is identified, among others.
More importantly, this approach is agnostic to the source type or resolution of the data, making it applicable to exoplanet spectra 
(cf.\ \citealt{2018MNRAS.481.4698F,2024A&A...687A.110L})
and high-resolution data (cf.\ \citealt{2020AJ....159..192F,2022MNRAS.512.4618G}) with appropriate modifications.

\section{Summary} \label{sec:summary}

This study has assessed two approaches to fitting atmospheric models to low-temperature L and T dwarf spectra: Markov Chain Monte Carlo (MCMC) and Random Forest Retrieval (RFR). 
These approaches were applied using three sets of atmosphere models (Sonora Diamondback, Sonora Elf Owl, and SAND) and a sample of near-infrared spectra for eleven benchmark companions for which independent assessments of age and metallicity were available from their primaries and estimates of temperature and radius were available from evolutionary models.
The goal of this study was to determine the relative tradeoffs in these approaches in terms of accuracy, precision, and efficiency, and their future application to large spectral samples from current or future surveys. Our primary findings are as follows:

\begin{itemize}
    \item All eleven sources had a least one model set which provided a robust fit to near-infrared spectroscopy. In general, Diamondback provided the best fits for early L dwarfs, 
    Elf Owl for mid- and late-T dwarfs, and 
    SAND for late-L and early-T dwarfs and the young L dwarf G~196-3B.
    Absence of condensate opacity and errors in specific molecular gas opacities (e.g., CH$_4$ for SAND) were the general causes of poor fits.
    \item MCMC fits generally matched the observed spectra better than RFR fits, and both fitting methods generally converged to the same best model set, with some individual exceptions (e.g., G~196-3B, HN~Peg~B).
    \item MCMC also provided more accurate parameter estimates (smaller uncertainties) than RFR, although the actual parameters were generally consistent between the methods. In both cases, there was mixed agreement between the fit parameters and expected values based on the primaries (metallicity, surface gravity) or evolutionary models (radius, effective temperature).
    \item The RFR approach is a far more efficient algorithm, with fits taking seconds as compared to minutes for the MCMC approach. This approach also identifies features  of importance that align with temperature, surface gravity, and metallicity sensitive features, potentially useful for more directed parameter fitting schemes. 
    \item Based on this analysis, we propose a method for combining the strengths of these approaches to address the variances inherent in existing pre-computed models. Specifically, we advocate for a first-pass RFR fit to identify the optimal model set and initial parameters, followed by a more precise evaluation using MCMC fits when needed. This hybrid approach is flexible and can be adapted to different machine learning and Bayesian analysis approaches, as well as different datasets and source types. 
\end{itemize}

Our combined approach joins a growing set of ML-assisted strategies that aim to significantly increase the efficiency of fitting pre-computed models to stellar, brown dwarf, and exoplanet spectra.
Continued advancements in these methodologies, coupled with improved interpolation techniques and more finely-computed model grids, are needed to enhance our ability to accurately and precisely constrain the multiple physical parameters and processes that shape the spectra of low-temperature atmospheres.

\begin{acknowledgments}
The authors recognize and acknowledge the significant cultural role
and reverence that the summit of Maunakea has with the
indigenous Hawaiian community, and that the NASA Infrared Telescope Facility
stands on Crown and Government Lands that the
State of Hawai’i is obligated to protect and preserve for future
generations of indigenous Hawaiians. 
Portions of this work
were conducted at the University of California, San Diego,
which was built on the unceded territory of the Kumeyaay
Nation, whose people continue to maintain their political
sovereignty and cultural traditions as vital members of the San
Diego community.
A.L. acknowledges partial financial support from the Swiss National Science Foundation and the European Research Council (via a Consolidator Grant to Kevin Heng; grant number 771620), as well as administrative support from the Centre for Space and Habitability (CSH). This study would not have been possible without the open-source code HELA \citep{2018NatAs...2..719M}, which can be found on the Exoclimes Simulation Platform: https://github.com/exoclime. 
\end{acknowledgments}

\vspace{5mm}
\facilities{Infrared Telescope Facility(SpeX)}

\software{
astropy \citep{2013AandA...558A..33A,2018AJ....156..123A},  
Matplotlib \citep{2007CSE.....9...90H},
NumPy \citep{2011CSE....13b..22V},
pandas \citep{mckinney-proc-scipy-2010},
SciPy \citep{2020NatMe..17..261V},
SPLAT \citep{2017ASInC..14....7B}, and
scikit.learn \citep{scikit-learn2011}
}

\appendix

\begin{figure}[h]
\centering
\includegraphics[width=\textwidth]{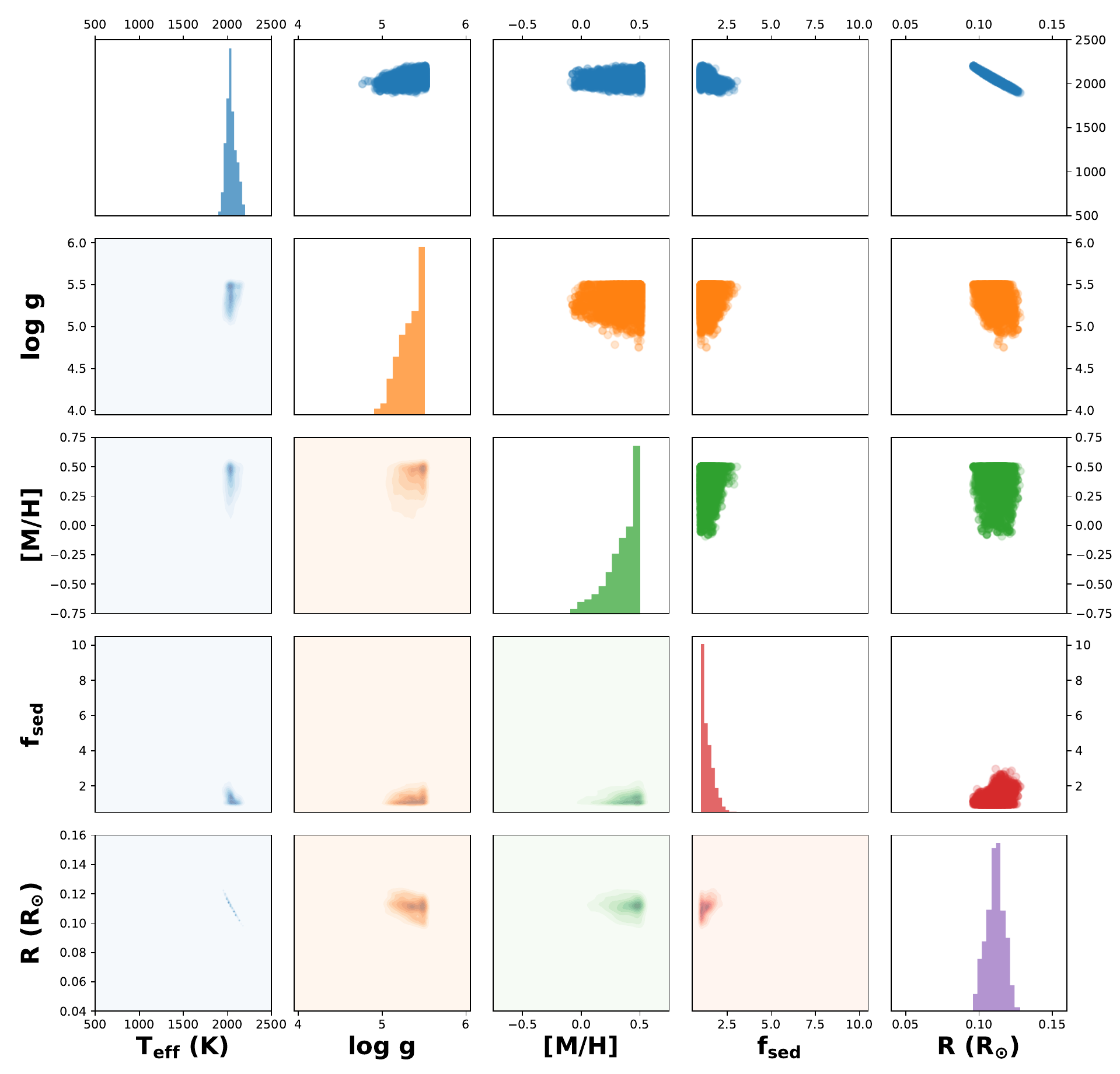}
\caption{Posterior parameter distributions from our MCMC Diamondback fits of the near-infrared spectrum of the L0 dwarf LP 465-70B. 
Diagonal plots display the marginalized 1D parameter distributions
for temperature T$_\mathrm{eff}$ (blue), surface gravity $\log{g}$ (orange), metallicity [$M/H$] (green), cloud sedimentation efficiency $f_\mathrm{sed}$ (red), and radius R (purple).
The other panel display joint parameters distributions; the 
lower left panels display contour plots of probability, the upper right panels display individual parameters in the MCMC chain.
} \label{fig:mcmc-corner-LP465-70B}
\end{figure}

\begin{figure}
    \centering
    \includegraphics[width=\textwidth]{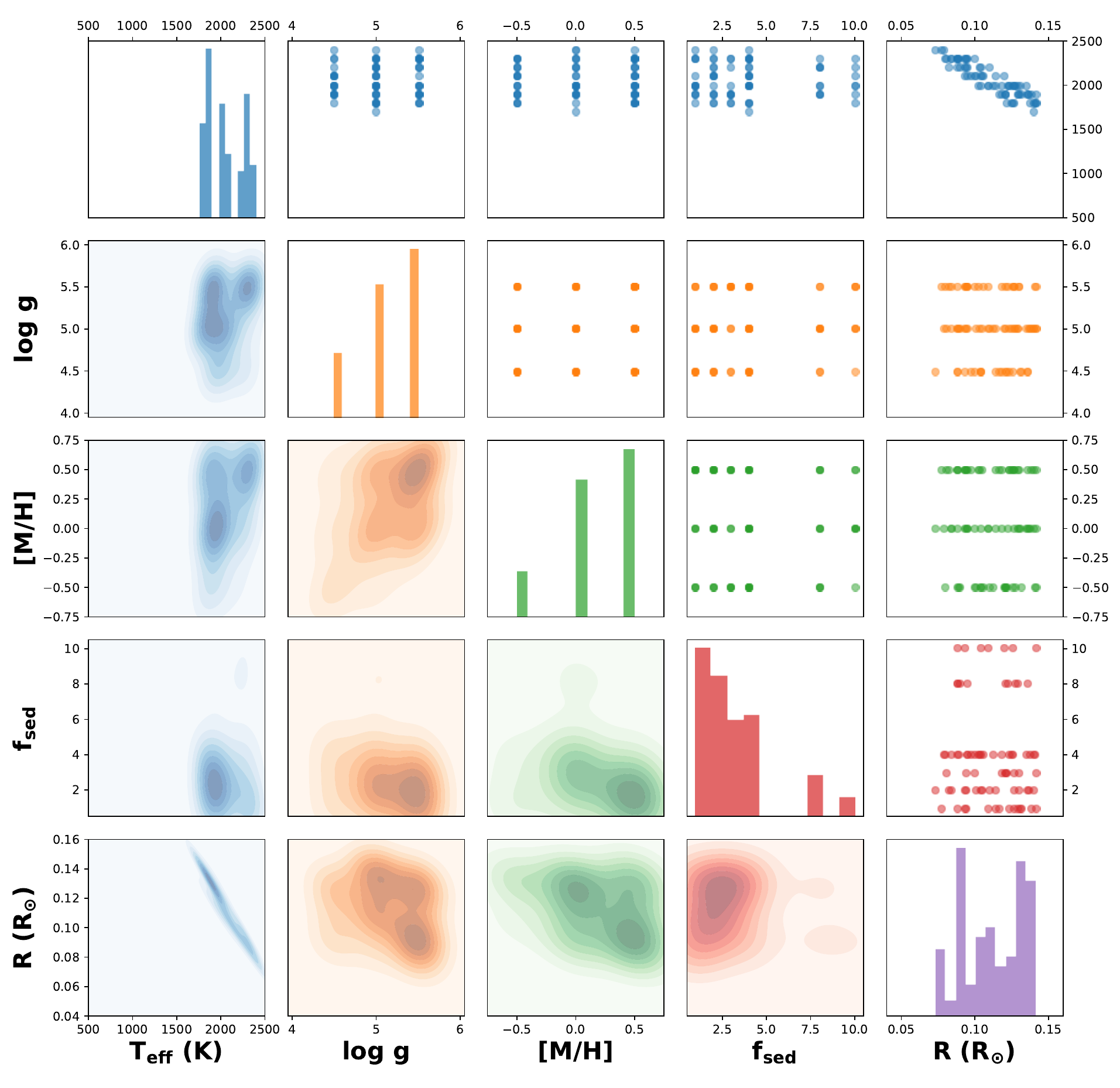}
    \caption{Posterior parameter distributions for our RFR Diamondback retrieval of the L0 dwarf LP 465-70B. Paramaters and panels are as defined as Figure~\ref{fig:mcmc-corner-LP465-70B}, with the upper right panels showing the predictions of individual trees within the random forest ensemble.}   
   \label{fig:Dback_posteriors_LP465-70B}
\end{figure}

\begin{figure}[h]
\centering
\includegraphics[width=\textwidth]{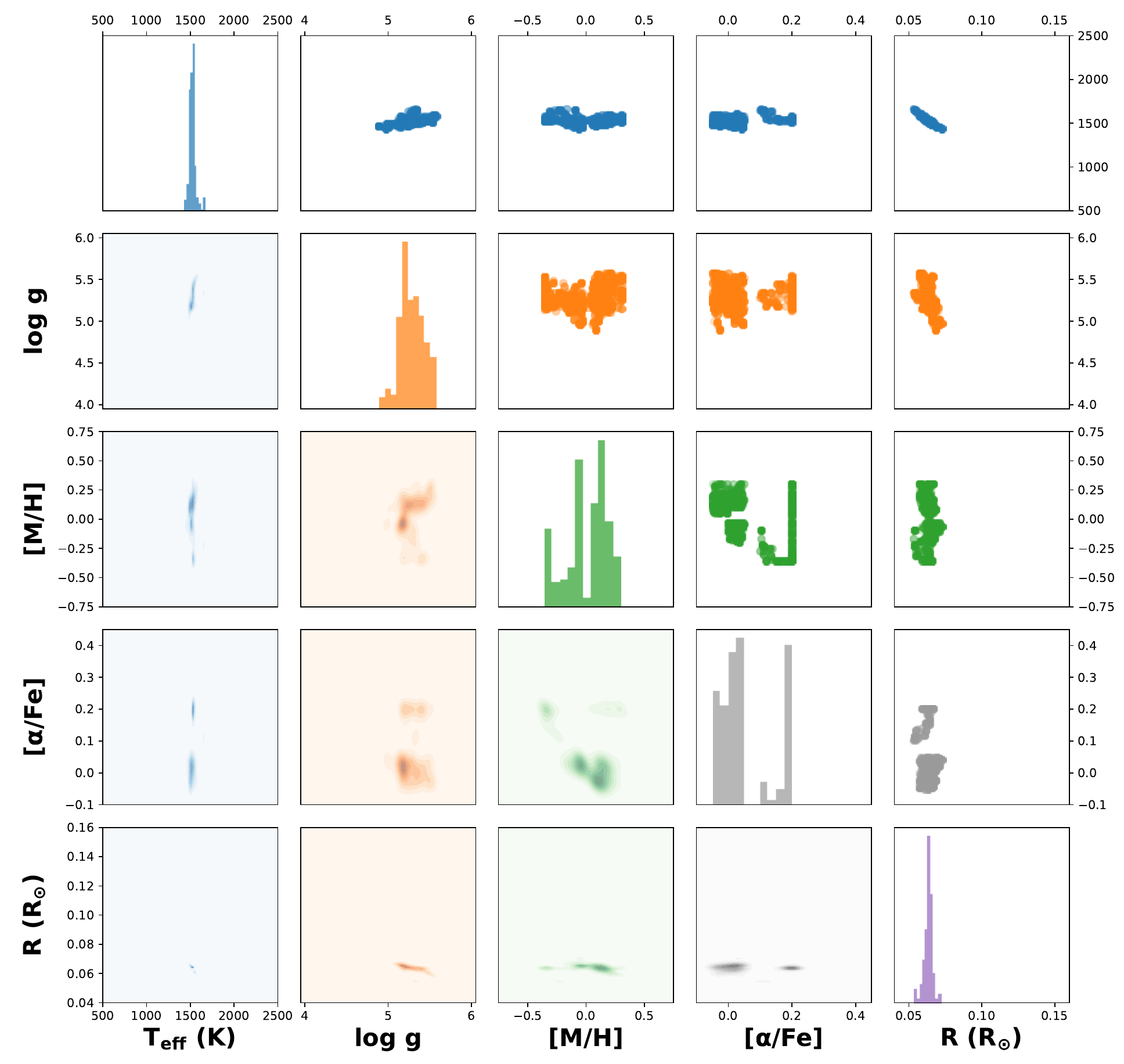}
\caption{Same as Figure~\ref{fig:mcmc-corner-LP465-70B} for MCMC SAND fits of the near-infrared spectrum of the L8 dwarf Gliese 584C.
Parameters shown are temperature T$_\mathrm{eff}$ (blue), surface gravity $\log{g}$ (orange), metallicity [$M/H$] (green), the alpha element enrichment [$\alpha$/\ch{Fe}] (gray), and R (purple).} \label{fig:mcmc-corner-gl584C}
\end{figure}

\begin{figure}
    \centering
    \includegraphics[width=\textwidth]{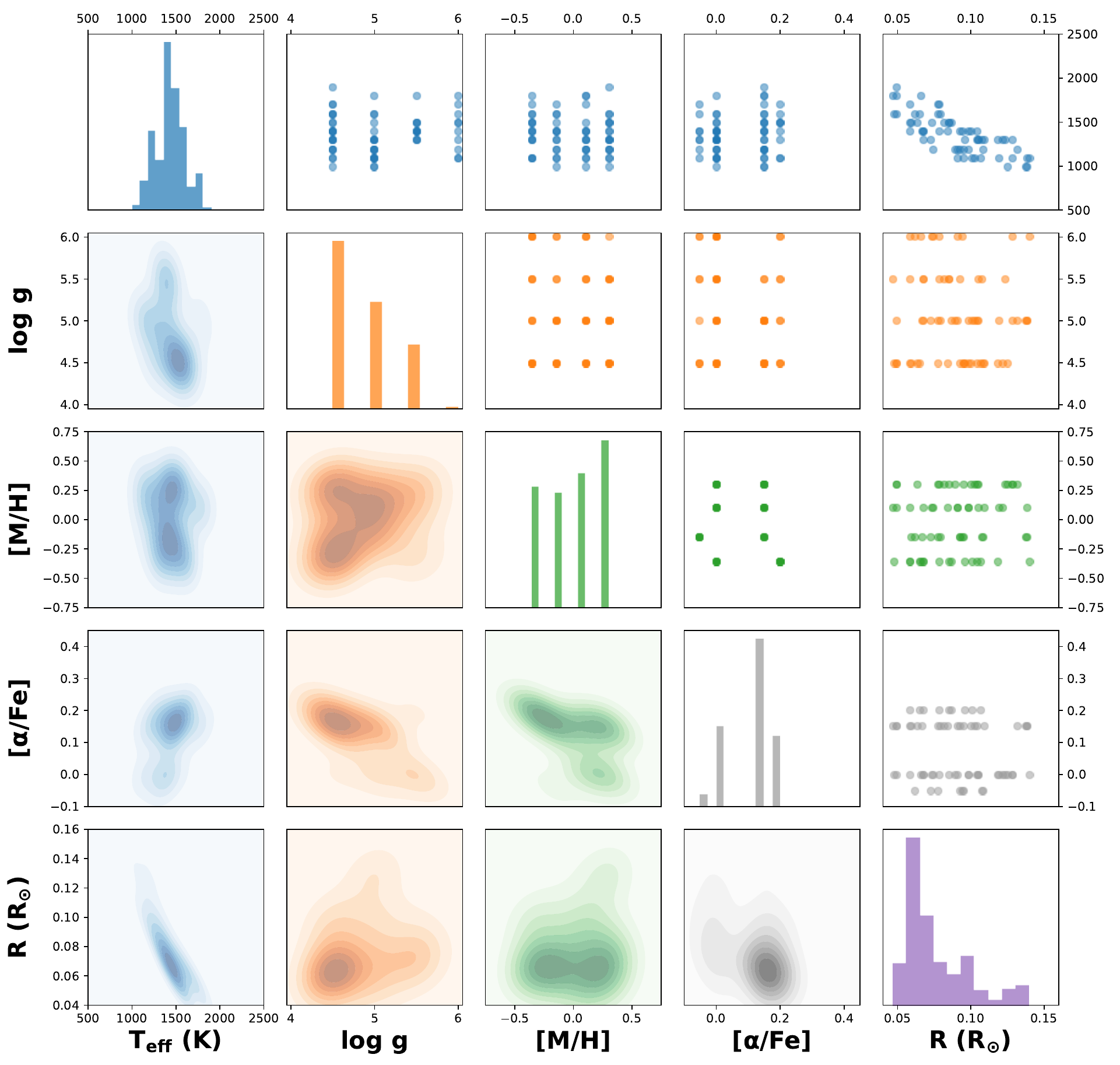}
    \caption{Same as Figure~\ref{fig:mcmc-corner-LP465-70B} for the RFR SAND retrieval of the near-infrared spectrum of the L8 dwarf Gliese 584C. }
    \label{fig:Sand_posteriors_Gliese584C}
\end{figure}

\begin{figure}[h]
\centering
\includegraphics[width=\textwidth]{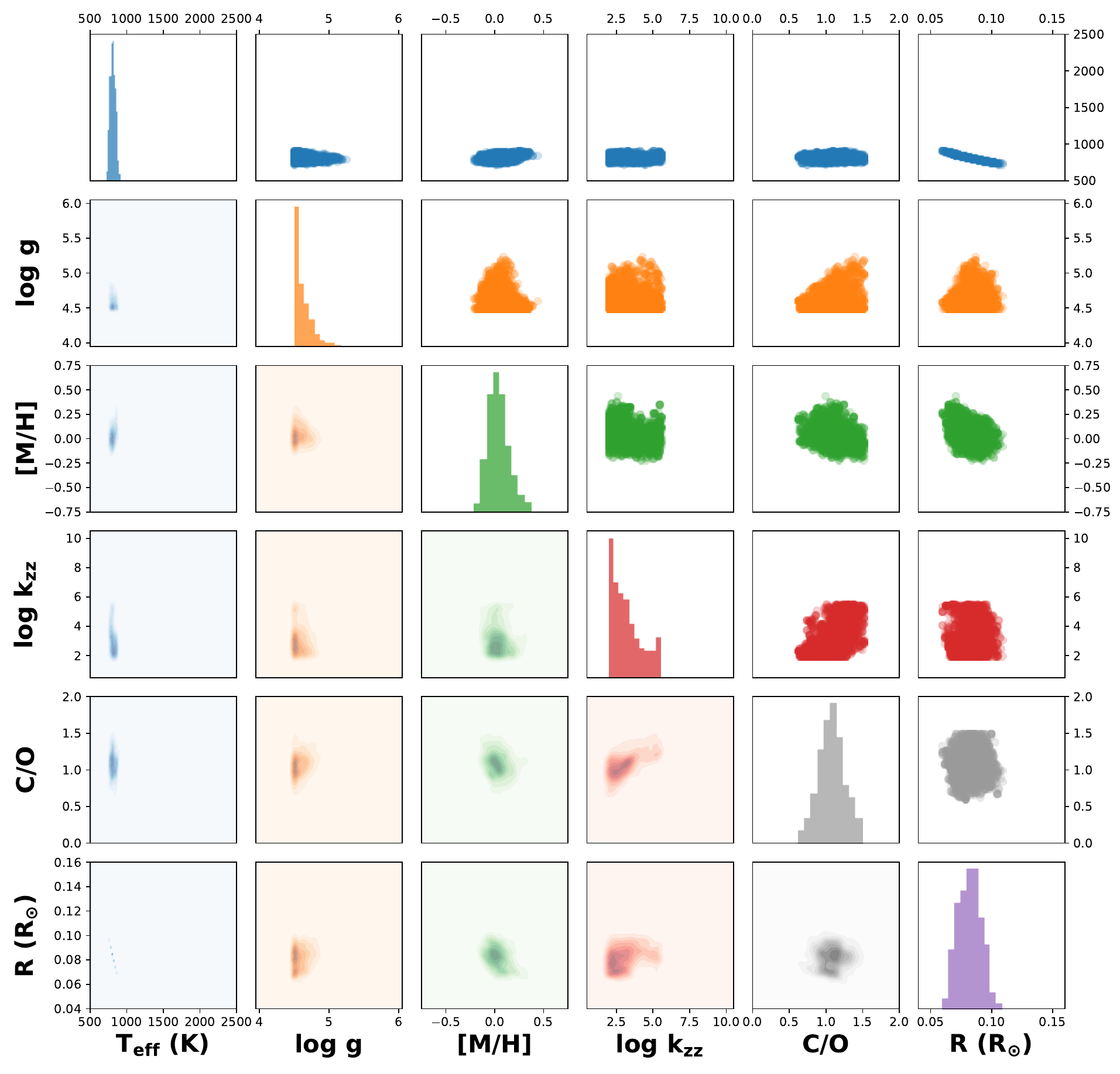}
\caption{Same as Figure~\ref{fig:mcmc-corner-LP465-70B} for MCMC Elf Owl fits of the near-infrared spectrum of the T8 dwarf Gliese 570D.
Parameters shown are temperature T$_\mathrm{eff}$ (blue), surface gravity $\log{g}$ (orange), metallicity [$M/H$] (green), vertical eddy diffusion coefficient $\log k_{\rm zz}$ (red), carbon-to-oxygen ratio C/O (gray), and radius R (purple).} 
\label{fig:mcmc-corner-gliese570D}
\end{figure}

\begin{figure}
    \centering
    \includegraphics[width=\textwidth]{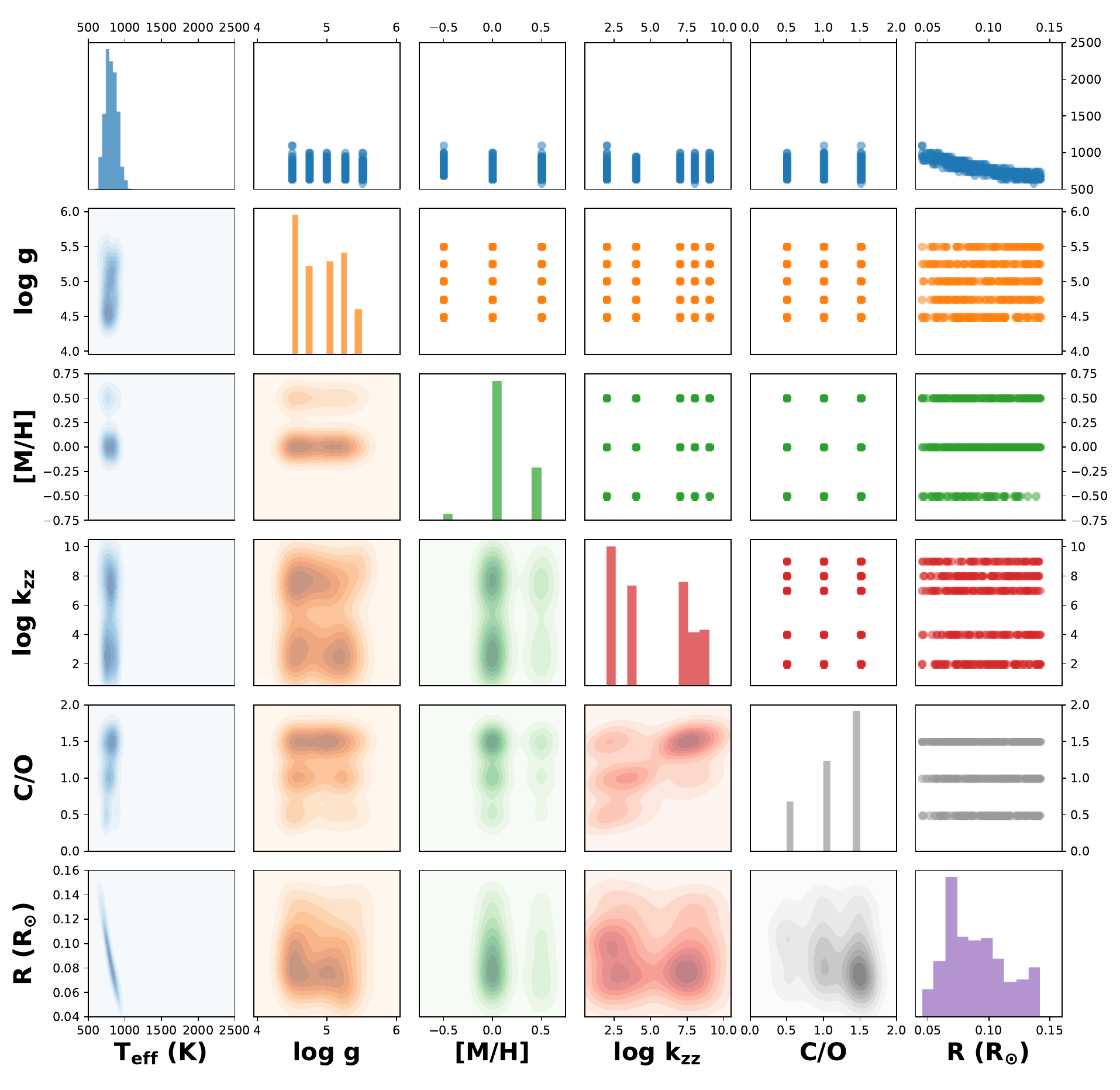}
    \caption{Same as Figure~\ref{fig:mcmc-corner-gliese570D} for the RFR Elf Owl retrieval of the near-infrared spectrum of the T8 dwarf Gliese 570D.} 
    \label{fig:Elfowl_posteriors_Gliese570D}
\end{figure}

Here, we provide in Figures~\ref{fig:mcmc-corner-LP465-70B} through~\ref{fig:Elfowl_posteriors_Gliese570D} the joint posterior parameter distributions for our best-fit MCMC and RFR analyses of three exemplary brown dwarfs in our sample:
the early-type L dwarf LP 465-70B,
the L/T transition object Gliese 584C, 
and the late T dwarf Gliese 570D (Section~\ref{sec:casestudies}). 
Each figure shows the marginalized individual parameters distributions as histograms, and joint distributions as 2D contour plots, with common colors assigned to each parameter. The optimal parameter values and estimated uncertainties are provided in Tables~\ref{tab:MCMCresults} for MCMC and~\ref{tab:RFRresults} for RFR.

\clearpage

\bibliography{bib}{}

\begin{thebibliography}{}
\expandafter\ifx\csname natexlab\endcsname\relax\def\natexlab#1{#1}\fi
\providecommand{\url}[1]{\href{#1}{#1}}
\providecommand{\dodoi}[1]{doi:~\href{http://doi.org/#1}{\nolinkurl{#1}}}
\providecommand{\doeprint}[1]{\href{http://ascl.net/#1}{\nolinkurl{http://ascl.net/#1}}}
\providecommand{\doarXiv}[1]{\href{https://arxiv.org/abs/#1}{\nolinkurl{https://arxiv.org/abs/#1}}}

\bibitem[{{Ackerman} \& {Marley}(2001)}]{2001ApJ...556..872A}
{Ackerman}, A.~S., \& {Marley}, M.~S. 2001, \apj, 556, 872, \dodoi{10.1086/321540}

\bibitem[{{Aganze} {et~al.}(2016){Aganze}, {Burgasser}, {Faherty}, {Choban}, {Escala}, {Lopez}, {Jin}, {Tamiya}, {Tallis}, \& {Rockward}}]{2016AJ....151...46A}
{Aganze}, C., {Burgasser}, A.~J., {Faherty}, J.~K., {et~al.} 2016, \aj, 151, 46, \dodoi{10.3847/0004-6256/151/2/46}

\bibitem[{{Allard} {et~al.}(1997){Allard}, {Hauschildt}, {Alexander}, \& {Starrfield}}]{1997ARA&A..35..137A}
{Allard}, F., {Hauschildt}, P.~H., {Alexander}, D.~R., \& {Starrfield}, S. 1997, \araa, 35, 137, \dodoi{10.1146/annurev.astro.35.1.137}

\bibitem[{{Allard} {et~al.}(2001){Allard}, {Hauschildt}, {Alexander}, {Tamanai}, \& {Schweitzer}}]{2001ApJ...556..357A}
{Allard}, F., {Hauschildt}, P.~H., {Alexander}, D.~R., {Tamanai}, A., \& {Schweitzer}, A. 2001, \apj, 556, 357, \dodoi{10.1086/321547}

\bibitem[{{Allard} {et~al.}(2011){Allard}, {Homeier}, \& {Freytag}}]{2011ASPC..448...91A}
{Allard}, F., {Homeier}, D., \& {Freytag}, B. 2011, in Astronomical Society of the Pacific Conference Series, Vol. 448, 16th Cambridge Workshop on Cool Stars, Stellar Systems, and the Sun, ed. C.~{Johns-Krull}, M.~K. {Browning}, \& A.~A. {West}, 91, \dodoi{10.48550/arXiv.1011.5405}

\bibitem[{{Allard} {et~al.}(2012){Allard}, {Homeier}, \& {Freytag}}]{2012RSPTA.370.2765A}
{Allard}, F., {Homeier}, D., \& {Freytag}, B. 2012, Philosophical Transactions of the Royal Society A, 370, 2765, \dodoi{10.1098/rsta.2011.0269}

\bibitem[{{Allard} {et~al.}(2005){Allard}, {Allard}, \& {Kielkopf}}]{2005A&A...440.1195A}
{Allard}, N.~F., {Allard}, F., \& {Kielkopf}, J.~F. 2005, \aap, 440, 1195, \dodoi{10.1051/0004-6361:20053162}

\bibitem[{{Alvarado} {et~al.}(2024){Alvarado}, {Gerasimov}, {Burgasser}, {Brooks}, {Aganze}, \& {Theissen}}]{SAND}
{Alvarado}, E., {Gerasimov}, R., {Burgasser}, A.~J., {et~al.} 2024, Research Notes of the American Astronomical Society, 8, 134, \dodoi{10.3847/2515-5172/ad4bd7}

\bibitem[{{Ard{\'e}vol Mart{\'\i}nez} {et~al.}(2022){Ard{\'e}vol Mart{\'\i}nez}, {Min}, {Kamp}, \& {Palmer}}]{2022A&A...662A.108A}
{Ard{\'e}vol Mart{\'\i}nez}, F., {Min}, M., {Kamp}, I., \& {Palmer}, P.~I. 2022, \aap, 662, A108, \dodoi{10.1051/0004-6361/202142976}

\bibitem[{{Astropy Collaboration} {et~al.}(2013){Astropy Collaboration}, {Robitaille}, {Tollerud}, {Greenfield}, {Droettboom}, {Bray}, {Aldcroft}, {Davis}, {Ginsburg}, {Price-Whelan}, {Kerzendorf}, {Conley}, {Crighton}, {Barbary}, {Muna}, {Ferguson}, {Grollier}, {Parikh}, {Nair}, {Unther}, {Deil}, {Woillez}, {Conseil}, {Kramer}, {Turner}, {Singer}, {Fox}, {Weaver}, {Zabalza}, {Edwards}, {Azalee Bostroem}, {Burke}, {Casey}, {Crawford}, {Dencheva}, {Ely}, {Jenness}, {Labrie}, {Lim}, {Pierfederici}, {Pontzen}, {Ptak}, {Refsdal}, {Servillat}, \& {Streicher}}]{2013AandA...558A..33A}
{Astropy Collaboration}, {Robitaille}, T.~P., {Tollerud}, E.~J., {et~al.} 2013, \aap, 558, A33, \dodoi{10.1051/0004-6361/201322068}

\bibitem[{{Astropy Collaboration} {et~al.}(2018){Astropy Collaboration}, {Price-Whelan}, {Sip{\H{o}}cz}, {G{\"u}nther}, {Lim}, {Crawford}, {Conseil}, {Shupe}, {Craig}, {Dencheva}, {Ginsburg}, {VanderPlas}, {Bradley}, {P{\'e}rez-Su{\'a}rez}, {de Val-Borro}, {Aldcroft}, {Cruz}, {Robitaille}, {Tollerud}, {Ardelean}, {Babej}, {Bach}, {Bachetti}, {Bakanov}, {Bamford}, {Barentsen}, {Barmby}, {Baumbach}, {Berry}, {Biscani}, {Boquien}, {Bostroem}, {Bouma}, {Brammer}, {Bray}, {Breytenbach}, {Buddelmeijer}, {Burke}, {Calderone}, {Cano Rodr{\'\i}guez}, {Cara}, {Cardoso}, {Cheedella}, {Copin}, {Corrales}, {Crichton}, {D'Avella}, {Deil}, {Depagne}, {Dietrich}, {Donath}, {Droettboom}, {Earl}, {Erben}, {Fabbro}, {Ferreira}, {Finethy}, {Fox}, {Garrison}, {Gibbons}, {Goldstein}, {Gommers}, {Greco}, {Greenfield}, {Groener}, {Grollier}, {Hagen}, {Hirst}, {Homeier}, {Horton}, {Hosseinzadeh}, {Hu}, {Hunkeler}, {Ivezi{\'c}}, {Jain}, {Jenness}, {Kanarek}, {Kendrew}, {Kern}, {Kerzendorf}, {Khvalko}, {King}, {Kirkby}, {Kulkarni},
  {Kumar}, {Lee}, {Lenz}, {Littlefair}, {Ma}, {Macleod}, {Mastropietro}, {McCully}, {Montagnac}, {Morris}, {Mueller}, {Mumford}, {Muna}, {Murphy}, {Nelson}, {Nguyen}, {Ninan}, {N{\"o}the}, {Ogaz}, {Oh}, {Parejko}, {Parley}, {Pascual}, {Patil}, {Patil}, {Plunkett}, {Prochaska}, {Rastogi}, {Reddy Janga}, {Sabater}, {Sakurikar}, {Seifert}, {Sherbert}, {Sherwood-Taylor}, {Shih}, {Sick}, {Silbiger}, {Singanamalla}, {Singer}, {Sladen}, {Sooley}, {Sornarajah}, {Streicher}, {Teuben}, {Thomas}, {Tremblay}, {Turner}, {Terr{\'o}n}, {van Kerkwijk}, {de la Vega}, {Watkins}, {Weaver}, {Whitmore}, {Woillez}, {Zabalza}, \& {Astropy Contributors}}]{2018AJ....156..123A}
{Astropy Collaboration}, {Price-Whelan}, A.~M., {Sip{\H{o}}cz}, B.~M., {et~al.} 2018, \aj, 156, 123, \dodoi{10.3847/1538-3881/aabc4f}

\bibitem[{{Baraffe} {et~al.}(2003){Baraffe}, {Chabrier}, {Barman}, {Allard}, \& {Hauschildt}}]{2003A&A...402..701B}
{Baraffe}, I., {Chabrier}, G., {Barman}, T.~S., {Allard}, F., \& {Hauschildt}, P.~H. 2003, \aap, 402, 701, \dodoi{10.1051/0004-6361:20030252}

\bibitem[{{Baraffe} {et~al.}(2015){Baraffe}, {Homeier}, {Allard}, \& {Chabrier}}]{2015AA...577A..42B}
{Baraffe}, I., {Homeier}, D., {Allard}, F., \& {Chabrier}, G. 2015, \aap, 577, A42, \dodoi{10.1051/0004-6361/201425481}

\bibitem[{{Bardalez Gagliuffi} {et~al.}(2014){Bardalez Gagliuffi}, {Burgasser}, {Gelino}, {Looper}, {Nicholls}, {Schmidt}, {Cruz}, {West}, {Gizis}, \& {Metchev}}]{2014ApJ...794..143B}
{Bardalez Gagliuffi}, D.~C., {Burgasser}, A.~J., {Gelino}, C.~R., {et~al.} 2014, \apj, 794, 143, \dodoi{10.1088/0004-637X/794/2/143}

\bibitem[{{Barnes}(2007)}]{2007ApJ...669.1167B}
{Barnes}, S.~A. 2007, \apj, 669, 1167, \dodoi{10.1086/519295}

\bibitem[{{Beiler} {et~al.}(2024){Beiler}, {Mukherjee}, {Cushing}, {Kirkpatrick}, {Schneider}, {Kothari}, {Marley}, \& {Visscher}}]{2024ApJ...973...60B}
{Beiler}, S.~A., {Mukherjee}, S., {Cushing}, M.~C., {et~al.} 2024, \apj, 973, 60, \dodoi{10.3847/1538-4357/ad6759}

\bibitem[{{Best} {et~al.}(2021){Best}, {Liu}, {Magnier}, \& {Dupuy}}]{2021AJ....161...42B}
{Best}, W. M.~J., {Liu}, M.~C., {Magnier}, E.~A., \& {Dupuy}, T.~J. 2021, \aj, 161, 42, \dodoi{10.3847/1538-3881/abc893}

\bibitem[{{Bonfanti} {et~al.}(2016){Bonfanti}, {Ortolani}, \& {Nascimbeni}}]{2016AA...585A...5B}
{Bonfanti}, A., {Ortolani}, S., \& {Nascimbeni}, V. 2016, \aap, 585, A5, \dodoi{10.1051/0004-6361/201527297}

\bibitem[{{Breiman}(2001)}]{2001MachL..45....5B}
{Breiman}, L. 2001, Machine Learning, 45, 5, \dodoi{10.1023/A:1010933404324}

\bibitem[{Breiman(2001)}]{Breiman2001random}
Breiman, L. 2001, Machine learning, 45, 5

\bibitem[{{Burgasser} {et~al.}(2006{\natexlab{a}}){Burgasser}, {Burrows}, \& {Kirkpatrick}}]{2006ApJ...639.1095B}
{Burgasser}, A.~J., {Burrows}, A., \& {Kirkpatrick}, J.~D. 2006{\natexlab{a}}, \apj, 639, 1095, \dodoi{10.1086/499344}

\bibitem[{{Burgasser} {et~al.}(2006{\natexlab{b}}){Burgasser}, {Geballe}, {Leggett}, {Kirkpatrick}, \& {Golimowski}}]{2006ApJ...637.1067B}
{Burgasser}, A.~J., {Geballe}, T.~R., {Leggett}, S.~K., {Kirkpatrick}, J.~D., \& {Golimowski}, D.~A. 2006{\natexlab{b}}, \apj, 637, 1067, \dodoi{10.1086/498563}

\bibitem[{{Burgasser} {et~al.}(2002{\natexlab{a}}){Burgasser}, {Marley}, {Ackerman}, {Saumon}, {Lodders}, {Dahn}, {Harris}, \& {Kirkpatrick}}]{2002ApJ...571L.151B}
{Burgasser}, A.~J., {Marley}, M.~S., {Ackerman}, A.~S., {et~al.} 2002{\natexlab{a}}, \apjl, 571, L151, \dodoi{10.1086/341343}

\bibitem[{{Burgasser} \& {Splat Development Team}(2017)}]{2017ASInC..14....7B}
{Burgasser}, A.~J., \& {Splat Development Team}. 2017, in Astronomical Society of India Conference Series, Vol.~14, Astronomical Society of India Conference Series, 7--12.
\newblock \doarXiv{1707.00062}

\bibitem[{{Burgasser} {et~al.}(2009){Burgasser}, {Witte}, {Helling}, {Sanderson}, {Bochanski}, \& {Hauschildt}}]{2009ApJ...697..148B}
{Burgasser}, A.~J., {Witte}, S., {Helling}, C., {et~al.} 2009, \apj, 697, 148, \dodoi{10.1088/0004-637X/697/1/148}

\bibitem[{{Burgasser} {et~al.}(2002{\natexlab{b}}){Burgasser}, {Kirkpatrick}, {Brown}, {Reid}, {Burrows}, {Liebert}, {Matthews}, {Gizis}, {Dahn}, {Monet}, {Cutri}, \& {Skrutskie}}]{2002ApJ...564..421B}
{Burgasser}, A.~J., {Kirkpatrick}, J.~D., {Brown}, M.~E., {et~al.} 2002{\natexlab{b}}, \apj, 564, 421, \dodoi{10.1086/324033}

\bibitem[{{Burgasser} {et~al.}(2024{\natexlab{a}}){Burgasser}, {Gerasimov}, {Kremer}, {Brooks}, {Alvarado}, {Schneider}, {Meisner}, {Theissen}, {Softich}, {Karpoor}, {Bickle}, {Kabatnik}, {Rothermich}, {Caselden}, {Kirkpatrick}, {Faherty}, {Casewell}, {Kuchner}, \& {The Backyard Worlds: Planet 9 Collaboration}}]{2024ApJ...971L..25B}
{Burgasser}, A.~J., {Gerasimov}, R., {Kremer}, K., {et~al.} 2024{\natexlab{a}}, \apjl, 971, L25, \dodoi{10.3847/2041-8213/ad6607}

\bibitem[{{Burgasser} {et~al.}(2024{\natexlab{b}}){Burgasser}, {Schneider}, {Meisner}, {Caselden}, {Hsu}, {Gerasimov}, {Aganze}, {Softich}, {Karpoor}, {Theissen}, {Brooks}, {Bickle}, {Gagn{\'e}}, {Artigau}, {Marsset}, {Rothermich}, {Faherty}, {Kirkpatrick}, {Kuchner}, {Stevnbak Andersen}, {Beaulieu}, {Colin}, {Gantier}, {Gramaize}, {Hamlet}, {Hinckley}, {Kabatnik}, {Kiwy}, {Martin}, {Massat}, {Pendrill}, {Sainio}, {Sch{\"u}mann}, {Th{\'e}venot}, {Walla}, {W{\k{e}}dracki}, {Backyard Worlds}, {:}, \& {Planet 9 Collaboration}}]{2024arXiv241101378B}
{Burgasser}, A.~J., {Schneider}, A.~C., {Meisner}, A.~M., {et~al.} 2024{\natexlab{b}}, arXiv e-prints, arXiv:2411.01378, \dodoi{10.48550/arXiv.2411.01378}

\bibitem[{{Burningham} {et~al.}(2017){Burningham}, {Marley}, {Line}, {Lupu}, {Visscher}, {Morley}, {Saumon}, \& {Freedman}}]{2017MNRAS.470.1177B}
{Burningham}, B., {Marley}, M.~S., {Line}, M.~R., {et~al.} 2017, \mnras, 470, 1177, \dodoi{10.1093/mnras/stx1246}

\bibitem[{{Burrows} {et~al.}(2001){Burrows}, {Hubbard}, {Lunine}, \& {Liebert}}]{2001RvMP...73..719B}
{Burrows}, A., {Hubbard}, W.~B., {Lunine}, J.~I., \& {Liebert}, J. 2001, Reviews of Modern Physics, 73, 719

\bibitem[{{Burrows} {et~al.}(2006){Burrows}, {Sudarsky}, \& {Hubeny}}]{2006ApJ...640.1063B}
{Burrows}, A., {Sudarsky}, D., \& {Hubeny}, I. 2006, \apj, 640, 1063, \dodoi{10.1086/500293}

\bibitem[{{Charnay} {et~al.}(2018){Charnay}, {B{\'e}zard}, {Baudino}, {Bonnefoy}, {Boccaletti}, \& {Galicher}}]{2018ApJ...854..172C}
{Charnay}, B., {B{\'e}zard}, B., {Baudino}, J.~L., {et~al.} 2018, \apj, 854, 172, \dodoi{10.3847/1538-4357/aaac7d}

\bibitem[{{Cobb} {et~al.}(2019){Cobb}, {Himes}, {Soboczenski}, {Zorzan}, {O'Beirne}, {G{\"u}ne{\c{s}} Baydin}, {Gal}, {Domagal-Goldman}, {Arney}, {Angerhausen}, \& {2018 NASA FDL Astrobiology Team}}]{2019AJ....158...33C}
{Cobb}, A.~D., {Himes}, M.~D., {Soboczenski}, F., {et~al.} 2019, \aj, 158, 33, \dodoi{10.3847/1538-3881/ab2390}

\bibitem[{Cochran(1977)}]{Cochran1977}
Cochran, W.~G. 1977, Sampling Techniques, 3rd edn. (New York: Wiley)

\bibitem[{{Crill} {et~al.}(2020){Crill}, {Werner}, {Akeson}, {Ashby}, {Bleem}, {Bock}, {Bryan}, {Burnham}, {Byunh}, {Chang}, {Chiang}, {Cook}, {Cooray}, {Davis}, {Dor{\'e}}, {Dowell}, {Dubois-Felsmann}, {Eifler}, {Faisst}, {Habib}, {Heinrich}, {Heitmann}, {Heaton}, {Hirata}, {Hristov}, {Hui}, {Jeong}, {Kang}, {Kecman}, {Kirkpatrick}, {Korngut}, {Krause}, {Lee}, {Lisse}, {Masters}, {Mauskopf}, {Melnick}, {Miyasaka}, {Nayyeri}, {Nguyen}, {{\"O}berg}, {Padin}, {Paladini}, {Pourrahmani}, {Pyo}, {Smith}, {Song}, {Symons}, {Teplitz}, {Tolls}, {Unwin}, {Windhorst}, {Yang}, \& {Zemcov}}]{2020SPIE11443E..0IC}
{Crill}, B.~P., {Werner}, M., {Akeson}, R., {et~al.} 2020, in Society of Photo-Optical Instrumentation Engineers (SPIE) Conference Series, Vol. 11443, Space Telescopes and Instrumentation 2020: Optical, Infrared, and Millimeter Wave, ed. M.~{Lystrup} \& M.~D. {Perrin}, 114430I, \dodoi{10.1117/12.2567224}

\bibitem[{{Cruz} {et~al.}(2009){Cruz}, {Kirkpatrick}, \& {Burgasser}}]{2009AJ....137.3345C}
{Cruz}, K.~L., {Kirkpatrick}, J.~D., \& {Burgasser}, A.~J. 2009, \aj, 137, 3345, \dodoi{10.1088/0004-6256/137/2/3345}

\bibitem[{{Cushing} {et~al.}(2008){Cushing}, {Marley}, {Saumon}, {Kelly}, {Vacca}, {Rayner}, {Freedman}, {Lodders}, \& {Roellig}}]{2008ApJ...678.1372C}
{Cushing}, M.~C., {Marley}, M.~S., {Saumon}, D., {et~al.} 2008, \apj, 678, 1372, \dodoi{10.1086/526489}

\bibitem[{{Dupuy} \& {Liu}(2012)}]{2012ApJS..201...19D}
{Dupuy}, T.~J., \& {Liu}, M.~C. 2012, \apjs, 201, 19, \dodoi{10.1088/0067-0049/201/2/19}

\bibitem[{{Faherty} {et~al.}(2010){Faherty}, {Burgasser}, {West}, {Bochanski}, {Cruz}, {Shara}, \& {Walter}}]{2010AJ....139..176F}
{Faherty}, J.~K., {Burgasser}, A.~J., {West}, A.~A., {et~al.} 2010, \aj, 139, 176, \dodoi{10.1088/0004-6256/139/1/176}

\bibitem[{{Faherty} {et~al.}(2012){Faherty}, {Burgasser}, {Walter}, {Van der Bliek}, {Shara}, {Cruz}, {West}, {Vrba}, \& {Anglada-Escud{\'e}}}]{2012ApJ...752...56F}
{Faherty}, J.~K., {Burgasser}, A.~J., {Walter}, F.~M., {et~al.} 2012, \apj, 752, 56, \dodoi{10.1088/0004-637X/752/1/56}

\bibitem[{Fang {et~al.}(2005)Fang, Li, \& Sudjianto}]{fang2005design}
Fang, K.-T., Li, R., \& Sudjianto, A. 2005, Design and modeling for computer experiments (Chapman and Hall/CRC)

\bibitem[{{Fegley} \& {Prinn}(1985)}]{1985ApJ...299.1067F}
{Fegley}, B., J., \& {Prinn}, R.~G. 1985, \apj, 299, 1067, \dodoi{10.1086/163775}

\bibitem[{{Fisher} \& {Heng}(2018)}]{2018MNRAS.481.4698F}
{Fisher}, C., \& {Heng}, K. 2018, \mnras, 481, 4698, \dodoi{10.1093/mnras/sty2550}

\bibitem[{{Fisher} \& {Heng}(2022)}]{2022ApJ...934...31F}
---. 2022, \apj, 934, 31, \dodoi{10.3847/1538-4357/ac7801}

\bibitem[{{Fisher} {et~al.}(2020){Fisher}, {Hoeijmakers}, {Kitzmann}, {M{\'a}rquez-Neila}, {Grimm}, {Sznitman}, \& {Heng}}]{2020AJ....159..192F}
{Fisher}, C., {Hoeijmakers}, H.~J., {Kitzmann}, D., {et~al.} 2020, \aj, 159, 192, \dodoi{10.3847/1538-3881/ab7a92}

\bibitem[{{Gaia Collaboration} {et~al.}(2023){Gaia Collaboration}, {Vallenari}, {Brown}, {Prusti}, {de Bruijne}, {Arenou}, {Babusiaux}, {Biermann}, {Creevey}, {Ducourant}, {Evans}, {Eyer}, {Guerra}, {Hutton}, {Jordi}, {Klioner}, {Lammers}, {Lindegren}, {Luri}, {Mignard}, {Panem}, {Pourbaix}, {Randich}, {Sartoretti}, {Soubiran}, {Tanga}, {Walton}, {Bailer-Jones}, {Bastian}, {Drimmel}, {Jansen}, {Katz}, {Lattanzi}, {van Leeuwen}, {Bakker}, {Cacciari}, {Casta{\~n}eda}, {De Angeli}, {Fabricius}, {Fouesneau}, {Fr{\'e}mat}, {Galluccio}, {Guerrier}, {Heiter}, {Masana}, {Messineo}, {Mowlavi}, {Nicolas}, {Nienartowicz}, {Pailler}, {Panuzzo}, {Riclet}, {Roux}, {Seabroke}, {Sordo}, {Th{\'e}venin}, {Gracia-Abril}, {Portell}, {Teyssier}, {Altmann}, {Andrae}, {Audard}, {Bellas-Velidis}, {Benson}, {Berthier}, {Blomme}, {Burgess}, {Busonero}, {Busso}, {C{\'a}novas}, {Carry}, {Cellino}, {Cheek}, {Clementini}, {Damerdji}, {Davidson}, {de Teodoro}, {Nu{\~n}ez Campos}, {Delchambre}, {Dell'Oro}, {Esquej},
  {Fern{\'a}ndez-Hern{\'a}ndez}, {Fraile}, {Garabato}, {Garc{\'\i}a-Lario}, {Gosset}, {Haigron}, {Halbwachs}, {Hambly}, {Harrison}, {Hern{\'a}ndez}, {Hestroffer}, {Hodgkin}, {Holl}, {Jan{\ss}en}, {Jevardat de Fombelle}, {Jordan}, {Krone-Martins}, {Lanzafame}, {L{\"o}ffler}, {Marchal}, {Marrese}, {Moitinho}, {Muinonen}, {Osborne}, {Pancino}, {Pauwels}, {Recio-Blanco}, {Reyl{\'e}}, {Riello}, {Rimoldini}, {Roegiers}, {Rybizki}, {Sarro}, {Siopis}, {Smith}, {Sozzetti}, {Utrilla}, {van Leeuwen}, {Abbas}, {{\'A}brah{\'a}m}, {Abreu Aramburu}, {Aerts}, {Aguado}, {Ajaj}, {Aldea-Montero}, {Altavilla}, {{\'A}lvarez}, {Alves}, {Anders}, {Anderson}, {Anglada Varela}, {Antoja}, {Baines}, {Baker}, {Balaguer-N{\'u}{\~n}ez}, {Balbinot}, {Balog}, {Barache}, {Barbato}, {Barros}, {Barstow}, {Bartolom{\'e}}, {Bassilana}, {Bauchet}, {Becciani}, {Bellazzini}, {Berihuete}, {Bernet}, {Bertone}, {Bianchi}, {Binnenfeld}, {Blanco-Cuaresma}, {Blazere}, {Boch}, {Bombrun}, {Bossini}, {Bouquillon}, {Bragaglia}, {Bramante}, {Breedt},
  {Bressan}, {Brouillet}, {Brugaletta}, {Bucciarelli}, {Burlacu}, {Butkevich}, {Buzzi}, {Caffau}, {Cancelliere}, {Cantat-Gaudin}, {Carballo}, {Carlucci}, {Carnerero}, {Carrasco}, {Casamiquela}, {Castellani}, {Castro-Ginard}, {Chaoul}, {Charlot}, {Chemin}, {Chiaramida}, {Chiavassa}, {Chornay}, {Comoretto}, {Contursi}, {Cooper}, {Cornez}, {Cowell}, {Crifo}, {Cropper}, {Crosta}, {Crowley}, {Dafonte}, {Dapergolas}, {David}, {David}, {de Laverny}, {De Luise}, {De March}, {De Ridder}, {de Souza}, {de Torres}, {del Peloso}, {del Pozo}, {Delbo}, {Delgado}, {Delisle}, {Demouchy}, {Dharmawardena}, {Di Matteo}, {Diakite}, {Diener}, {Distefano}, {Dolding}, {Edvardsson}, {Enke}, {Fabre}, {Fabrizio}, {Faigler}, {Fedorets}, {Fernique}, {Fienga}, {Figueras}, {Fournier}, {Fouron}, {Fragkoudi}, {Gai}, {Garcia-Gutierrez}, {Garcia-Reinaldos}, {Garc{\'\i}a-Torres}, {Garofalo}, {Gavel}, {Gavras}, {Gerlach}, {Geyer}, {Giacobbe}, {Gilmore}, {Girona}, {Giuffrida}, {Gomel}, {Gomez}, {Gonz{\'a}lez-N{\'u}{\~n}ez},
  {Gonz{\'a}lez-Santamar{\'\i}a}, {Gonz{\'a}lez-Vidal}, {Granvik}, {Guillout}, {Guiraud}, {Guti{\'e}rrez-S{\'a}nchez}, {Guy}, {Hatzidimitriou}, {Hauser}, {Haywood}, {Helmer}, {Helmi}, {Sarmiento}, {Hidalgo}, {Hilger}, {H{\l}adczuk}, {Hobbs}, {Holland}, {Huckle}, {Jardine}, {Jasniewicz}, {Jean-Antoine Piccolo}, {Jim{\'e}nez-Arranz}, {Jorissen}, {Juaristi Campillo}, {Julbe}, {Karbevska}, {Kervella}, {Khanna}, {Kontizas}, {Kordopatis}, {Korn}, {K{\'o}sp{\'a}l}, {Kostrzewa-Rutkowska}, {Kruszy{\'n}ska}, {Kun}, {Laizeau}, {Lambert}, {Lanza}, {Lasne}, {Le Campion}, {Lebreton}, {Lebzelter}, {Leccia}, {Leclerc}, {Lecoeur-Taibi}, {Liao}, {Licata}, {Lindstr{\o}m}, {Lister}, {Livanou}, {Lobel}, {Lorca}, {Loup}, {Madrero Pardo}, {Magdaleno Romeo}, {Managau}, {Mann}, {Manteiga}, {Marchant}, {Marconi}, {Marcos}, {Marcos Santos}, {Mar{\'\i}n Pina}, {Marinoni}, {Marocco}, {Marshall}, {Martin Polo}, {Mart{\'\i}n-Fleitas}, {Marton}, {Mary}, {Masip}, {Massari}, {Mastrobuono-Battisti}, {Mazeh}, {McMillan}, {Messina}, {Michalik},
  {Millar}, {Mints}, {Molina}, {Molinaro}, {Moln{\'a}r}, {Monari}, {Mongui{\'o}}, {Montegriffo}, {Montero}, {Mor}, {Mora}, {Morbidelli}, {Morel}, {Morris}, {Muraveva}, {Murphy}, {Musella}, {Nagy}, {Noval}, {Oca{\~n}a}, {Ogden}, {Ordenovic}, {Osinde}, {Pagani}, {Pagano}, {Palaversa}, {Palicio}, {Pallas-Quintela}, {Panahi}, {Payne-Wardenaar}, {Pe{\~n}alosa Esteller}, {Penttil{\"a}}, {Pichon}, {Piersimoni}, {Pineau}, {Plachy}, {Plum}, {Poggio}, {Pr{\v{s}}a}, {Pulone}, {Racero}, {Ragaini}, {Rainer}, {Raiteri}, {Rambaux}, {Ramos}, {Ramos-Lerate}, {Re Fiorentin}, {Regibo}, {Richards}, {Rios Diaz}, {Ripepi}, {Riva}, {Rix}, {Rixon}, {Robichon}, {Robin}, {Robin}, {Roelens}, {Rogues}, {Rohrbasser}, {Romero-G{\'o}mez}, {Rowell}, {Royer}, {Ruz Mieres}, {Rybicki}, {Sadowski}, {S{\'a}ez N{\'u}{\~n}ez}, {Sagrist{\`a} Sell{\'e}s}, {Sahlmann}, {Salguero}, {Samaras}, {Sanchez Gimenez}, {Sanna}, {Santove{\~n}a}, {Sarasso}, {Schultheis}, {Sciacca}, {Segol}, {Segovia}, {S{\'e}gransan}, {Semeux}, {Shahaf}, {Siddiqui}, {Siebert},
  {Siltala}, {Silvelo}, {Slezak}, {Slezak}, {Smart}, {Snaith}, {Solano}, {Solitro}, {Souami}, {Souchay}, {Spagna}, {Spina}, {Spoto}, {Steele}, {Steidelm{\"u}ller}, {Stephenson}, {S{\"u}veges}, {Surdej}, {Szabados}, {Szegedi-Elek}, {Taris}, {Taylor}, {Teixeira}, {Tolomei}, {Tonello}, {Torra}, {Torra}, {Torralba Elipe}, {Trabucchi}, {Tsounis}, {Turon}, {Ulla}, {Unger}, {Vaillant}, {van Dillen}, {van Reeven}, {Vanel}, {Vecchiato}, {Viala}, {Vicente}, {Voutsinas}, {Weiler}, {Wevers}, {Wyrzykowski}, {Yoldas}, {Yvard}, {Zhao}, {Zorec}, {Zucker}, \& {Zwitter}}]{2023AA...674A...1G}
{Gaia Collaboration}, {Vallenari}, A., {Brown}, A.~G.~A., {et~al.} 2023, \aap, 674, A1, \dodoi{10.1051/0004-6361/202243940}

\bibitem[{Galassi {et~al.}(2011)Galassi, Davies, Theiler, Gough, Jungman, Booth, \& Rossi}]{Galassi2011}
Galassi, M., Davies, J., Theiler, J., {et~al.} 2011, GNU Scientific Library - Reference manual, Version 1.15.
\newblock \url{https://www.gnu.org/software/gsl/doc/html/intro.html}

\bibitem[{{Geballe} {et~al.}(2001){Geballe}, {Saumon}, {Leggett}, {Knapp}, {Marley}, \& {Lodders}}]{2001ApJ...556..373G}
{Geballe}, T.~R., {Saumon}, D., {Leggett}, S.~K., {et~al.} 2001, \apj, 556, 373, \dodoi{10.1086/321575}

\bibitem[{{Gebhard} {et~al.}(2024){Gebhard}, {Angerhausen}, {Konrad}, {Alei}, {Quanz}, \& {Sch{\"o}lkopf}}]{2024A&A...681A...3G}
{Gebhard}, T.~D., {Angerhausen}, D., {Konrad}, B.~S., {et~al.} 2024, \aap, 681, A3, \dodoi{10.1051/0004-6361/202346390}

\bibitem[{{Gerasimov} {et~al.}(2024){Gerasimov}, {Burgasser}, {Caiazzo}, {Homeier}, {Richer}, {Correnti}, \& {Heyl}}]{2024ApJ...961..139G}
{Gerasimov}, R., {Burgasser}, A.~J., {Caiazzo}, I., {et~al.} 2024, \apj, 961, 139, \dodoi{10.3847/1538-4357/ad08bf}

\bibitem[{{Gerasimov} {et~al.}(2022){Gerasimov}, {Burgasser}, {Homeier}, {Bedin}, {Rees}, {Scalco}, {Anderson}, \& {Salaris}}]{2022ApJ...930...24G}
{Gerasimov}, R., {Burgasser}, A.~J., {Homeier}, D., {et~al.} 2022, \apj, 930, 24, \dodoi{10.3847/1538-4357/ac61e5}

\bibitem[{{Gerasimov} {et~al.}(2020){Gerasimov}, {Homeier}, {Burgasser}, \& {Bedin}}]{2020RNAAS...4..214G}
{Gerasimov}, R., {Homeier}, D., {Burgasser}, A., \& {Bedin}, L.~R. 2020, Research Notes of the American Astronomical Society, 4, 214, \dodoi{10.3847/2515-5172/abcf2c}

\bibitem[{{Geweke}(1992)}]{geweke1992}
{Geweke}, J. 1992, in Bayesian Statistics 4, ed. A.~D. J.M.~Bernardo, J.O.~Berger \& A.~Smith (Oxford: Oxford University Press)

\bibitem[{{Gibson} {et~al.}(2022){Gibson}, {Nugroho}, {Lothringer}, {Maguire}, \& {Sing}}]{2022MNRAS.512.4618G}
{Gibson}, N.~P., {Nugroho}, S.~K., {Lothringer}, J., {Maguire}, C., \& {Sing}, D.~K. 2022, \mnras, 512, 4618, \dodoi{10.1093/mnras/stac091}

\bibitem[{{Gizis} {et~al.}(2001{\natexlab{a}}){Gizis}, {Kirkpatrick}, {Burgasser}, {Reid}, {Monet}, {Liebert}, \& {Wilson}}]{2001ApJ...551L.163G}
{Gizis}, J.~E., {Kirkpatrick}, J.~D., {Burgasser}, A., {et~al.} 2001{\natexlab{a}}, \apjl, 551, L163, \dodoi{10.1086/320017}

\bibitem[{{Gizis} {et~al.}(2001{\natexlab{b}}){Gizis}, {Kirkpatrick}, \& {Wilson}}]{2001AJ....121.2185G}
{Gizis}, J.~E., {Kirkpatrick}, J.~D., \& {Wilson}, J.~C. 2001{\natexlab{b}}, \aj, 121, 2185, \dodoi{10.1086/319937}

\bibitem[{{Gomes} {et~al.}(2013){Gomes}, {Pinfield}, {Marocco}, {Day-Jones}, {Burningham}, {Zhang}, {Jones}, {van Spaandonk}, \& {Weights}}]{2013MNRAS.431.2745G}
{Gomes}, J.~I., {Pinfield}, D.~J., {Marocco}, F., {et~al.} 2013, \mnras, 431, 2745, \dodoi{10.1093/mnras/stt371}

\bibitem[{{Gray} {et~al.}(2006){Gray}, {Corbally}, {Garrison}, {McFadden}, {Bubar}, {McGahee}, {O'Donoghue}, \& {Knox}}]{2006AJ....132..161G}
{Gray}, R.~O., {Corbally}, C.~J., {Garrison}, R.~F., {et~al.} 2006, \aj, 132, 161, \dodoi{10.1086/504637}

\bibitem[{{Gray} {et~al.}(2001){Gray}, {Napier}, \& {Winkler}}]{2001AJ....121.2148G}
{Gray}, R.~O., {Napier}, M.~G., \& {Winkler}, L.~I. 2001, \aj, 121, 2148, \dodoi{10.1086/319956}

\bibitem[{{Griffith} \& {Yelle}(1999)}]{1999ApJ...519L..85G}
{Griffith}, C.~A., \& {Yelle}, R.~V. 1999, \apjl, 519, L85, \dodoi{10.1086/312103}

\bibitem[{{Guzm{\'a}n-Mesa} {et~al.}(2020){Guzm{\'a}n-Mesa}, {Kitzmann}, {Fisher}, {Burgasser}, {Hoeijmakers}, {M{\'a}rquez-Neila}, {Grimm}, {Mandell}, {Sznitman}, \& {Heng}}]{2020AJ....160...15G}
{Guzm{\'a}n-Mesa}, A., {Kitzmann}, D., {Fisher}, C., {et~al.} 2020, \aj, 160, 15, \dodoi{10.3847/1538-3881/ab9176}

\bibitem[{{Hastings}(1970)}]{HASTINGS01041970}
{Hastings}, W.~K. 1970, Biometrika, 57, 97, \dodoi{10.1093/biomet/57.1.97}

\bibitem[{{Hauschildt} {et~al.}(1997){Hauschildt}, {Baron}, \& {Allard}}]{1997ApJ...483..390H}
{Hauschildt}, P.~H., {Baron}, E., \& {Allard}, F. 1997, \apj, 483, 390, \dodoi{10.1086/304233}

\bibitem[{{Hawley} {et~al.}(2002){Hawley}, {Covey}, {Knapp}, {Golimowski}, {Fan}, {Anderson}, {Gunn}, {Harris}, {Ivezi{\'c}}, {Long}, {Lupton}, {McGehee}, {Narayanan}, {Peng}, {Schlegel}, {Schneider}, {Spahn}, {Strauss}, {Szkody}, {Tsvetanov}, {Walkowicz}, {Brinkmann}, {Harvanek}, {Hennessy}, {Kleinman}, {Krzesinski}, {Long}, {Neilsen}, {Newman}, {Nitta}, {Snedden}, \& {York}}]{2002AJ....123.3409H}
{Hawley}, S.~L., {Covey}, K.~R., {Knapp}, G.~R., {et~al.} 2002, \aj, 123, 3409, \dodoi{10.1086/340697}

\bibitem[{{Helling}(2005)}]{2005AN....326..627H}
{Helling}, C. 2005, Astronomische Nachrichten, 326, 627

\bibitem[{{Helling} {et~al.}(2008){Helling}, {Dehn}, {Woitke}, \& {Hauschildt}}]{2008ApJ...675L.105H}
{Helling}, C., {Dehn}, M., {Woitke}, P., \& {Hauschildt}, P.~H. 2008, \apjl, 675, L105, \dodoi{10.1086/533462}

\bibitem[{{Helling} {et~al.}(2001){Helling}, {Oevermann}, {L{\"u}ttke}, {Klein}, \& {Sedlmayr}}]{2001A&A...376..194H}
{Helling}, C., {Oevermann}, M., {L{\"u}ttke}, M.~J.~H., {Klein}, R., \& {Sedlmayr}, E. 2001, \aap, 376, 194, \dodoi{10.1051/0004-6361:20010937}

\bibitem[{Ho(1998)}]{Ho1998random}
Ho, T.~K. 1998, IEEE transactions on pattern analysis and machine intelligence, 20, 832

\bibitem[{{Houk} \& {Swift}(1999)}]{1999MSS...C05....0H}
{Houk}, N., \& {Swift}, C. 1999, Michigan Spectral Survey, 5, 0

\bibitem[{{Hunter}(2007)}]{2007CSE.....9...90H}
{Hunter}, J.~D. 2007, Computing in Science and Engineering, 9, 90, \dodoi{10.1109/MCSE.2007.55}

\bibitem[{{Irwin} {et~al.}(2008){Irwin}, {Teanby}, {de Kok}, {Fletcher}, {Howett}, {Tsang}, {Wilson}, {Calcutt}, {Nixon}, \& {Parrish}}]{2008JQSRT.109.1136I}
{Irwin}, P.~G.~J., {Teanby}, N.~A., {de Kok}, R., {et~al.} 2008, \jqsrt, 109, 1136, \dodoi{10.1016/j.jqsrt.2007.11.006}

\bibitem[{{Keenan} \& {McNeil}(1989)}]{1989ApJS...71..245K}
{Keenan}, P.~C., \& {McNeil}, R.~C. 1989, \apjs, 71, 245, \dodoi{10.1086/191373}

\bibitem[{{Kirkpatrick} {et~al.}(2000){Kirkpatrick}, {Reid}, {Liebert}, {Gizis}, {Burgasser}, {Monet}, {Dahn}, {Nelson}, \& {Williams}}]{2000AJ....120..447K}
{Kirkpatrick}, J.~D., {Reid}, I.~N., {Liebert}, J., {et~al.} 2000, \aj, 120, 447, \dodoi{10.1086/301427}

\bibitem[{{Kitzmann} {et~al.}(2020){Kitzmann}, {Heng}, {Oreshenko}, {Grimm}, {Apai}, {Bowler}, {Burgasser}, \& {Marley}}]{2020ApJ...890..174K}
{Kitzmann}, D., {Heng}, K., {Oreshenko}, M., {et~al.} 2020, \apj, 890, 174, \dodoi{10.3847/1538-4357/ab6d71}

\bibitem[{Kleijnen(2015)}]{kleijnen2015design}
Kleijnen, J.~P. 2015, in International workshop on simulation, Springer, 3--22

\bibitem[{{Knapp} {et~al.}(2004){Knapp}, {Leggett}, {Fan}, {Marley}, {Geballe}, {Golimowski}, {Finkbeiner}, {Gunn}, {Hennawi}, {Ivezi{\'c}}, {Lupton}, {Schlegel}, {Strauss}, {Tsvetanov}, {Chiu}, {Hoversten}, {Glazebrook}, {Zheng}, {Hendrickson}, {Williams}, {Uomoto}, {Vrba}, {Henden}, {Luginbuhl}, {Guetter}, {Munn}, {Canzian}, {Schneider}, \& {Brinkmann}}]{2004AJ....127.3553K}
{Knapp}, G.~R., {Leggett}, S.~K., {Fan}, X., {et~al.} 2004, \aj, 127, 3553, \dodoi{10.1086/420707}

\bibitem[{{Laureijs} {et~al.}(2010){Laureijs}, {Duvet}, {Escudero Sanz}, {Gondoin}, {Lumb}, {Oosterbroek}, \& {Saavedra Criado}}]{2010SPIE.7731E..1HL}
{Laureijs}, R.~J., {Duvet}, L., {Escudero Sanz}, I., {et~al.} 2010, in Society of Photo-Optical Instrumentation Engineers (SPIE) Conference Series, Vol. 7731, Space Telescopes and Instrumentation 2010: Optical, Infrared, and Millimeter Wave, 77311H, \dodoi{10.1117/12.857123}

\bibitem[{{Lawrence} {et~al.}(2007){Lawrence}, {Warren}, {Almaini}, {Edge}, {Hambly}, {Jameson}, {Lucas}, {Casali}, {Adamson}, {Dye}, {Emerson}, {Foucaud}, {Hewett}, {Hirst}, {Hodgkin}, {Irwin}, {Lodieu}, {McMahon}, {Simpson}, {Smail}, {Mortlock}, \& {Folger}}]{2007MNRAS.379.1599L}
{Lawrence}, A., {Warren}, S.~J., {Almaini}, O., {et~al.} 2007, \mnras, 379, 1599, \dodoi{10.1111/j.1365-2966.2007.12040.x}

\bibitem[{{Leggett} {et~al.}(2007){Leggett}, {Saumon}, {Marley}, {Geballe}, {Golimowski}, {Stephens}, \& {Fan}}]{2007ApJ...655.1079L}
{Leggett}, S.~K., {Saumon}, D., {Marley}, M.~S., {et~al.} 2007, \apj, 655, 1079, \dodoi{10.1086/510014}

\bibitem[{{Leggett} {et~al.}(2010){Leggett}, {Burningham}, {Saumon}, {Marley}, {Warren}, {Smart}, {Jones}, {Lucas}, {Pinfield}, \& {Tamura}}]{2010ApJ...710.1627L}
{Leggett}, S.~K., {Burningham}, B., {Saumon}, D., {et~al.} 2010, \apj, 710, 1627, \dodoi{10.1088/0004-637X/710/2/1627}

\bibitem[{{Leggett} {et~al.}(2021){Leggett}, {Tremblin}, {Phillips}, {Dupuy}, {Marley}, {Morley}, {Schneider}, {Caselden}, {Guillaume}, \& {Logsdon}}]{2021ApJ...918...11L}
{Leggett}, S.~K., {Tremblin}, P., {Phillips}, M.~W., {et~al.} 2021, \apj, 918, 11, \dodoi{10.3847/1538-4357/ac0cfe}

\bibitem[{{Line} {et~al.}(2014){Line}, {Fortney}, {Marley}, \& {Sorahana}}]{2014ApJ...793...33L}
{Line}, M.~R., {Fortney}, J.~J., {Marley}, M.~S., \& {Sorahana}, S. 2014, \apj, 793, 33, \dodoi{10.1088/0004-637X/793/1/33}

\bibitem[{{Line} {et~al.}(2015){Line}, {Teske}, {Burningham}, {Fortney}, \& {Marley}}]{2015ApJ...807..183L}
{Line}, M.~R., {Teske}, J., {Burningham}, B., {Fortney}, J.~J., \& {Marley}, M.~S. 2015, \apj, 807, 183, \dodoi{10.1088/0004-637X/807/2/183}

\bibitem[{{Linsky}(1969)}]{1969ApJ...156..989L}
{Linsky}, J.~L. 1969, \apj, 156, 989, \dodoi{10.1086/150030}

\bibitem[{{Lodders} {et~al.}(2009){Lodders}, {Palme}, \& {Gail}}]{2009LanB...4B..712L}
{Lodders}, K., {Palme}, H., \& {Gail}, H.~P. 2009, Landolt B{\"o}rnstein, 4B, 712, \dodoi{10.1007/978-3-540-88055-4_34}

\bibitem[{{Lodieu} {et~al.}(2022){Lodieu}, {Zapatero Osorio}, {Mart{\'\i}n}, {Rebolo L{\'o}pez}, \& {Gauza}}]{2022A&A...663A..84L}
{Lodieu}, N., {Zapatero Osorio}, M.~R., {Mart{\'\i}n}, E.~L., {Rebolo L{\'o}pez}, R., \& {Gauza}, B. 2022, \aap, 663, A84, \dodoi{10.1051/0004-6361/202243516}

\bibitem[{{Lueber} {et~al.}(2024{\natexlab{a}}){Lueber}, {Heng}, {Bowler}, {Kitzmann}, {Vos}, \& {Zhou}}]{2024A&A...690A.357L}
{Lueber}, A., {Heng}, K., {Bowler}, B.~P., {et~al.} 2024{\natexlab{a}}, \aap, 690, A357, \dodoi{10.1051/0004-6361/202451301}

\bibitem[{{Lueber} {et~al.}(2023){Lueber}, {Kitzmann}, {Fisher}, {Bowler}, {Burgasser}, {Marley}, \& {Heng}}]{2023ApJ...954...22L}
{Lueber}, A., {Kitzmann}, D., {Fisher}, C.~E., {et~al.} 2023, \apj, 954, 22, \dodoi{10.3847/1538-4357/ace530}

\bibitem[{{Lueber} {et~al.}(2024{\natexlab{b}}){Lueber}, {Novais}, {Fisher}, \& {Heng}}]{2024A&A...687A.110L}
{Lueber}, A., {Novais}, A., {Fisher}, C., \& {Heng}, K. 2024{\natexlab{b}}, \aap, 687, A110, \dodoi{10.1051/0004-6361/202348802}

\bibitem[{{Luhman} {et~al.}(2007)}]{2007ApJ...654..570L}
{Luhman}, K.~L., {et~al.} 2007, \apj, 654, 570, \dodoi{10.1086/509073}

\bibitem[{{Madhusudhan}(2012)}]{2012ApJ...758...36M}
{Madhusudhan}, N. 2012, \apj, 758, 36, \dodoi{10.1088/0004-637X/758/1/36}

\bibitem[{{Madhusudhan} \& {Seager}(2009)}]{2009ApJ...707...24M}
{Madhusudhan}, N., \& {Seager}, S. 2009, \apj, 707, 24, \dodoi{10.1088/0004-637X/707/1/24}

\bibitem[{{Marfil} {et~al.}(2021){Marfil}, {Tabernero}, {Montes}, {Caballero}, {L{\'a}zaro}, {Gonz{\'a}lez Hern{\'a}ndez}, {Nagel}, {Passegger}, {Schweitzer}, {Ribas}, {Reiners}, {Quirrenbach}, {Amado}, {Cifuentes}, {Cort{\'e}s-Contreras}, {Dreizler}, {Duque-Arribas}, {Galad{\'\i}-Enr{\'\i}quez}, {Henning}, {Jeffers}, {Kaminski}, {K{\"u}rster}, {Lafarga}, {L{\'o}pez-Gallifa}, {Morales}, {Shan}, \& {Zechmeister}}]{2021AA...656A.162M}
{Marfil}, E., {Tabernero}, H.~M., {Montes}, D., {et~al.} 2021, \aap, 656, A162, \dodoi{10.1051/0004-6361/202141980}

\bibitem[{{Marley} \& {Robinson}(2015)}]{2015ARA&A..53..279M}
{Marley}, M.~S., \& {Robinson}, T.~D. 2015, \araa, 53, 279, \dodoi{10.1146/annurev-astro-082214-122522}

\bibitem[{{M{\'a}rquez-Neila} {et~al.}(2018){M{\'a}rquez-Neila}, {Fisher}, {Sznitman}, \& {Heng}}]{2018NatAs...2..719M}
{M{\'a}rquez-Neila}, P., {Fisher}, C., {Sznitman}, R., \& {Heng}, K. 2018, Nature Astronomy, 2, 719, \dodoi{10.1038/s41550-018-0504-2}

\bibitem[{{McKay} {et~al.}(1989){McKay}, {Pollack}, \& {Courtin}}]{1989Icar...80...23M}
{McKay}, C.~P., {Pollack}, J.~B., \& {Courtin}, R. 1989, \icarus, 80, 23, \dodoi{10.1016/0019-1035(89)90160-7}

\bibitem[{{Metropolis} {et~al.}(1953){Metropolis}, {Rosenbluth}, {Rosenbluth}, {Teller}, \& {Teller}}]{1953JChPh..21.1087M}
{Metropolis}, N., {Rosenbluth}, A.~W., {Rosenbluth}, M.~N., {Teller}, A.~H., \& {Teller}, E. 1953, \jcp, 21, 1087, \dodoi{10.1063/1.1699114}

\bibitem[{{Morley} {et~al.}(2024){Morley}, {Mukherjee}, {Marley}, {Fortney}, {Visscher}, {Lupu}, {Gharib-Nezhad}, {Thorngren}, {Freedman}, \& {Batalha}}]{2024ApJ...975...59M}
{Morley}, C.~V., {Mukherjee}, S., {Marley}, M.~S., {et~al.} 2024, \apj, 975, 59, \dodoi{10.3847/1538-4357/ad71d5}

\bibitem[{{Mukherjee} {et~al.}(2023){Mukherjee}, {Batalha}, {Fortney}, \& {Marley}}]{2023ApJ...942...71M}
{Mukherjee}, S., {Batalha}, N.~E., {Fortney}, J.~J., \& {Marley}, M.~S. 2023, \apj, 942, 71, \dodoi{10.3847/1538-4357/ac9f48}

\bibitem[{{Mukherjee} {et~al.}(2024){Mukherjee}, {Fortney}, {Morley}, {Batalha}, {Marley}, {Karalidi}, {Visscher}, {Lupu}, {Freedman}, \& {Gharib-Nezhad}}]{ELFOWL}
{Mukherjee}, S., {Fortney}, J.~J., {Morley}, C.~V., {et~al.} 2024, \apj, 963, 73, \dodoi{10.3847/1538-4357/ad18c2}

\bibitem[{{Nakajima} {et~al.}(2004){Nakajima}, {Tsuji}, \& {Yanagisawa}}]{2004ApJ...607..499N}
{Nakajima}, T., {Tsuji}, T., \& {Yanagisawa}, K. 2004, \apj, 607, 499, \dodoi{10.1086/383299}

\bibitem[{{Oreshenko} {et~al.}(2020){Oreshenko}, {Kitzmann}, {M{\'a}rquez-Neila}, {Malik}, {Bowler}, {Burgasser}, {Sznitman}, {Fisher}, \& {Heng}}]{2020AJ....159....6O}
{Oreshenko}, M., {Kitzmann}, D., {M{\'a}rquez-Neila}, P., {et~al.} 2020, \aj, 159, 6, \dodoi{10.3847/1538-3881/ab5955}

\bibitem[{{Patience} {et~al.}(2012){Patience}, {King}, {De Rosa}, {Vigan}, {Witte}, {Rice}, {Helling}, \& {Hauschildt}}]{2012A&A...540A..85P}
{Patience}, J., {King}, R.~R., {De Rosa}, R.~J., {et~al.} 2012, \aap, 540, A85, \dodoi{10.1051/0004-6361/201118058}

\bibitem[{Pedregosa {et~al.}(2011)Pedregosa, Varoquaux, Gramfort, Michel, Thirion, Grisel, Blondel, Prettenhofer, Weiss, Dubourg, Vanderplas, Passos, Cournapeau, Brucher, Perrot, \& Duchesnay}]{scikit-learn2011}
Pedregosa, F., Varoquaux, G., Gramfort, A., {et~al.} 2011, Journal of Machine Learning Research, 12, 2825

\bibitem[{{Rayner} {et~al.}(2003){Rayner}, {Toomey}, {Onaka}, {Denault}, {Stahlberger}, {Vacca}, {Cushing}, \& {Wang}}]{2003PASP..115..362R}
{Rayner}, J.~T., {Toomey}, D.~W., {Onaka}, P.~M., {et~al.} 2003, \pasp, 115, 362

\bibitem[{{Reis} {et~al.}(2019){Reis}, {Baron}, \& {Shahaf}}]{Reis2019AJ....157...16R}
{Reis}, I., {Baron}, D., \& {Shahaf}, S. 2019, \aj, 157, 16, \dodoi{10.3847/1538-3881/aaf101}

\bibitem[{{Sanghi} {et~al.}(2023){Sanghi}, {Liu}, {Best}, {Dupuy}, {Siverd}, {Zhang}, {Hurt}, {Magnier}, {Aller}, \& {Deacon}}]{2023ApJ...959...63S}
{Sanghi}, A., {Liu}, M.~C., {Best}, W. M.~J., {et~al.} 2023, \apj, 959, 63, \dodoi{10.3847/1538-4357/acff66}

\bibitem[{Santner {et~al.}(2003)Santner, Williams, Notz, \& Williams}]{santner2003design}
Santner, T.~J., Williams, B.~J., Notz, W.~I., \& Williams, B.~J. 2003, The design and analysis of computer experiments, Vol.~1 (Springer)

\bibitem[{{Saumon} \& {Marley}(2008)}]{2008ApJ...689.1327S}
{Saumon}, D., \& {Marley}, M.~S. 2008, \apj, 689, 1327, \dodoi{10.1086/592734}

\bibitem[{{Saumon} {et~al.}(2006){Saumon}, {Marley}, {Cushing}, {Leggett}, {Roellig}, {Lodders}, \& {Freedman}}]{2006ApJ...647..552S}
{Saumon}, D., {Marley}, M.~S., {Cushing}, M.~C., {et~al.} 2006, \apj, 647, 552, \dodoi{10.1086/505419}

\bibitem[{{Schneider} {et~al.}(2023{\natexlab{a}}){Schneider}, {Munn}, {Vrba}, {Bruursema}, {Dahm}, {Williams}, {Liu}, \& {Dorland}}]{2023AJ....166..103S}
{Schneider}, A.~C., {Munn}, J.~A., {Vrba}, F.~J., {et~al.} 2023{\natexlab{a}}, \aj, 166, 103, \dodoi{10.3847/1538-3881/ace9bf}

\bibitem[{{Schneider} {et~al.}(2020){Schneider}, {Burgasser}, {Gerasimov}, {Marocco}, {Gagn{\'e}}, {Goodman}, {Beaulieu}, {Pendrill}, {Rothermich}, {Sainio}, {Kuchner}, {Caselden}, {Meisner}, {Faherty}, {Mamajek}, {Hsu}, {Greco}, {Cushing}, {Kirkpatrick}, {Bardalez-Gagliuffi}, {Logsdon}, {Allers}, {Debes}, \& {Backyard Worlds: Planet 9 Collaboration}}]{2020ApJ...898...77S}
{Schneider}, A.~C., {Burgasser}, A.~J., {Gerasimov}, R., {et~al.} 2020, \apj, 898, 77, \dodoi{10.3847/1538-4357/ab9a40}

\bibitem[{{Schneider} {et~al.}(2023{\natexlab{b}}){Schneider}, {Burgasser}, {Bruursema}, {Munn}, {Vrba}, {Caselden}, {Kabatnik}, {Rothermich}, {Sainio}, {Bickle}, {Dahm}, {Meisner}, {Kirkpatrick}, {Su{\'a}rez}, {Gagn{\'e}}, {Faherty}, {Vos}, {Kuchner}, {Williams}, {Bardalez Gagliuffi}, {Aganze}, {Hsu}, {Theissen}, {Cushing}, {Marocco}, {Casewell}, \& {Backyard Worlds: Planet 9 Collaboration}}]{2023ApJ...943L..16S}
{Schneider}, A.~C., {Burgasser}, A.~J., {Bruursema}, J., {et~al.} 2023{\natexlab{b}}, \apjl, 943, L16, \dodoi{10.3847/2041-8213/acb0cd}

\bibitem[{{Shkolnik} {et~al.}(2009){Shkolnik}, {Liu}, \& {Reid}}]{2009ApJ...699..649S}
{Shkolnik}, E., {Liu}, M.~C., \& {Reid}, I.~N. 2009, \apj, 699, 649, \dodoi{10.1088/0004-637X/699/1/649}

\bibitem[{Sisson {et~al.}(2018)Sisson, Fan, \& Beaumont}]{Sisson2018handbook}
Sisson, S.~A., Fan, Y., \& Beaumont, M. 2018, Handbook of approximate Bayesian computation (CRC Press)

\bibitem[{{Skrutskie} {et~al.}(2006){Skrutskie}, {Cutri}, {Stiening}, {Weinberg}, {Schneider}, {Carpenter}, {Beichman}, {Capps}, {Chester}, {Elias}, {Huchra}, {Liebert}, {Lonsdale}, {Monet}, {Price}, {Seitzer}, {Jarrett}, {Kirkpatrick}, {Gizis}, {Howard}, {Evans}, {Fowler}, {Fullmer}, {Hurt}, {Light}, {Kopan}, {Marsh}, {McCallon}, {Tam}, {Van Dyk}, \& {Wheelock}}]{2006AJ....131.1163S}
{Skrutskie}, M.~F., {Cutri}, R.~M., {Stiening}, R., {et~al.} 2006, \aj, 131, 1163, \dodoi{10.1086/498708}

\bibitem[{{Sorahana} \& {Yamamura}(2014)}]{2014ApJ...793...47S}
{Sorahana}, S., \& {Yamamura}, I. 2014, \apj, 793, 47, \dodoi{10.1088/0004-637X/793/1/47}

\bibitem[{{Sprague} {et~al.}(2022){Sprague}, {Culhane}, {Kounkel}, {Olney}, {Covey}, {Hutchinson}, {Lingg}, {Stassun}, {Rom{\'a}n-Z{\'u}{\~n}iga}, {Roman-Lopes}, {Nidever}, {Beaton}, {Borissova}, {Stutz}, {Stringfellow}, {Ram{\'\i}rez}, {Ram{\'\i}rez-Preciado}, {Hern{\'a}ndez}, {Kim}, \& {Lane}}]{2022AJ....163..152S}
{Sprague}, D., {Culhane}, C., {Kounkel}, M., {et~al.} 2022, \aj, 163, 152, \dodoi{10.3847/1538-3881/ac4de7}

\bibitem[{{Stephens} {et~al.}(2009){Stephens}, {Leggett}, {Cushing}, {Marley}, {Saumon}, {Geballe}, {Golimowski}, {Fan}, \& {Noll}}]{2009ApJ...702..154S}
{Stephens}, D.~C., {Leggett}, S.~K., {Cushing}, M.~C., {et~al.} 2009, \apj, 702, 154, \dodoi{10.1088/0004-637X/702/1/154}

\bibitem[{{Stephenson}(1986)}]{1986AJ.....92..139S}
{Stephenson}, C.~B. 1986, \aj, 92, 139, \dodoi{10.1086/114146}

\bibitem[{{Struve} \& {Franklin}(1955)}]{1955ApJ...121..337S}
{Struve}, O., \& {Franklin}, K.~L. 1955, \apj, 121, 337, \dodoi{10.1086/145993}

\bibitem[{{Tremblin} {et~al.}(2016){Tremblin}, {Amundsen}, {Chabrier}, {Baraffe}, {Drummond}, {Hinkley}, {Mourier}, \& {Venot}}]{2016ApJ...817L..19T}
{Tremblin}, P., {Amundsen}, D.~S., {Chabrier}, G., {et~al.} 2016, \apjl, 817, L19, \dodoi{10.3847/2041-8205/817/2/L19}

\bibitem[{{Tremblin} {et~al.}(2015){Tremblin}, {Amundsen}, {Mourier}, {Baraffe}, {Chabrier}, {Drummond}, {Homeier}, \& {Venot}}]{2015ApJ...804L..17T}
{Tremblin}, P., {Amundsen}, D.~S., {Mourier}, P., {et~al.} 2015, \apjl, 804, L17, \dodoi{10.1088/2041-8205/804/1/L17}

\bibitem[{{van der Walt} {et~al.}(2011){van der Walt}, {Colbert}, \& {Varoquaux}}]{2011CSE....13b..22V}
{van der Walt}, S., {Colbert}, S.~C., \& {Varoquaux}, G. 2011, Computing in Science and Engineering, 13, 22, \dodoi{10.1109/MCSE.2011.37}

\bibitem[{{van Leeuwen}(2007)}]{2007AA...474..653V}
{van Leeuwen}, F. 2007, \aap, 474, 653, \dodoi{10.1051/0004-6361:20078357}

\bibitem[{{Vasist} {et~al.}(2023){Vasist}, {Rozet}, {Absil}, {Molli{\`e}re}, {Nasedkin}, \& {Louppe}}]{2023A&A...672A.147V}
{Vasist}, M., {Rozet}, F., {Absil}, O., {et~al.} 2023, \aap, 672, A147, \dodoi{10.1051/0004-6361/202245263}

\bibitem[{{Veyette} {et~al.}(2016){Veyette}, {Muirhead}, {Mann}, \& {Allard}}]{2016ApJ...828...95V}
{Veyette}, M.~J., {Muirhead}, P.~S., {Mann}, A.~W., \& {Allard}, F. 2016, \apj, 828, 95, \dodoi{10.3847/0004-637X/828/2/95}

\bibitem[{{Virtanen} {et~al.}(2020){Virtanen}, {Gommers}, {Oliphant}, {Haberland}, {Reddy}, {Cournapeau}, {Burovski}, {Peterson}, {Weckesser}, {Bright}, {van der Walt}, {Brett}, {Wilson}, {Millman}, {Mayorov}, {Nelson}, {Jones}, {Kern}, {Larson}, {Carey}, {Polat}, {Feng}, {Moore}, {VanderPlas}, {Laxalde}, {Perktold}, {Cimrman}, {Henriksen}, {Quintero}, {Harris}, {Archibald}, {Ribeiro}, {Pedregosa}, {van Mulbregt}, \& {SciPy 1. 0 Contributors}}]{2020NatMe..17..261V}
{Virtanen}, P., {Gommers}, R., {Oliphant}, T.~E., {et~al.} 2020, Nature Methods, 17, 261, \dodoi{10.1038/s41592-019-0686-2}

\bibitem[{{Waldmann}(2016)}]{2016ApJ...820..107W}
{Waldmann}, I.~P. 2016, \apj, 820, 107, \dodoi{10.3847/0004-637X/820/2/107}

\bibitem[{{W}es {M}c{K}inney(2010)}]{mckinney-proc-scipy-2010}
{W}es {M}c{K}inney. 2010, in {P}roceedings of the 9th {P}ython in {S}cience {C}onference, ed. {S}t\'efan van~der {W}alt \& {J}arrod {M}illman, 56 -- 61, \dodoi{10.25080/Majora-92bf1922-00a}

\bibitem[{{Xuan} {et~al.}(2024){Xuan}, {Hsu}, {Finnerty}, {Wang}, {Ruffio}, {Zhang}, {Knutson}, {Mawet}, {Mamajek}, {Inglis}, {Wallack}, {Bryan}, {Blake}, {Molli{\`e}re}, {Hejazi}, {Baker}, {Bartos}, {Calvin}, {Cetre}, {Delorme}, {Doppmann}, {Echeverri}, {Fitzgerald}, {Jovanovic}, {Liberman}, {L{\'o}pez}, {Morris}, {Pezzato}, {Sappey}, {Schofield}, {Skemer}, {Wallace}, {Wang}, {Agrawal}, \& {Horstman}}]{2024ApJ...970...71X}
{Xuan}, J.~W., {Hsu}, C.-C., {Finnerty}, L., {et~al.} 2024, \apj, 970, 71, \dodoi{10.3847/1538-4357/ad4796}

\bibitem[{{York} {et~al.}(2000){York}, {Adelman}, {Anderson}, {Anderson}, {Annis}, {Bahcall}, {Bakken}, {Barkhouser}, {Bastian}, {Berman}, {Boroski}, {Bracker}, {Briegel}, {Briggs}, {Brinkmann}, {Brunner}, {Burles}, {Carey}, {Carr}, {Castander}, {Chen}, {Colestock}, {Connolly}, {Crocker}, {Csabai}, {Czarapata}, {Davis}, {Doi}, {Dombeck}, {Eisenstein}, {Ellman}, {Elms}, {Evans}, {Fan}, {Federwitz}, {Fiscelli}, {Friedman}, {Frieman}, {Fukugita}, {Gillespie}, {Gunn}, {Gurbani}, {de Haas}, {Haldeman}, {Harris}, {Hayes}, {Heckman}, {Hennessy}, {Hindsley}, {Holm}, {Holmgren}, {Huang}, {Hull}, {Husby}, {Ichikawa}, {Ichikawa}, {Ivezi{\'c}}, {Kent}, {Kim}, {Kinney}, {Klaene}, {Kleinman}, {Kleinman}, {Knapp}, {Korienek}, {Kron}, {Kunszt}, {Lamb}, {Lee}, {Leger}, {Limmongkol}, {Lindenmeyer}, {Long}, {Loomis}, {Loveday}, {Lucinio}, {Lupton}, {MacKinnon}, {Mannery}, {Mantsch}, {Margon}, {McGehee}, {McKay}, {Meiksin}, {Merelli}, {Monet}, {Munn}, {Narayanan}, {Nash}, {Neilsen}, {Neswold}, {Newberg}, {Nichol}, {Nicinski},
  {Nonino}, {Okada}, {Okamura}, {Ostriker}, {Owen}, {Pauls}, {Peoples}, {Peterson}, {Petravick}, {Pier}, {Pope}, {Pordes}, {Prosapio}, {Rechenmacher}, {Quinn}, {Richards}, {Richmond}, {Rivetta}, {Rockosi}, {Ruthmansdorfer}, {Sandford}, {Schlegel}, {Schneider}, {Sekiguchi}, {Sergey}, {Shimasaku}, {Siegmund}, {Smee}, {Smith}, {Snedden}, {Stone}, {Stoughton}, {Strauss}, {Stubbs}, {SubbaRao}, {Szalay}, {Szapudi}, {Szokoly}, {Thakar}, {Tremonti}, {Tucker}, {Uomoto}, {Vanden Berk}, {Vogeley}, {Waddell}, {Wang}, {Watanabe}, {Weinberg}, {Yanny}, {Yasuda}, \& {SDSS Collaboration}}]{2000AJ....120.1579Y}
{York}, D.~G., {Adelman}, J., {Anderson}, Jr., J.~E., {et~al.} 2000, \aj, 120, 1579, \dodoi{10.1086/301513}

\bibitem[{{Zakhozhay} {et~al.}(2023){Zakhozhay}, {Osorio}, {B{\'e}jar}, {Climent}, {Guirado}, {Gauza}, {Lodieu}, {Semenov}, {Perez-Torres}, {Azulay}, {Rebolo}, {Mart{\'\i}n-Pintado}, \& {Lef{\`e}vre}}]{2023AA...674A..66Z}
{Zakhozhay}, O.~V., {Osorio}, M. R.~Z., {B{\'e}jar}, V. J.~S., {et~al.} 2023, \aap, 674, A66, \dodoi{10.1051/0004-6361/202345944}

\bibitem[{{Zhang} {et~al.}(2024){Zhang}, {Liu}, \& {Zhang}}]{2024ApJ...960..105Z}
{Zhang}, R., {Liu}, M.~C., \& {Zhang}, Z. 2024, \apj, 960, 105, \dodoi{10.3847/1538-4357/ad083c}

\bibitem[{{Zhang} {et~al.}(2021{\natexlab{a}}){Zhang}, {Liu}, {Marley}, {Line}, \& {Best}}]{2021ApJ...916...53Z}
{Zhang}, Z., {Liu}, M.~C., {Marley}, M.~S., {Line}, M.~R., \& {Best}, W. M.~J. 2021{\natexlab{a}}, \apj, 916, 53, \dodoi{10.3847/1538-4357/abf8b2}

\bibitem[{{Zhang} {et~al.}(2021{\natexlab{b}}){Zhang}, {Liu}, {Marley}, {Line}, \& {Best}}]{2021ApJ...921...95Z}
---. 2021{\natexlab{b}}, \apj, 921, 95, \dodoi{10.3847/1538-4357/ac0af7}

\bibitem[{{Zhang} {et~al.}(2020){Zhang}, {Liu}, {Hermes}, {Magnier}, {Marley}, {Tremblay}, {Tucker}, {Do}, {Payne}, \& {Shappee}}]{2020ApJ...891..171Z}
{Zhang}, Z., {Liu}, M.~C., {Hermes}, J.~J., {et~al.} 2020, \apj, 891, 171, \dodoi{10.3847/1538-4357/ab765c}

\end{thebibliography}
\bibliographystyle{aasjournal}

\end{document}